\pdfoutput=1
\documentclass[11pt,a4paper]{article}

\RequirePackage[utf8]{inputenc}

\usepackage{amsfonts}
\usepackage{amssymb}
\usepackage{comment}
\usepackage{graphicx}
\usepackage{amsmath}
\usepackage{tabu}
\usepackage{dsfont}
\usepackage{float}
\usepackage{amsthm}
\usepackage{slashed}
\usepackage{setspace}
\usepackage[labelformat=simple]{subcaption}

\usepackage{mathrsfs}

\usepackage[utf8]{inputenc}
\usepackage{amsmath, amsthm, amssymb, amsfonts}
\usepackage[font=small, labelfont=bf]{caption}
\usepackage{amsmath, amsthm, amssymb, amsfonts}
\usepackage{multirow}
\usepackage{graphicx}
\usepackage{float}
\usepackage[numbers,sort&compress]{natbib}
\usepackage{jheppub}
\usepackage{enumerate}

\def\bR{\mathbb{R}}

\def\bZ{\mathbb{Z}}

\newcommand{\nn}{\nonumber}

\newcommand{\p}{\partial}

\newcommand{\cY}{\mathcal{Y}}

\newcommand{\dif}{\,\text{d}}

\newcommand{\cT}{\mathcal{T}}
\newcommand{\cD}{\mathcal{D}}
\newcommand{\cA}{\mathcal{A}}
\newcommand{\raw}{\rightarrow}
\newcommand{\cN}{\mathcal{N}}

\newcommand{\cJ}{\mathcal{J}}
\newcommand{\cL}{\mathcal{L}}
\newcommand{\cV}{\mathcal{V}}
\newcommand{\cR}{\mathcal{R}}

\newcommand{\cO}{\mathcal{O}}
\newcommand{\cP}{\mathcal{P}}
\newcommand{\ee}{\mathrm{e}}
\newcommand{\tr}{\mathrm{Tr}}
\newcommand{\I}{\mathrm{i}}
\newcommand{\im}{\mathrm{Im}}
\newcommand{\gs}{g_{\mathrm{s}}}
\newcommand{\kom}{\, ,\quad }

\newcommand{\fft}[2]{\frac{#1}{#2}}

\newcommand{\cK}{\mathcal{K}}
\newcommand{\cW}{\mathcal{W}}

\newcommand{\nc}{\newcommand}
\nc{\beq}{\begin{equation}}
\nc{\eeq}{\end{equation}}
\nc{\bea}{\begin{eqnarray}}
\nc{\eea}{\end{eqnarray}}

\def\ov{\overline}

\allowdisplaybreaks[2]
\numberwithin{equation}{section}

\usepackage[framemethod=TikZ]{mdframed}
\mdfsetup{skipabove=\topskip, skipbelow=\topskip}

\usepackage{jheppub}

\newcounter{equ}[section]

\newcounter{Boxequ}[section]
\newenvironment{Boxequ}[1][]{%
\mdfsetup{innertopmargin=10pt,linecolor=blue!20,%
middlelinewidth=2pt,topline=true,}
\begin{mdframed}[]\relax%
\vspace{-0.1cm}
}{\end{mdframed}}

\title{Type IIB at eight derivatives: insights from Superstrings, Superfields and Superparticles}

\author[a]{James T. Liu,}
\author[b]{Ruben Minasian,}
\author[c]{Raffaele Savelli,}
\author[d]{Andreas Schachner}

\affiliation[a]{Leinweber Center for Theoretical Physics, Randall Laboratory of Physics, University of Michigan,
Ann Arbor, MI 48109-1040, USA}
\affiliation[b]{Institut de Physique Th{\'e}orique, Universit{\'e} Paris Saclay, CNRS, CEA, 91191 Gif-sur-Yvette Cedex, France}
\affiliation[c]{Dipartimento di Fisica and Sezione INFN, Universit{\`a} di Roma ``Tor Vergata'', via della Ricerca
Scientifica 1, I-00133 Roma, Italy}
\affiliation[d]{DAMTP, Centre for Mathematical Sciences, Wilberforce Road, Cambridge, CB3 0WA, UK.}

\emailAdd{jimliu@umich.edu}
\emailAdd{ruben.minasian@ipht.fr}
\emailAdd{raffaele.savelli@roma2.infn.it}
\emailAdd{as2673@maths.cam.ac.uk}

\abstract{
We study the non-linear structure of Type IIB eight-derivative couplings involving the metric and the complexified three-form $G_3$.
We show that, at the level of five-point string amplitudes, the kinematics in the maximally R-symmetry-violating sector is fully matched  by  standard superspace integrals and by superparticle amplitudes in M-theory on a two-torus. The latter approach is used to determine the complete effective action in this sector and to verify its invariance under SL$(2,\mathbb{Z})$ duality. We further comment on the general structure of the higher-point kinematics. Compactifications to lower dimensions provide both tests for our results and the arena for their applications. We verify that K3 reductions are fully consistent with the constraints of six-dimensional supersymmetry, and derive the four-dimensional flux scalar potential and axion kinetic terms at order $(\alpha^{\prime})^{3}$ in Calabi-Yau threefold reductions.

}


\usepackage{todonotes}

\begin{document}

\maketitle

\bigskip

\newpage

\section{Introduction and summary}

Despite the long history of studying perturbative higher-derivative corrections to the ten-dimensional effective actions \cite{Gross:1986mw, Gross:1986iv}, understanding their complete structure, notably non-linear completions arising form higher-point functions, remains a key challenge for string theory.
While recent years were marked by conceptual and computational breakthroughs in our ability to compute string amplitudes,  a unifying approach or a guiding principle towards completing the effective action at higher order in the $\alpha^{\prime}$ and loop expansion has not emerged yet.
The main outstanding issue is the construction of local effective actions reproducing the precise amplitude, where a plethora of kinematical structures, intricate pole-subtraction procedures and field redefinitions complicate any bootstrapping attempt.

In the NSNS sector, a partial set of quintic and some sextic higher-derivative terms in the Type II effective action have been identified  \cite{Peeters:2001ub, Liu:2013dna,Liu:2019ses}.
For the Type IIB effective actions, 
the authors of \cite{Policastro:2006vt,Policastro:2008hg} completed the quartic couplings for all fields but the four-form tensor field with anti-self-dual tensor field strength $F_5$. In the absence of the dilaton and the complex $3$-form on the Type IIB side, the full action has been determined in \cite{Green:2003an,Paulos:2008tn}.
Here, the supersymmetric completion of $R^{4}$ including the $F_{5}$ field strength \cite{deHaro:2002vk,Green:2003an,Rajaraman:2005ag} has been inferred from the $\cN=2$ superspace approach of \cite{Howe:1983sra}.
Particularly noteworthy are applications of these results to multi D3-brane backgrounds \cite{Green:2003an}
and black-hole solutions with $\mathrm{AdS}_{5}\times S^{5}$ asymptotics in AdS/CFT \cite{Paulos:2008tn,Melo:2020amq}.
Recently a full  completion of the NSNS-sector of the effective action at tree level has been obtained in \cite{Garousi:2020lof} using constraints imposed by T-duality invariance, if not directly string theory.
However, given that the choice of field basis is rather unnatural from the point of view of tensor structures like $t_{8}$ or $\epsilon_{10}$ appearing in string vertex operators or superspace integrals,
it is hard to make a straightforward comparison, let alone extend the results to the RR-sector or to $1$-loop order.

\medskip

Up to now, these different developments have not been put together in a systematic fashion, and
no unifying approach towards completing string effective actions has been proposed. One would hope that a proper framing of effective actions with quantum corrections in terms of some generalised or super-geometry should emerge and eventually be helpful in constraining if not predicting the higher-order interactions. So far, the usefulness of connections with torsion given by the NSNS three-form $H_3$, $\Omega_{\pm} = \Omega^{\mbox{\tiny LC}} \pm \frac12 H$ in terms of the Levi-Civita connection $\Omega^{\mbox{\tiny LC}}$, observed at the linearised level \cite{Gross:1986iv}, has been confirmed at the non-linear level as well  \cite{Liu:2013dna,Liu:2019ses}. However it has also been shown that a simple replacement of $\Omega^{\mbox{\tiny LC}}$ by a connection with torsion fails to capture the full kinematics of eight-derivative terms at one-loop and especially at tree-level. 

In this paper, we make progress towards developing such an approach  by scrutinising the kinematical structures discovered in \cite{Liu:2019ses}. We compare the results of conventional ten-dimensional ($10$D) superstring amplitudes with eleven-dimensional ($11$D) superparticle amplitudes compactified on a two-torus \cite{Green:1997as,Green:1997me,Green:1999by}, and superspace approaches to Type IIB $\cN=2$ supergravity \cite{Howe:1983sra}.
In the effective action, we find that couplings in the maximally $\mathrm{U}(1)$-violating (MUV) sector of $\mathrm{U}(1)$ (or R-symmetry) charge $|Q_{\text{max}}|=2(P-4)$ at the level of $P$-point amplitudes are captured by simple fundamental higher-dimensional index structures generalising the well-known $t_{8}$ or $\epsilon_{D}$ tensors.
Our results for the 10D action at 5-point level including the metric and complexified three-form flux $G_3$ (up to $\nabla G_{3}$ terms) can then be summarised as 
\begin{align}\label{eq:FullResultG2R3} 
\cL&=f_{0}(\tau,\bar{\tau})t_{16}R^{4}+\dfrac{3}{2} \left (f_{1}(\tau,\bar{\tau})t_{18}G^{2}_{3}R^{3}+f_{-1}(\tau,\bar{\tau})t_{18}\ov G^{2}_{3}R^{3}\right )\nn\\[0.5em]
&\quad +f_{0}(\tau,\bar{\tau})\left [T(\epsilon_{10},t_{8})-t_{18}  \right ]|G_{3}|^{2}R^{3}\, .
\end{align}
The first line encodes the MUV couplings which are nicely repackaged into a single index structure $t_{N}$ (with $N= 16+2w$  for couplings of the form $G_{3}^{2w}R^{4-w}$).
In comparison to \cite{Liu:2019ses},
this means an extraordinary simplification of the $5$-point results.

We have chosen to split  the non-MUV sector in \eqref{eq:FullResultG2R3}, and write separately 
the kinematical structure $T(\epsilon_{10},t_{8})$, which is a function of the standard tensors $t_{8}$ and $\epsilon_{10}$ and  a factor of $-t_{18}$. The latter cannot be written in terms of $t_{8}$ or $\epsilon_{10}$, as observed in \cite{Liu:2019ses} where these terms were simply given by the expansion in the full basis of $H^2R^3$ terms. Stated differently,
$t_{18}$ is not directly seen in string amplitudes, given that it is not directly built into the vertex operators. As we will see, an a posteriori justification of such a split in the non-MUV sector comes from the fact that the $t_{18}$ piece plays an important role in Calabi-Yau threefold reductions to four dimensions.

\medskip 

In Type IIB,
$\mathrm{SL}(2,\mathbb{Z})$ invariance dictates that the coefficients of higher-derivative terms are written in terms of modular forms.
The appearance of the modular function $f_{0}$, the non-holomorphic Eisenstein series of weight $3/2$, was first observed in \cite{Green:1997tv,Green:1997as}, while other non-holomorphic modular forms $f_{w}$ first appeared in \cite{Kehagias:1997jg,Green:1997me}, see also \cite{Alessio:2021krn} for a more recent discussion.
For $R^{4}$,
leading order D-instanton calculations \cite{Sen:2021tpp,Sen:2021jbr} confirmed this result.
Moreover, the amount of supersymmetry in 10D is a powerful tool to relate various higher-derivative terms in the $\alpha^{\prime}$ expansion \cite{Green:1998by}.
While linearised SUSY is powerful enough to predict the existence of higher-derivative terms,
it is incapable of explaining either the presence of the coefficient functions $f_{w}$ or the tensor structures in the non-MUV sector.
These coefficient functions can be derived, instead, by studying their origin in M-theory, integrating out towers of winding modes on $T^{2}$ of vanishing volume, in a light-cone worldline formalism for the 11D superparticle \cite{Green:1997as,Green:1997me,Green:1999by}.
We compute such amplitudes in 11D and, besides deriving explicitly the axio-dilaton dependent coefficients, we are able to reproduce exactly the kinematics in the MUV sector of \eqref{eq:FullResultG2R3}, as expected from the superspace approach.



\medskip

Going beyond $5$ points,
we prove that MUV couplings (as well as a specific subset of non-MUV couplings) of the $3$-form, the $5$-form, and the metric in the effective action are given by
\begin{equation}\label{eq:MUVResultIntroFull} 
\cL^{\text{MUV}}= \sum_{w=0}^{4}\, C_{w}\, f_{w}(\tau,\bar{\tau})\; t_{24}G_{3}^{2w}\cR^{4-w}+\text{c.c.}
\end{equation}
in terms of numerical coefficients $C_{w}$ and the 6-index tensor $\cR$
\begin{equation} \label{eq:CalR}
\cR=R+\I \nabla F_{5}+F_{5}^{2}+|G_{3}|^{2}\, .
\end{equation}
The tensor $\cR$ is tightly constrained by non-linear supersymmetry by appearing at $\Theta^{4}$ in a (non-linear) scalar superfield.
The derivation of \eqref{eq:MUVResultIntroFull} is based on all the three aforementioned approaches:
\begin{itemize}
\item \emph{superstring amplitudes}: up to $5$-points, \eqref{eq:MUVResultIntroFull} essentially reduces to the full first and a subset of terms in the second line of \eqref{eq:FullResultG2R3}.
At $6$-points, \cite{Green:2019rhz} provides the coefficient $C_{2}$ from tree-level pure spinor amplitudes.
Higher amplitudes are in principle available, but determining the structure of contact terms in \eqref{eq:MUVResultIntroFull} is currently out of reach.
\item \emph{superfields}: the string kinematics is easily determined from $16$-fermion integrals, thereby making the existence of a single unifying index structure obvious.
In principle, the coefficients $C_{w}$ can be determined from supersymmetry/geometry following e.g.~\cite{Green:1998by}, but we do not follow this approach here.
\item \emph{superparticles}: the structure of MUV amplitudes of M-theory compactified on $T^{2}$ is actually simple enough not only to reproduce \eqref{eq:MUVResultIntroFull} kinematically,
but also to derive the $f_{w}(\tau,\bar{\tau})$ alongside the $C_{w}$ from first principles. This proves the higher power of this approach in determining the MUV effective couplings at any order.
\end{itemize}

\noindent
The most significant takeaway message from \eqref{eq:MUVResultIntroFull} is that it unifies $46$ individual tensor structures in such a way that they are kinematically captured by \emph{a single index structure $t_{24}$}.
Indeed,
$t_{24}$ is the largest structure from which all other tensors $t_{N}$ with $N<24$ relevant for this paper can be constructed upon suitable metric contractions.
For a given weight $w$,
we find
\begin{enumerate}
\item $w=0$: The generalised $R^{4}$ term corresponding to $f_{0}t_{24}\cR^{4}$ was inferred in \cite{deHaro:2002vk,Green:2003an,Rajaraman:2005ag,Paulos:2008tn} in the absence of the $|G_{3}|^{2}$ term.
At the level of $5$-point contact terms,
we found evidence for the coupling $|G_{3}|^{2}$ inside $\cR$, which is again obtained from $f_{0}t_{24}\cR^{4}$ upon expanding to quadratic order in the $3$-form.
Given that the relative coefficients inside $\cR$ are determined by supersymmetry,
we provide further contact with a supersymmetric completion of the $R^{4}$ coupling in the presence of a non-trivial $G_{3}$ background.
\item $w=1$: The part $f_{1}t_{24} G_{3}^{2}\cR^{3}$ reduces at 5 points to the second term in the first line of \eqref{eq:FullResultG2R3}.
From the superfield perspective,
the replacement $R\raw \cR$ is completely justified by supersymmetry, even though there might be further higher-order contributions in the non-MUV sector, just as for $|G_{3}|^{2}R^{3}$.
\item $w\geq 2$: We utilise the 11D superparticle to predict the string coefficients of MUV amplitudes beyond five points.
The $w=2$ coefficient $C_{2}$ matches the predictions of \cite{Green:2019rhz} at the level of six-point pure spinor amplitudes.
Moreover,
the higher order coefficients are in agreement with expectations from modular invariance of the Type IIB superstring.
\end{enumerate}

\noindent
Our results are reminiscent of the MUV amplitudes computed in \cite{Green:2019rhz} based on the spinor helicity formalism of \cite{Caron-Huot:2010nes,Boels:2012ie}.
It was argued there that general MUV amplitudes appear without any poles (i.e.~they \emph{are} contact terms),
are represented by a superfield which matches the linearised on-shell superfield of Type IIB supergravity,
and violate the $\mathrm{U}(1)$ by $2(P-4)$ units of charge (see also \cite{Boels:2012zr,Boels:2013jua,Wang:2015jna}).

\medskip
We conclude our analysis by two basic lower-dimensional consistency checks of our findings. 
When compactifying our proposed 10D action on a K3 to six dimensions, we show how
 several non-trivial cancellations among the various $5$-point index structures ensure consistency with the constraints imposed by $\cN=(2,0)$ supersymmetry in 6D.
 When reducing the 10D action to four dimensions on a Calabi-Yau threefold we show that constraints on the $(\alpha^{\prime})^{3}$-corrected flux scalar potential from 4D supersymmetry are also  perfectly matched.
Furthermore,
we derive the 4D kinetic terms for the hypermultiplet scalars, in particular the $C_{2}$/$B_{2}$-axions, at order $(\alpha^{\prime})^{3}$ at string tree and 1-loop level.

\medskip
The paper is organised as follows.
In  Sect.~\ref{sec:TypeIIBIntroAPExp} we review the systematics of $8$-derivative terms in the Type IIB effective action with a particular focus on $\mathrm{SL}(2,\bZ)$-invariance and sixteen fermionic integrals giving rise to higher-dimensional index structures.
Subsequently,  in Sect.~\ref{sec:EightDerFivePointAction},
we demonstrate that such structures play an outstanding role also for $5$-point contact terms of the form $|G_{3}|^{2}R^{3}, G_{3}^{2}R^{3}$ and $\ov G_{3}^{2}R^{3}$.
In Sect.~\ref{sec:EightDerBeyondFivePoint},
we argue that the entire eight-derivative action for couplings of the form $G_{3}^{m}\ov G_{3}^{n}R^{4-w}$, $w=(m+n)/2$, in the MUV sector (i.e. for $m\cdot n =0$) 
is determined by a single index structure obtained from a sixteen fermion integral.
Using supersymmetry, these terms can be partially generalised to include also some non-MUV couplings through replacing $R\raw \cR$. In Sect.~\ref{sec:Compactifications} we apply our proposal for the 10D effective action to compactifications to 6D and to 4D. Finally we list a number of open question that should hopefully be addressed in the near future in Sect.~\ref{sec:Conclusions}. Some technical material is collected in five appendices.

\section{Type IIB supergravity and its $\alpha^{\prime}$-expansion}\label{sec:TypeIIBIntroAPExp}

The classical $10$D effective action reads in Einstein frame
\begin{align}\label{eq:ch1:ActionIIB10D} 
S^{(0)}&=\dfrac{1}{2\kappa_{10}^{2}}\,\int\, \left ( R-2\cP_{M}\overline{\cP}^{M}-\dfrac{ |G_{3}|^{2}}{2\cdot 3!}-\dfrac{ | F_{5}|^{2}}{4\cdot 5!}\right )\star_{10}\mathds{1}+\dfrac{1}{8\I\kappa_{10}^{2}}\int\,  C_{4}\wedge  G_{3}\wedge \overline{ G}_{3}
\end{align}
in terms of the complexified fields 
\begin{equation}\label{eq:TauPGT} 
\tau= C_{0}+\I\mathrm{e}^{-\phi}\kom \cP_{M}=\dfrac{\I \nabla_{M}\tau}{2\im(\tau)}\kom  G_{3}=\dfrac{1}{\sqrt{\im(\tau)}}\left ( F_{3}-\tau H_{3}\right )= \dfrac{\tilde{G}_{3}}{\sqrt{\im(\tau)}}\, .
\end{equation}
where the $p$-form field strengths are defined as
\begin{align}
\label{eq:ch1:FieldStrengthH3}& H_{3}=\dif  B_{2}\kom  F_{1}=\dif  C_{0}\kom F_{3}=\dif  C_{2}\kom F_{5}= \dif C_{4}-\dfrac{1}{2} H_{3}\wedge  C_{2}+\dfrac{1}{2} F_{3}\wedge  B_{2}\, .
\end{align}
In addition to the standard equations of motion, the $5$-form flux must satisfy the self-duality condition $ F_{5}=\star_{10} F_{5}$.

The Type IIB fields form representations under $\mathrm{SL}(2,\bR)\times \mathrm{U}(1)$ with the first being a global, the second being a local symmetry denoted by $Q_{\text{IIB}}$. The various fields have in our convention $\mathrm{U}(1)$-charges 
\begin{equation}\label{eq:UOneChargesTypeIIBFields} 
Q_{\text{IIB}}(\cP)=+2\kom Q_{\text{IIB}}( G_{3})=+1\kom Q_{\text{IIB}}(g_{MN})=Q_{\text{IIB}}( F_{5})=0
\end{equation}
with the opposite charges for the complex conjugates. The complexified scalars parametrise the coset (or moduli) space $\mathrm{SL}(2,\bR)/\mathrm{SO}(2)\cong \mathrm{SU}(1,1)/\mathrm{U}(1)$ \cite{Nahm:1977tg,Green:1982tk,Schwarz:1983wa,Schwarz:1983qr,Howe:1983sra}.
In string theory,
the $\mathrm{U}(1)$ subgroup of $\mathrm{SL}(2,\bR)$ rotating the two supercharges into each other does not leave the superstring invariant: The S-duality $\mathrm{SL}(2,\bZ)$ group survives \cite{Green:1997tv,Green:1997as}.
Under $\mathrm{SL}(2,\bZ)$,
the axio-dilaton transforms according to
\begin{equation}
\tau\raw \dfrac{a\tau+b}{c\tau+d}\kom ad-bc=1\, .
\end{equation}
This implies that
\begin{equation}
\tilde{G}_{3}\raw \dfrac{\tilde{G}_{3}}{c\tau+d}\kom \tau_{2}\raw \dfrac{\tau_{2}}{|c\tau+d|^{2}}
\end{equation}
as well as
\begin{equation}
\cP\raw \dfrac{c\bar{\tau}+d}{c\tau+d}\cP\kom G_{3}\raw \left (\dfrac{c\bar{\tau}+d}{c\tau+d}\right )^{\frac{1}{2}}G_{3}\, .
\end{equation}
More generally,
a combination $\Phi$ of fields with $\mathrm{U}(1)$-charge $Q_{\text{IIB}}=2k$ transforms with weight $k$ so that
\begin{equation}
\Phi\raw\left ( \dfrac{c\bar{\tau}+d}{c\tau+d}\right )^{k} \Phi\, .
\end{equation}
The individual contact terms in the effective action must be invariant under these $\mathrm{SL}(2,\bZ)$ transformations.
For higher derivative terms,
this highly constrains the coefficient functions to be appropriate $\mathrm{SL}(2,\bZ)$-covariant modular forms.

\subsection{Perturbative corrections in $10$D at order $(\alpha^{\prime})^{3}$}

The effective action enjoys a double expansion in terms of $\gs$ (worldsheet topologies / loops in the spacetime theory) and $\alpha^{\prime}$ (loops in the worldsheet theory / higher-derivative terms).
Given that $\alpha^{\prime}$ parametrises the way string theory deviates from a theory of point-like objects,
it is of critical importance for our understand of quantum gravity.

The low-energy description of superstrings is traditionally obtained from string scattering amplitudes of massless string excitations giving rise to an effective field theory description in the limit $\alpha^{\prime}\raw 0$.
Below, we argue that other approaches can be equally effective by employing duality considerations to M-theory and a superspace formalism.

Throughout this paper,
we focus on $8$-derivative couplings where the bosonic action can schematically be written as
\begin{align}\label{eq:AP3Action} 
S^{(3)}&\sim \int\star_{10}\mathds{1}\biggl \{R^{4}+R^{3}\left ( G_{3}^{2}+| G_{3}|^{2}+\overline{ G}_{3}^{2}+ F_{5}^{2}+\ldots\right )\\
&\quad+R^{2}\left (|\nabla G_{3}|^{2}+(\nabla F_{5})^{2}+ G_{3}^{4}+\ldots\right )+R\left ( G_{3}^{6}+\ldots\right )+\left ( G_{3}^{8}+|\nabla G_{3}|^{4}+\ldots\right )\biggl \}\nn\, .
\end{align}
To ensure invariance of the individual terms in \eqref{eq:AP3Action} under $\mathrm{SL}(2,\bZ)\times \mathrm{U}(1)$, each of the contact terms must be multiplied by an appropriate $\mathrm{SL}(2,\bZ)$ covariant function of opposite charge. For our purposes, it suffices to consider the \emph{modular functions}\index{Modular functions}\footnote{At higher orders in the $\alpha^{\prime}$ expansion, more general modular forms have to be introduced, see \cite{Green:2019rhz} for a recent discussions and for further references.}
\begin{equation}\label{eq:DefModFuncWeightK} 
f_{w}(\tau,\bar{\tau})=\sum_{(\hat{l}_{1},\hat{l}_{2})\neq (0,0)}\, \dfrac{\text{Im}(\tau)^{\frac{3}{2}}}{(\hat{l}_{1}+\tau\hat{l}_{2})^{\frac{3}{2}+k}(\hat{l}_{1}+\bar{\tau}\hat{l}_{2})^{\frac{3}{2}-k}}\kom  Q_{\text{IIB}}(f_{w})=-2w\, .
\end{equation}
These forms have special properties collected in App.~\ref{app:ModFunctions} which are quintessential for our investigations. By counting the total $\mathrm{U}(1)$ charge, we can determine which modular function needs to be supplemented to each of the terms in \eqref{eq:AP3Action}. For instance, the uncharged term $R^{4}$ is multiplied by $f_{0}$ which is the non-holomorphic Eisenstein series of weight $3/2$ \cite{Douglas:1996yp,Green:1997tv,Green:1997di,Green:1997tn,Green:1998yf}.
Further constraints on the structure of \eqref{eq:AP3Action} arise from supersymmetry \cite{deRoo:1992sm,Green:1998by}.

\subsection{The quartic effective action}

A complete assessment of string $4$-point amplitudes \cite{Policastro:2006vt,Policastro:2008hg} leads to the quartic action\footnote{Notice that the last term in the first line of eq.~(3.3) in \cite{Policastro:2008hg} is in fact wrong due to $\mathrm{U}(1)$ violation, see in particular \cite{Liu:2019ses} for the corrected result.}
\begin{align}\label{eq:FourPointCouplingsAP3} 
\cL_{4-\text{pt}}^{(3)}&=\alpha\, f_{0}(\tau,\bar{\tau})\biggl \{\cJ_{0}+ \left ( t_{8}t_{8}-\frac{1}{4}\epsilon_{8}\epsilon_{8}\right )\biggl [6R^{2}\left (4|\nabla\cP|^{2}+|\nabla G_{3}|^{2}\right )+24|\nabla \cP|^{2}|\nabla G_{3}|^{2}\nn\\[0.4em]
&\quad+12R\left (\nabla\cP(\nabla\overline{G}_{3})^{2}+\nabla\overline{\cP}(\nabla{G}_{3})^{2}\right )\biggl ]+\cO_{1}\left ((|\nabla\cP|^{2})^{2}\right )+\cO_{2}\left ((|\nabla G_{3}|^{2})^{2}\right )
\biggl\}
\end{align}
where
\begin{equation}\label{eq:DefAlpha} 
 \alpha=\dfrac{(\alpha^{\prime})^{3}}{3\cdot 2^{12}}\, .
\end{equation}
For details concerning the definition of the operators $\cO_{1}$ and $\cO_{2}$, we refer the reader to \cite{Policastro:2008hg,Liu:2019ses} which can also be recovered from an effective $12$D lift \cite{Minasian:2015bxa}.

The well-known $R^{4}$ structure is defined as \cite{Schwarz:1982jn,Gross:1986iv,Sakai:1986bi}
\begin{equation}\label{eq:SuperInvRF} 
\cJ_{0}=\left (t_{8}t_{8}-\dfrac{1}{4}\epsilon_{8}\epsilon_{8}\right )R^{4}\, .
\end{equation}
It is obtained from four closed-string scattering or directly from the worldsheet $\sigma$-model.
While at the level of the $4$-point amplitude only the $t_{8}t_{8}\tilde{R}^{4}$ part in terms of the linearised Riemann tensor $\tilde{R}_{\mu\nu\rho\sigma}=-2\p_{[\mu}h_{\nu][\rho,\sigma]}$ is non-vanishing,
the additional $\epsilon_{8}\epsilon_{8}$ piece can already be inferred from the structure of the $4$-point amplitude and directly verified by computing the odd-odd $5$-point function, cf.~\cite{Liu:2013dna} for a summary.
General covariance dictates that in the purely gravitational sector higher-point graviton amplitudes replace the linearised Riemann tensor $\tilde{R}$ by the full Riemann tensor.

One can show by computing the RHS of \eqref{eq:SuperInvRF} explicitly that $\cJ_{0}$ can be written in terms of the Weyl tensor $C_{MNPQ}$ as \cite{Gross:1986iv,Green:2005qr}
\begin{equation}\label{eq:R4WeylTensor} 
\dfrac{\cJ_{0}}{3\cdot 2^{8}}=-\dfrac{1}{4}C^{MNPQ}C_{MN}\,^{RS}C_{PR}\,^{TU}C_{QSTU}+C^{MNPQ}C_{M}\,^{R}\,_{P}\,^{S}C_{R}\,^{T}\,_{N}\,^{U}C_{STQU}\, .
\end{equation}
The fact that only the Weyl tensor appears as part of \eqref{eq:SuperInvRF} is due to the symmetries in the linearised scalar superfield constructed in App.~\ref{app:Superfield}.

\subsection{Going beyond four points}\label{sec:SixteenFermionIntegral} 

Beyond four points,
the gravitational part of the action is fixed by general covariance.
The full completion including the anti-symmetric tensors and the axio-dilaton remains however an open task.
Generalised geometry provides a hint in this direction by introducing a torsionful connection involving the $B_{2}$-field, cf.~Sect.~\ref{sec:ReviewFlux}.
However, it is confirmed that such an approach does not capture the complete string-theoretic result as verified by $1$-loop $5$-point \cite{Peeters:2001ub,Richards:2008sa} and $6$-point \cite{Liu:2013dna} function computations as well as more recently tree level $5$-point results \cite{Liu:2019ses}.
It is therefore desirable to introduce a new ordering principle that incorporates ideas from supersymmetry and $\mathrm{SL}(2,\mathbb{Z})$-invariance.

The first crucial feature that appears in the higher-point effective action is the presence of non-$\mathrm{U}(1)$-invariant terms. Indeed, for a given number of $P$ fields,
the maximal $\mathrm{U}(1)$ charge satisfies the bound \cite{Green:2019rhz}
\begin{equation}
|Q_{\text{IIB}}|\leq 2(P-4)\, ,
\end{equation}
which is compatible with the fact that the quartic action \eqref{eq:FourPointCouplingsAP3} contains only 
$\mathrm{U}(1)$-preserving terms.

Moreover, although the tensor structures $t_{8}$ and $\epsilon_{10}$ in \eqref{eq:SuperInvRF}, which are very natural from the perspective of string amplitudes, seem to be appropriate representations for the kinematics at 4 points, when one goes to higher points, more fundamental higher-dimensional tensors, generalising $t_8$, make their appearance in the kinematics. 

The higher-derivative action can be constructed from fundamental superspace integrals as established in the 1980s by the seminal works \cite{Nilsson:1981bn,Howe:1983sra,Nilsson:1986rh,Kallosh:1987mb}.
We provide a review of this approach in App.~\ref{app:Superfield}.
The general outcome of this formalism is that the couplings in the effective action are obtained from a single superspace integral
\begin{equation}\label{eq:FermIntegrals} 
t_{3n+2m}=\int\dif^{16}\Theta (\Theta \Gamma^{(3)}\Theta)^{n}(\Theta \Gamma^{(2)}\Theta)^{m}\kom 2n+2m=16\, .
\end{equation}
Here, $\Gamma^{(3)}$ denotes the anti-symmetric product of $3$ $\mathrm{SO}(1,9)$ $\Gamma$-matrices and
\begin{equation}
(\Theta \Gamma^{(2)}\Theta)^{2}=(\Theta \Gamma^{M_{1}M_{2}k}\Theta)(\Theta \Gamma^{N_{1}N_{2}}\,_{k}\Theta)
\end{equation}
so that $t_{3n+2m}$ carries $3n+2m$ indices.
The role of such index structures has previously been discussed in \cite{Peeters:2000qj,Peeters:2001ub,Green:2003an,Green:2005qr,Peeters:2005tb,Policastro:2006vt,Paulos:2008tn}, though their direct manifestation for non-trivial $G_{3}$ and $\nabla \tau$ backgrounds in the string effective action remains largely unexplored.

Before we get to that,
let us review some established aspects of the linearised superspace approach.
Interestingly,
in this formalism the complete kinematics for $R^{4}$ is obtained from a single elementary superspace integral
\begin{equation}
t_{16}R^{4}=\int\dif^{16}\Theta \left [(\Theta\Gamma^{M_{1}M_{2}k}\Theta)(\Theta\Gamma_{k}\,^{M_{3}M_{4}}\Theta)R_{M_{1}M_{2}M_{3}M_{4}}\right ]^{4}\, .
\end{equation}
We find indeed that
\begin{equation}\label{eq:J0T16R4} 
\cJ_{0}=t_{16}R^{4}
\end{equation}
which is consistent with appendix B.2 of \cite{Peeters:2000qj}.
One arrives at a similar result by studying $4$-point functions of light-cone supermembrane vertex operators \cite{Dasgupta:2000df} or in the Green-Schwarz formalism \cite{Green:1997tv}.
Crucially, the above result cannot be extended to the torsionful Riemann tensor as we show in more detail in Sect.~\ref{sec:ExpandingRiemannTorsion}.

The formalism of \cite{Howe:1983sra} already allows to infer additional non-linear couplings.
In fact, the authors of \cite{deHaro:2002vk,Green:2003an} constructed the entire effective action at order $(\alpha^{\prime})^{3}$ for vanishing $G_{3}$ and $\nabla \tau$ from a \emph{single superspace integral}.
To this end,
one defines the $6$-index tensor
\begin{align}\label{eq:DefForR6ModF5} 
\tilde{\cR}_{M_{1}M_{2}M_{3}M_{4}M_{5}M_{6}}&=\dfrac{g_{M_{3}M_{6}}}{8}C_{M_{1}M_{2}M_{4}M_{5}}+\dfrac{\I}{48}\nabla_{M_{1}}F_{M_{2}M_{3}M_{4}M_{5}M_{6}}\nn\\
&\quad+\dfrac{1}{768}\left (F_{M_{1}M_{2}M_{3}kl}F_{M_{4}M_{5}M_{6}}\,^{kl}-3F_{M_{1}M_{2}M_{6}kl}F_{M_{4}M_{5}M_{3}}\,^{kl}\right )
\end{align}
associated with the $\Theta^{4}$-term in the superfield language.\footnote{To arrive at \eqref{eq:DefForR6ModF5},
one uses non-linear SUSY constraints \cite{deHaro:2002vk,Green:2003an} and applies straightforward rules for the decomposition of $\Gamma$-matrices for which we employed the \texttt{Gamma} software package \cite{Gran:2001yh}.
Afterwards,
one utilises the corresponding projection operators as detailed in App.~\ref{sec:NonLinSuperfieldApp}.
These methods can be employed in a similar fashion to construct the corresponding expression for $|G_{3}|^{2}$ which we leave for the future.}
This term enters the superfield $\Phi$ at order $\Theta^{4}$ in such a way that
\begin{equation}\label{eq:NonLinSuperTheta4F5Rev} 
\Phi\supset (\Theta\Gamma^{M_{1}M_{2}M_{3}}\Theta)(\Theta\Gamma^{M_{4}M_{5}M_{6}}\Theta)\left (\tilde{\cR}_{M_{1}M_{2}M_{3}M_{4}M_{5}M_{6}} +\ldots \right )
\end{equation}
where $\ldots$ denotes further non-linear terms $\sim |G_{3}|^{2}$ or $\sim |\cP|^{2}$.
This clearly implies that $\tilde{\cR}$ as defined in \eqref{eq:DefForR6ModF5} is symmetric under the exchange $(M_{1},M_{2},M_{3})\leftrightarrow(M_{4},M_{5},M_{6})$, anti-symmetric in $(M_{1},M_{2},M_{3})$ and $(M_{4},M_{5},M_{6})$ and enjoys additional symmetries collected in Sect.~\ref{sec:NonLinSuperfieldApp}.

After having identified the non-linear piece at order $\Theta^{4}$ in \eqref{eq:NonLinSuperTheta4F5Rev},
it is straightforward to determine the contribution to the effective action.
It is encoded in the integral
\begin{equation}\label{eq:T24DefinitionF5} 
t_{24}\tilde{\cR}^{4}=\int\dif^{16}\Theta \left [(\Theta\Gamma^{M_{1}M_{2}M_{3}}\Theta)(\Theta\Gamma^{M_{4}M_{5}M_{6}}\Theta)\tilde{\cR}_{M_{1}M_{2}M_{3}M_{4}M_{5}M_{6}}\right ]^{4}\subset \int\dif^{16}\Theta\,\Phi^{4}
\end{equation}
which was explicitly computed\footnote{We refer the reader to \cite{Melo:2020amq} for the corrected results of \cite{Paulos:2008tn}.} in \cite{Paulos:2008tn} using the results of \cite{Green:2005qr}.
For instance, applying the results of \cite{Green:2005qr},
we can show that\footnote{Notice that our normalisation differs from \cite{Green:2005qr} where they found $3^{4}\cdot 2^{24}$ on the RHS.}
\begin{equation}\label{eq:T24J0F5} 
t_{24}\tilde{\cR}^{4}\bigl |_{gR}=\dfrac{1}{3^{2}\cdot 2^{5}}\, \cJ_{0}\, .
\end{equation}
By expanding \eqref{eq:T24DefinitionF5} to higher order in $F_{5}$,
one finds schematically \cite{Paulos:2008tn,Melo:2020amq}
\begin{equation}
t_{24}\tilde{\cR}^{4}=\cJ_{0}+F_{5}^{2}R^{3}+(\nabla F_{5})^{2}R^{2}+F_{5}^{4}R^{2}+F_{5}^{6}R+(\nabla F_{5})^{2}F_{5}^{2}R+(\nabla F_{5})^{4}+F_{5}^{8}\, .
\end{equation}
The odd powers of $F_{5}$ have to be absent because the action is necessarily real.

With regard to $8$-derivative terms involving $G_{3}$ or $\tau$, much less is known about the structure of the action \eqref{eq:AP3Action}. Partial one-loop results for terms like $H_{3}^{2}R^{3}$ at the $5$-point level have been computed \cite{Peeters:2001ub,Richards:2008sa,Liu:2013dna} with the tree-level counterparts obtained in \cite{Liu:2019ses}. The authors of \cite{Bonetti:2016dqh} succeeded in restricting terms of the form $(\nabla\phi)^{2}R^{3}$ through consistency with supersymmetry in $4$D which is equivalent to the earlier work \cite{Becker:2002nn}.

At tree level, the complete $8$-derivative action in the NSNS-sector was inferred in \cite{Garousi:2020mqn,Garousi:2020gio,Garousi:2020lof} upon using constraints of T-duality and double geometry. However, the exorbitant use of field redefinitions and the missing representation of the final result in terms of fundamental index structures makes it virtually impossible to compare the results to the other literature on this subject. This motivates initiating a more unifying approach.

Although RR-sector couplings can be partially inferred from NSNS-sector results at tree and 1-loop level \cite{Liu:2019ses}, a concise definition of manifestly $\mathrm{SL}(2,\mathbb{Z})$-invariant quintic vertices demands a more unifying approach.
For instance, one expects further contributions to \eqref{eq:DefForR6ModF5} of the schematic for $|G_{3}|^{2}$ and $|\cP|^{2}$ which would relate a subset of higher derivative terms $|G_{3}|^{2m}|\cP|^{2n}R^{4-m-n}$ to the famous $R^{4}$ structure by means of a single superspace integral \eqref{eq:T24DefinitionF5}.
We will have more to say about this in Sect.~\ref{sec:ExpandingRiemannTorsion}.

\section{Eight-derivative couplings at five points}\label{sec:EightDerFivePointAction} 

Throughout this paper, we are particularly interested in the $5$-point structure $G_{3}^{2}R^{3}$ together with its variants $|G_{3}|^{2}R^{3}$ and $\overline{G}_{3}^{2}R^{3}$ which can be treated in a similar fashion.
These couplings contribute e.g.~to the leading order $(\alpha^{\prime})^{3}$-correction to the $4$D $F$-term scalar potential \cite{Becker:2002nn,Conlon:2005ki,Cicoli:2021rub}. In this section we discuss them, and in general the full structure arising at five point, according to the behavior of the various terms under the U(1) R-symmetry.

\subsection{Couplings from superstring amplitudes}\label{sec:ReviewFlux} 

Up to $5$-points and including only $R$ and $G_{3}$,
the effective action up to quadratic\footnote{The NSNS-sector couplings $H_{3}^{2}(\nabla H_{3})^{2}R$ have been completely specified at $1$-loop where, in addition to the piece coming from expanding $R(\Omega_{+})^{4}$, one finds an additional contribution $4/9\,  \epsilon_{9}\epsilon_{9}H_{3}^{2}(\nabla H_{3})^{2}R$.
The tree-level counterparts could in principle be determined following the procedures outlined in \cite{Liu:2019ses}.
Obtaining the equivalent expressions in terms of $G_{3},\ov G_{3}$ is slightly more complicated given that the pure NSNS-sector terms do not fully determine cross terms with the RR $3$-form $F_{3}$ beyond quadratic order.
This becomes already evident at $4$-points in \cite{Policastro:2008hg} which led to a new operator $\cO_{2}((|\nabla G_{3}|^{2})^{2})$ reducing to $t_{8}t_{8}$ only in the pure NSNS- or RR-sector.} order in the flux may be written as \cite{Liu:2019ses}
\begin{equation}\label{eq:QuinticActionFluxComp}
\cL(R,G_{3},\ov G_{3})=\cL_{R(\Omega_{+})^{4}}+\cL_{|G_{3}|^{2}R^{3}}+\cL_{G_{3}^{2}R^{3}+\text{c.c.}}+\cL_{\text{CP-odd}}
\end{equation}
where
\begin{align}
\label{eq:QuinticActionFluxCompV1}\cL_{|G_{3}|^{2}R^{3}}&=\alpha f_{0}\biggl \{-\dfrac{1}{2} t_{8}t_{8} |G_{3}|^{2}R^{3}-\dfrac{7}{24}\epsilon_{9}\epsilon_{9} |G_{3}|^{2}R^{3}+2\cdot 4!\,\sum_{i=1}^{8} \tilde{d}_{i}|G_{3}|^{2}\tilde{Q}^{i} \biggl\}\, ,\\[0.4em]
\label{eq:QuinticActionFluxCompV2}\cL_{G_{3}^{2}R^{3}+\text{c.c.}}&=\alpha f_{1}\biggl \{\dfrac{3}{4}t_{8}t_{8} G_{3}^{2}R^{3}-\dfrac{1}{16}\epsilon_{9}\epsilon_{9} G_{3}^{2}R^{3}-3\cdot 4!\,\sum_{i=1}^{8}\tilde{d}_{i} G_{3}^{2}\tilde{Q}^{i} \biggl\}+\text{c.c.}\, ,\\[0.4em]
\label{eq:QuinticActionFluxCompV3}\cL_{\text{CP-odd}}&=3^{2}\cdot 2^{4}\alpha \biggl \{G_{3}\wedge \left (f_{0}X_{7}(\Omega, \ov G_{3})+f_{1}X_{7}(\Omega,  G_{3})\right )+\text{c.c.}\biggl \}
\end{align}
in terms of
\begin{equation}\label{eq:DCoefficientsG2R3} 
(\tilde{d}_{1},\ldots ,\tilde{d}_{8})=4\left (1,-\dfrac{1}{4},0,\dfrac{1}{3},1,\dfrac{1}{4},-2,\dfrac{1}{8}\right )\, .
\end{equation}
We suppressed indices on the objects $\tilde{Q}^{i}$ corresponding to certain $6$-index elements of a basis for $R^{3}$ to be introduced below, see also App.~\ref{app:BasisRC} for definitions.
The index structure in the even-even sector is
\begin{equation}\label{eq:TTGTRC} 
t_{8}t_{8}G_{3}^{2}R^{3}=t_{ M_{1}\ldots M_{8}}t^{ N_{1}\ldots N_{8}}G^{ M_{1} M_{2}P}G_{ N_{1} N_{2}P} R^{ M_{3} M_{4}}\,_{ N_{3} N_{4}}\ldots R^{ M_{7} M_{8}}\,_{ N_{7} N_{8}}
\end{equation}
and the odd-odd sector couplings are \cite{Peeters:2001ub,Liu:2013dna,Liu:2019ses}
\begin{align}\label{eq:EEGTRC} 
\epsilon_{9}\epsilon_{9}G_{3}^{2}R^{3}&=-\epsilon_{P M_{0}\ldots M_{8}}\epsilon^{P N_{0}\ldots N_{8}}G^{ M_{1} M_{2}}\,_{ N_{0}}\, G_{ N_{1} N_{2}}\,^{ M_{0}}\, R^{ M_{3} M_{4}}\,_{ N_{3} N_{4}}\, R^{ M_{5} M_{6}}\,_{ N_{5} N_{6}}\, R^{ M_{7} M_{8}}\,_{ N_{7} N_{8}}\, .
\end{align}
The expressions \eqref{eq:QuinticActionFluxCompV1} and \eqref{eq:QuinticActionFluxCompV2} has the rather surprising feature that some contractions cannot be repacked into the conventional $t_{8}t_{8}$ or $\epsilon_{9}\epsilon_{9}$ structures.
We argue in the following that this should not come as a surprise, but rather as a clear indication that more fundamental index structures are prerequisites to encode the full string kinematics.

The remaining piece $\cL_{R(\Omega_{+})^{4}}$ in \eqref{eq:QuinticActionFluxComp} originates from a connection with torsion given by\footnote{In Einstein frame, the torsionful connection and likewise $R(\Omega_{+})$ include additional derivatives with respect to the dilaton.
For the purposes of this paper, we may ignore such terms by treating $\phi$ as a constant.}
\begin{equation}
\left (\Omega_{\pm}\right )_{M}\,^{KL}=\Omega_{M}\,^{KL}\pm \dfrac{1}{2}\ee^{-\phi/2} H_{M}\,^{KL}
\end{equation}
resulting in the $4$-index tensor
\begin{equation}\label{eq:RiemmTWTorsion} 
R(\Omega_{\pm})_{MN}\,^{KL}=R_{MN}\,^{KL}\pm\ee^{-\phi/2}\nabla_{[M} H_{N]}\,^{KL}+\dfrac{\ee^{-\phi}}{2}H_{[M}\,^{K P}H_{N] P}\,^{L}\, .
\end{equation}
In Einstein frame, the $R^{4}$ contribution is replaced by
\begin{equation}\label{eq:RiemmTWTorsionR4} 
\cL_{R(\Omega_{+})^{4}}=\alpha\, f_{0}(\tau,\bar{\tau})\,  \left (t_{8}t_{8}-\dfrac{1}{4}\epsilon_{8}\epsilon_{8}\right )R(\Omega_{+})^{4}\, .
\end{equation}
Up to $5$-point contact terms,
we may expand $R(\Omega_{+})^{4}$ as usual
\begin{align}\label{eq:ExpandingR4Beg} 
\cL_{R(\Omega_{+})^{4}}&=\alpha\, f_{0}\,  \left (t_{8}t_{8}-\dfrac{1}{4}\epsilon_{8}\epsilon_{8}\right )\biggl \{R^{4}+\ee^{-2\phi}(\nabla H_{3})^{4}+6\ee^{-\phi}(\nabla H_{3})^{2}R^{2}\nn\\
&\hphantom{=\alpha\, f_{0}\,  \left (t_{8}t_{8}-\dfrac{1}{4}\epsilon_{8}\epsilon_{8}\right )\biggl \{}+2\ee^{-\phi}H_{3}^{2}R^{3}+6\ee^{-2\phi}H_{3}^{2}(\nabla H_{3})^{2}R+\ldots\biggl \}
\end{align}
where the appropriate (anti-)symmetrisation of indices on $H_{3}$ and $\nabla H_{3}$ is implied.
The identity $R_{M_{1}M_{2}N_{1}N_{2}}(\Omega_{+})=R_{N_{1}N_{2}M_{1}M_{2}}(\Omega_{-})$ due to closure of $H_{3}$ implies the absence of odd powers of $H_{3}$ in the above expansion.
Given that all terms are multiplied by $f_{0}$, i.e.~the string kinematics is equivalent at tree and 1-loop level,
we can replace $H_{3}^{2}$ and $(\nabla H_{3})^{2}$ by the corresponding $\mathrm{U}(1)$-preserving combinations\footnote{One needs to take special care of $(|\nabla G_{3}|^{2})^{2}$ since the index structure is $\cO_{2}$ rather than $t_{8}t_{8}-\epsilon_{8}\epsilon_{8}/4$ in \eqref{eq:FourPointCouplingsAP3}.
Further, we expect that $H_{3}^{2}(\nabla H_{3})^{2}R \raw |G_{3}|^{2}|\nabla G_{3}|^{2}R$ based on the structure of superparticle amplitudes.
One again runs into the aforementioned issue that crossterms between $F_{3}$ and $H_{3}$ are not fully kinematically determined by the pure NSNS expressions.
We leave the study of such terms for future works.} such that
\begin{align}\label{eq:ExpandingR4} 
\cL_{R(\Omega_{+})^{4}}&=\cL_{4-\text{pt}}^{(3)}\bigl |_{\cP,\ov\cP=0}+2\alpha\, f_{0}\,  \left (\tilde{t}_{8}\tilde{t}_{8}-\dfrac{1}{4}\epsilon_{8}\epsilon_{8}\right ) |G_{3}|^{2}R^{3}+\ldots\, .
\end{align}
The quartic terms clearly reproduce \eqref{eq:FourPointCouplingsAP3} which was already observed in \cite{Policastro:2006vt,Policastro:2008hg}.
Further, we defined a second\footnote{In the NSNS sector, the $t_{8}t_{8}$-combination \eqref{eq:TTGTRC} was proposed in \cite{Grimm:2017okk} to recover supersymmetry in Type IIA Calabi-Yau compactifications to $4$ dimensions, but was also previously obtained in \cite{Peeters:2001ub} from a covariant RNS calculation at the 1-loop level. Similarly, \eqref{eq:TTGTRCV2} was found by computing one-loop string amplitudes at $5$-points in the light-cone gauge GS formalism in \cite{Richards:2008sa} and utilised in \cite{Liu:2013dna,Liu:2019ses}.}
$t_{8}t_{8}$ and $\epsilon_{8}\epsilon_{8}$ index structure
\begin{align}\label{eq:TTGTRCV2} 
\tilde{t}_{8}\tilde{t}_{8}|G_{3}|^{2}R^{3}&=t_{ M_{1}\ldots M_{8}}t^{ N_{1}\ldots N_{8}} G^{[ M_{1}}\,_{ N_{1}P}\ov G^{ M_{2}]P}\,_{ N_{2}} R^{ M_{3} M_{4}}\,_{ N_{3} N_{4}}\ldots R^{ M_{7} M_{8}}\,_{ N_{7} N_{8}}\, ,\\
\label{eq:EEGTRCV2} \epsilon_{8}\epsilon_{8}|G_{3}|^{2}R^{3}&=\epsilon^{M_{1}\ldots M_{8}}\epsilon^{N_{1}\ldots N_{8}} G_{[M_{1}| N_{1}k}\, \overline{G}_{|M_{2}]}\,^{k}\,_{N_{2}}\,  R_{M_{3}M_{4}N_{3}N_{4}} \ldots R_{M_{7}M_{8}N_{7}N_{8}}\, .
\end{align}
Before we proceed,
let us comment on the role of the object $R(\Omega_{+})$.
From the perspective of generalised geometry,
it seems natural to introduce torsion in the form of $H_{3}$ in order to capture a big part of the NSNS-sector kinematics.
As is evident from the additional contributions in \eqref{eq:QuinticActionFluxComp},
this is clearly not sufficient to specify the complete effective action, see also Sect.~\ref{sec:TreeOneLoopKinFiveP} and \cite{Liu:2019ses}.
Furthermore,
this approach is not manifestly $\mathrm{SL}(2,\mathrm{Z})$ invariant and might even fail when working with sixteen fermion integrals.
Below,
we argue that the terms obtained from \eqref{eq:RiemmTWTorsionR4} are actually highly non-trivial from the perspective of the superfield approach in the sense that such contributions seem to be (at least partially) associated with non-linear terms at order $\Theta^{6}$ rather than $\Theta^{4}$.

\begin{table}[t!]
\centering
{\small
\begin{tabular}{|m{1.45cm}|m{-0.1cm}m{0.25cm}|m{0.05cm}m{0.1cm}m{0.25cm}|m{0.15cm}m{0.15cm}m{0.15cm}m{0.05cm}m{0.15cm}m{0.15cm}m{0.15cm}m{0.05cm}m{0.15cm}m{0.15cm}m{0.3cm}|m{0.05cm}m{0.15cm}m{0.15cm}m{0.15cm}m{0.15cm}m{0.15cm}m{0.05cm}m{0.3cm}|}
\hline 
 & $a_{1}$ & $a_{2}$ & $b_{1}$ & $b_{2}$ & $b_{3}$ & $c_{1}$ & $c_{2}$ & $c_{3}$ & $c_{4}$ & $c_{5}$ & $c_{6}$ & $c_{7}$ & $c_{8}$ & $c_{9}$ & $c_{10}$ & $c_{11}$ & $d_{1}$ & $d_{2}$ & $d_{3}$ & $d_{4}$ & $d_{5}$ & $d_{6}$ & $d_{7}$ & $d_{8}$ \\ 
\hline 
\hline 
 &  &  &  &  &  &  &  &  &  &  &  &  &  &  &  &  &  &  &  &  &  &  &  &  \\ [-0.5em]
\hspace*{-0.15cm}$\frac{-t_{8}t_{8}|G_{3}|^{2}R^{3}}{16\cdot 4!}$ & 0 & 0 & 0 & 0 & 0 & $\frac{-1}{32}$ & $\frac{1}{2}$ & $\frac{-1}{16}$ & 0 & $\frac{-1}{2}$ & $\frac{-1}{2}$ & $\frac{1}{2}$ & 1 & ${\tiny-}1$ & $\frac{1}{4}$ & 0 & 0 & 0 & 0 & 0 & 0 & 0 & 0 & 0 \\ [0.5em]
\hline 
 &  &  &  &  &  &  &  &  &  &  &  &  &  &  &  &  &  &  &  &  &  &  &  &  \\ [-0.5em]
\hspace*{-0.15cm}$\frac{-\tilde{t}_{8}\tilde{t}_{8}|G_{3}|^{2}R^{3}}{16\cdot 4!}$  & 0 & 0 & 0 & 0 & 0 & $\frac{1}{64}$ & $\frac{-1}{4}$ & $\frac{1}{32}$ & 0 & $\frac{1}{4}$ & $\frac{-1}{4}$ & $\frac{-1}{4}$ & $0$ & $\frac{1}{2}$ & $\frac{1}{8}$ & $\frac{-1}{4}$ & 0 & 0 & 0 & 0 & 0 & 0 & 0 & 0 \\ [0.5em]
\hline 
 &  &  &  &  &  &  &  &  &  &  &  &  &  &  &  &  &  &  &  &  &  &  &  &  \\ [-0.5em]
$\frac{\epsilon_{9}\epsilon_{9}|G_{3}|^{2}R^{3}}{192\cdot 4!}$  & $\frac{1}{72}$ & $\frac{1}{36}$ & $\frac{1}{4}$ & $\frac{-1}{4}$ & $\frac{1}{2}$ & $\frac{1}{32}$ & $\frac{-1}{2}$ & $\frac{1}{16}$ & 0 & $\frac{1}{2}$ & $\frac{1}{2}$ & $\frac{-1}{2}$ & 0 & $-1$ & $\frac{1}{4}$ & $\frac{-1}{2}$ & 0 & $\frac{-1}{4}$ & $\frac{-1}{2}$ & $\frac{-1}{3}$ & $-1$ & $\frac{-1}{4}$ & $2$ & $\frac{-1}{8}$ \\ [0.5em]
\hline 
 &  &  &  &  &  &  &  &  &  &  &  &  &  &  &  &  &  &  &  &  &  &  &  &  \\ [-0.5em]
$\frac{-t_{18}|G_{3}|^{2}R^{3}}{8\cdot 4!}$  & $\frac{1}{72}$ & $\frac{1}{36}$ & $\frac{1}{4}$ & $\frac{-1}{4}$ & $\frac{1}{2}$ & $0$ &  $0$ & $0$ & 0 &  $0$ & $0$ & $0$ & $1$ & $-2$ & $\frac{1}{2}$ & $\frac{-1}{2}$ & $1$ & $\frac{-1}{2}$ & $\frac{-1}{2}$ &  $0$ & $0$ & $0$ & $0$ & $0$  \\ [0.5em]
\hline 
 &  &  &  &  &  &  &  &  &  &  &  &  &  &  &  &  &  &  &  &  &  &  &  &  \\ [-0.5em]
\hspace*{-0.15cm}{\tiny $\frac{-\epsilon_{8}\epsilon_{8}|G_{3}|^{2}R^{3}}{96\cdot 4!}$ } & $\frac{1}{72}$ & $\frac{1}{36}$ & $\frac{1}{6}$ & $\frac{-1}{6}$ & $\frac{1}{3}$ & $\frac{1}{96}$ & $\frac{-1}{6}$ & $\frac{1}{48}$ & 0 & $\frac{1}{6}$ & $\frac{1}{6}$ & $\frac{-1}{6}$ & 0 & $\frac{-1}{3}$ & $\frac{1}{12}$ & $\frac{-1}{6}$ & 0 & $0$ & $0$ & $0$ & $0$ & $0$ & $0$ & $0$  \\ [0.5em]
\hline 
\end{tabular} 
}\caption{Decomposition of index structures for $|G_{3}|^{2}R^{3}$ in the 24 component basis for $R^{3}$ defined in App.~\ref{app:BasisRC}.}\label{tab:CoeffKinStructGTRC} 
\end{table}

For later convenience,
we expand the above kinematical structures into independent Lorentz singlets that can be built from $|G_{3}|^{2}R^{3}$. This can be easily determined by utilising the software package \texttt{LiE} \cite{van1992lie,van1994lie} looking for all singlets under $\mathrm{SO}(1,9)$. In App.~\ref{app:BasisRC}, we define a $24$-dimensional basis for $R^{3}$ based on the conventions in \cite{Liu:2019ses} in order to write
{\small
\begin{align}\label{eq:BasisExpR310D} 
\cL_{|G_{3}|^{2}R^{3}}&=\sum_{i=1}^{2}a_{i}\,  |G_{3}|^{2}\, \tilde{S}^{i}+\sum_{i=1}^{3}b_{i}\, G^{M_{1}}\,_{N_{1}N_{2}}\ov G^{M_{2}N_{1}N_{2}}\, \tilde{W}_{M_{1}M_{2}}^{i}\\[0.5em]
&\quad+\sum_{i=1}^{11}c_{i}\, G^{M_{1}M_{2}}\,_{N_{1}}\ov G^{M_{3}M_{4}N_{1}}\, \tilde{X}^{i}_{M_{1}M_{2}M_{3}M_{4}}+\sum_{i=1}^{8}d_{i}\, G^{M_{1}M_{2}M_{3}}\ov G^{M_{4}M_{5}M_{6}}\, \tilde{Q}^{i}_{M_{1}M_{2}M_{3}M_{4}M_{5}M_{6}}\nn
\end{align}
}where we implicitly symmetrise $G_{3}$ and $\ov G_{3}$. In this basis, we may expand the kinematical structures appearing in \eqref{eq:TTGTRC}, \eqref{eq:EEGTRC}, \eqref{eq:TTGTRCV2} and \eqref{eq:EEGTRCV2} as summarised in Tab.~\ref{tab:CoeffKinStructGTRC}.

\subsection{The $\mathrm{U}(1)$-violating sector}

In general, for maximally $\mathrm{U}(1)$-violating (MUV) processes,
substantial progress has been made in \cite{Boels:2012zr,Green:2019rhz,Green:2020eyj} at the level of amplitudes and modular forms. This makes this sector of the effective action particularly well-behaved.
Moreover,
one is led to a similar conclusion by studying superparticle/supermembrane amplitudes in $11$ dimensions on a $T^{2}$ \cite{Green:1997as,Green:1997me,Green:1999by,Peeters:2005tb} as we argue in Sect.~\ref{sec:SuperparticleFivePG2R3}.

\subsubsection{The role of higher-dimensional index structures}

In Section \ref{sec:SixteenFermionIntegral},
we argued that the notion of higher-dimensional index tensors is natural from the effective-action point of view,
though unnatural from the string amplitude calculus.
In this section,
we show that the 5-point action can be dramatically simplified with the use of these tensors, thereby explaining a subset of relative coefficients.
Subsequently,
we provide a microscopic derivation of these results from M-theory loop amplitudes which we show to agree in the U(1)-violating sector with the linearised superfield expectation.

Specifically,
the linearised superfield suggests that the following tensor is expected to play an outstanding role in the string kinematics at 5 points
\begin{equation}\label{eq:DefT18} 
t_{18}G_{3}^{2}R^{3}=\int\dif^{16}\Theta\,  \left [(\Theta\Gamma^{M_{1}M_{2}M_{3}}\Theta)G_{M_{1}M_{2}M_{3}}\right ]^{2} \left [(\Theta\Gamma^{M_{1}M_{2}}\,_{k}\Theta)(\Theta\Gamma^{M_{3}M_{4}k}\Theta) R_{M_{1}M_{2}M_{3}M_{4}}\right ]^{3}\, .
\end{equation}
In principle, this expression can be computed via the methods of \cite{Green:2005qr}.
Instead,
we make use of two independent results available in the literature.
First, the authors of \cite{Peeters:2005tb} expanded $t_{18}$ in a basis of $26$ Lorentz singlets under $\mathrm{SO}(9)$. Equivalently,
a second way to compute \eqref{eq:DefT18} is using the tensor $t_{24}$ computed in \cite{Green:2005qr} as a generating object for lower-order tensors $t_{24-2n}$ upon appropriate contraction with $n$ metric factors.
We discuss this second option in more detail in Sect.~\ref{sec:ExpandingRiemannTorsion}.
Utilising the results of \cite{Peeters:2005tb},
we may simplify \eqref{eq:QuinticActionFluxCompV2} drastically by writing
\begin{align}\label{eq:PredG2R3FromStringsML}
\cL_{G_{3}^{2}R^{3}+\text{c.c.}}&=\dfrac{3}{2}\alpha \biggl \{f_{1} t_{18}G_{3}^{2}R^{3}+f_{-1}t_{18}\overline{G}_{3}^{2}R^{3} \biggl\}
\end{align}
where in our normalisation
\begin{equation}\label{eq:T18EVENEVENODDODD} 
t_{18}G_{3}^{2}R^{3} = \dfrac{1}{2}t_{8}t_{8}G_{3}^{2}R^{3}-\dfrac{1}{24}\epsilon_{9}\epsilon_{9}G_{3}^{2}R^{3}-2\cdot 4!\sum_{i=1}^{8}\, \tilde{d}_{i}G_{3}^{2}\tilde{Q}^{i}
\end{equation}
in terms of the $\tilde{d}_{i}$ coefficients \eqref{eq:DCoefficientsG2R3} (and equivalently for $\ov G_{3}^{2}R^{3}$). The overall coefficient $3/2$ in Eq.~\eqref{eq:PredG2R3FromStringsML} is expected from tree level $5$-point scattering in the pure spinor formalism \cite{Green:2019rhz}.

The result \eqref{eq:PredG2R3FromStringsML} has many striking implications.
The definition of $t_{18}G_{3}^{2}R^{3}$ in \eqref{eq:T18EVENEVENODDODD} resolves the apparent puzzle that some terms $\sim \tilde{Q}^{i}$ in \eqref{eq:QuinticActionFluxComp} do not repackage nicely into $t_{8}t_{8}$ or $\epsilon_{n}\epsilon_{n}$.
Even more interestingly,
it captures the entire string kinematics in the MUV sector.
In this sense,
the $t_{18}$ structure \eqref{eq:DefT18} is arguably a more suitable representation of the string kinematics at the level of the 5-point effective action.
While this has clearly been known for many years (at least implicitly in the linearised superfield approach \cite{Peeters:2001ub})\footnote{In essence, the authors of \cite{Peeters:2001ub} argued  that the full kinematics of the NSNS-sector coupling $H_{3}^{2}R^{3}$ obtained from string scattering amplitudes is \textbf{not} captured by only the index structure $t_{18}$ appearing in the linearised superfield calculus.
Our results demonstrate that $t_{18}$ nonetheless plays an important role for the tree level $H_{3}^{2}R^{3}$ couplings, cf.~Eq.~\eqref{eq:NSNSKinematicsH2R3Tree} below.},
the above provides the first direct proof that such higher-dimensional index structures appear in both the MUV and non-MUV (see the next subsection) sectors at the level of $5$-point string amplitudes.

For the MUV terms, we can be even more precise with regard to the overall coefficient.
In fact,
the modular forms satisfy (see Eq.~\eqref{eq:CovDerMoFunc})
\begin{equation}\label{eq:CovDerMoFuncF0} 
\cD_{0}f_{0}=\dfrac{3}{4}f_{1}\kom \ov\cD_{0}f_{0}=\dfrac{3}{4}f_{-1}
\end{equation}
which allows us to write
\begin{equation}
\cL_{G_{3}^{2}R^{3}+\text{c.c.}}=2\alpha \biggl \{(\cD_{0}f_{0}) t_{18}G_{3}^{2}R^{3}+(\ov\cD_{0}f_{0})t_{18}\overline{G}_{3}^{2}R^{3} \biggl\}\, .
\end{equation}


\subsubsection{A derivation from superparticles}\label{sec:SuperparticleFivePG2R3}

While the index structure $t_{18}$ naturally appears in the context of the linearised superfield approximation \cite{Peeters:2001ub},
it is not obvious at all from standard string amplitudes (see however \cite{Green:2019rhz} for MUV amplitudes and references therein).
In this part,
we show that the $t_{18}$ structure appears naturally in superparticle amplitudes in M-theory on $2$-tori $T^{2}$ \cite{Green:1997as,Green:1997me,Green:1999by}.
The calculation proceeds similar to the famous derivation of $R^{4}$ in \cite{Green:1997as} which matched not only the well-known $4$-graviton kinematics at tree and 1-loop level, but also provided evidence for modular functions in the Type IIB effective action \cite{Green:1997tv,Green:1997di}.

In the MUV sector,
the only non-vanishing $5$-point superparticle amplitude involving two $3$-forms and three gravitons in $9$D is given by\footnote{The normalisation of these amplitudes will be discussed further in Sect.~\ref{sec:MUVConjecture}.}
\begin{align}
\cA_{G_{3}^{2}R^{3}+\text{c.c.}}^{(\text{SP})}&=\dfrac{1}{2^{6}\pi^{9/2}\, \Gamma\left (\frac{3}{2}\right ) v_{0}}\int\, \dfrac{\dif t}{t} \int\dif^{9}\mathbf{p}\, \sum_{l_{1},l_{2}\in \bZ} \mathrm{e}^{-t(\mathbf{p}^{2}+g^{ab}l_{a}l_{b})}\; \nn\\
&\quad t^{5} \tr\left ( \left [2h_{ij} \cR^{il}\cR^{jm}k_{l}k_{m}\right ]^{3} \left [-\sqrt{2}G_{3} P^{z} \cR^{lmn}\right ]^{2}\right )
\end{align}
in terms of $v_{0}=\mathrm{Vol}(T^{2})$.
This contribution is associated with superparticles running in the loop carrying non-trivial KK-charges on the $T^{2}$ compensating for the $\mathrm{U}(1)$-charges of $G_{3}^{2}$ to give a real expression.\footnote{In general, in the context of superparticle amplitudes, the higher-dimensional index structures $t_{N}$ for the $G$-flux naturally arise (even in the non-MUV sector) whenever the superparticles in the loop carry non-trivial KK-momentum.}
Looking at the trace over fermions,
we clearly notice the resemblance with \eqref{eq:DefT18} in terms of the linearised Riemann tensor in 9 dimensions which allows us to simplify the expression to
\begin{align}
\cA_{G_{3}^{2}R^{3}+\text{c.c.}}^{(\text{SP})}&=\dfrac{t_{18}G_{3}^{2}R^{3}}{2^{2}\, \Gamma\left (\frac{3}{2}\right ) v_{0}} \,  \int\dfrac{\dif t}{\sqrt{t}}\sum_{l_{1},l_{2}\in \bZ}\, P_{z}^{2} \mathrm{e}^{-tg^{ab}l_{a}l_{b}}+\text{c.c.}\, .
\end{align}
In App.~\ref{sec:MUVSuperPartGRAmps},
we show that
\begin{equation}
\int\dfrac{\dif t}{\sqrt{t}}\sum_{l_{1},l_{2}\in \bZ}\, P_{z}^{2} \mathrm{e}^{-tg^{ab}l_{a}l_{b}}=\dfrac{4\Gamma\left (\frac{5}{2}\right )}{\sqrt{v_{0}}}\, f_{1}(\tau,\bar{\tau})\, .
\end{equation}
The volume scaling implies that the amplitude vanishes in the decompactification limit $v_{0}\rightarrow \infty$.
As opposed to $\mathrm{U}(1)$-preserving amplitudes like $t_{16}R^{4}$ \cite{Green:1997as},
$\mathrm{U}(1)$-violating effects such as the above are not present in $11$D supergravity \cite{Green:1997me,Green:1999by}.

Taking the limit to Type IIB,
we find
\begin{align}
v_{0}\cA_{G_{3}^{2}R^{3}+\text{c.c.}}^{(\text{SP})}\; &\xrightarrow{\; \; v_{0}\rightarrow 0\; \; }\;\cL_{G_{3}^{2}R^{3}+\text{c.c.}}^{(\text{SP})}
\end{align}
in terms of
\begin{equation}\label{eq:PredG2R3From11DSuperParticles} 
\cL_{G_{3}^{2}R^{3}+\text{c.c.}}^{(\text{SP})}=\dfrac{3}{2}\left (f_{1}(\tau,\bar{\tau})\, t_{18}G_{3}^{2}R^{3}+f_{-1}(\tau,\bar{\tau})\, t_{18}\ov{G}_{3}^{2}R^{3}\right )
\end{equation}
in agreement with \eqref{eq:QuinticActionFluxCompV2}.
We stress that, while the linearised superfield and perturbative superstring amplitudes typically only see a small subset of terms of the full modular forms $f_{w}$,
a single superparticle amplitude derives the full $f_{w}$ from first principles.
Critically,
this involves also non-perturbative D-instanton contributions which have only recently been derived for $R^{4}$ from string field theory \cite{Sen:2021tpp,Sen:2021jbr}.

\subsection{The $\mathrm{U}(1)$-preserving sector -- evidence for non-linear superfields}\label{sec:ExpandingRiemannTorsion}

Let us now move our attention to the $\mathrm{U}(1)$-preserving sector of the five-point effective action, whose structure turns out to be much more involved.
As described in Sect.~\ref{sec:ReviewFlux},
there is a contribution originating from the torsionful Riemann tensor \eqref{eq:ExpandingR4} as well as a remainder given by \eqref{eq:QuinticActionFluxCompV1}.
Inspecting the latter,
we notice a close resemblance to the MUV contact terms in \eqref{eq:QuinticActionFluxCompV2}.
Indeed,
\eqref{eq:T18EVENEVENODDODD} is equivalently defined for $|G_{3}|^{2}R^{3}$ allowing us to recast \eqref{eq:QuinticActionFluxCompV1} together with \eqref{eq:ExpandingR4} in the form
\begin{equation}\label{eq:NonMUVResultActionNT} 
\cL_{R(\Omega_{+})^{4}}+\cL_{|G_{3}|^{2}R^{3}} \biggl |_{\text{5-point}}=\alpha f_{0}\biggl (- t_{18}-\dfrac{1}{3}\epsilon_{9}\epsilon_{9}+2\tilde{t}_{8}\tilde{t}_{8}-\dfrac{1}{2}\epsilon_{8}\epsilon_{8}\biggl )|G_{3}|^{2}R^{3} \, .
\end{equation}
Contrary to above,
there is an odd-odd structure remaining which can in fact be traced back to the $1$-loop NSNS strcture $-1/3 \epsilon_{9}\epsilon_{9}H_{3}^{2}R^{3}$ of \cite{Richards:2008sa,Liu:2013dna}, while $t_{18}H_{3}^{2}R^{3}$ can only be seen at string tree level \cite{Liu:2019ses}.
We will have more to say about this in the next section.

The fact that $\cL_{|G_{3}|^{2}R^{3}}$ contains another $t_{18}$ piece with an overall factor of $-1$ will in fact play quite a crucial role in the reductions to $4$D in Sect.~\ref{sec:ScalarPotentialSLTZInvDer}.
While in the MUV sector such a term is completely specified by the linearised superfield,
we stress that \eqref{eq:NonMUVResultActionNT} can only be obtained from non-linear terms in the superfield such as through contributions $\sim |G_{3}|^{2}$ to \eqref{eq:DefForR6ModF5}.
In this way,
we expect a subset of the terms in \eqref{eq:NonMUVResultActionNT} to be related to $R^{4}$ through \eqref{eq:T24DefinitionF5} as we now demonstrate.
Ultimately, we will arrive at a similar conclusion as \cite{Peeters:2001ub} in the NSNS sector, namely that generating the odd-odd contribution $\epsilon_{9}\epsilon_{9}|G_{3}|^{2}R^{3}$ from a non-linear superfield requires corrections at order $\Theta^{6}$.

To recapitulate,
we observed that the linearised superfield indeed captures the complete string-theory result in the MUV sector which is encoded by a single superspace integral giving rise to the tensor structure $t_{18}$.
Even more importantly,
we obtained evidence that the same tensor also enters in the non-MUV sector of the action in such a way that it cancels out at 1-loop for $H_{3}^{2}R^{3}$.
We expect this to be a clear hint at potential non-linear couplings in the superfield.
Non-linear completions of the superfield \eqref{eq:LinSuperfield} are given by (schematically)
\begin{align}\label{eq:NonLinearModTheta4G3} 
\Delta&\supset \Theta^{2}G_{3}+\Theta^{4}(R+|G_{3}|^{2}+\ldots)+\Theta^{6}(\nabla^{2}\ov G_{3}+R\ov G_{3}+\ldots)+\Theta^{8}(R\ov G^{2}_{3}+\ldots)\, .
\end{align}
The terms entering at order $\Theta^{4}$ were discussed in Sect.~\ref{sec:SixteenFermionIntegral} for $F_{5}^{2}$.
Clearly,
one similarly expects terms of the form $|G_{3}|^{2}$ also to enter at this order \cite{deHaro:2002vk,Green:2003an}.
They can indeed be obtained utilising the results of \cite{Howe:1983sra}.
For now,
we work with a general parametrisation modifying \eqref{eq:DefForR6ModF5} in such a way that
\begin{align}
\label{eq:DefForR6ModG3} 
\cR_{M_{1} \ldots M_{6}}&=\tilde{\cR}_{M_{1}\ldots M_{6}}+\dfrac{1}{768}\biggl (\lambda_{1}\, G_{M_{1}M_{2}M_{3}}\ov G_{M_{4}M_{5}M_{6}}+\lambda_{2}\, G_{M_{1}M_{2}M_{6}}\ov G_{M_{4}M_{5}M_{3}}\nn\\
&\quad+\lambda_{3}\, g_{M_{3}M_{6}}G_{M_{1}M_{2}k}\ov G_{M_{4}M_{5}}\,^{k}+\lambda_{4}\, g_{M_{3}M_{6}}G_{M_{1}M_{5}k}\ov G_{M_{4}M_{2}}\,^{k}\nn\\
&\quad+\lambda_{5}\, \epsilon_{k_{1}\ldots k_{5} M_{2}\ldots M_{6}}\left (G^{k_{1}k_{2}}\,_{M_{1}}\overline{G}^{k_{3}k_{4}k_{5}}+ \ov G^{k_{1}k_{2}}\,_{M_{1}} {G}^{k_{3}k_{4}k_{5}}\right )\biggl )\, .
\end{align}
We generically expect $\lambda_{i}\neq 0$ for all $\lambda_{i}$.
The symmetries of $\cR$ are determined e.g. by Fierz identities implying the absence of double traces, cf.~App.~\ref{sec:NonLinSuperfieldApp}.

The contribution to the effective action may be written as
\begin{equation}
\cL_{\cR^{4}}=c\, \int\dif^{16}\Theta \left [(\Theta\Gamma^{M_{1}M_{2}M_{3}}\Theta)(\Theta\Gamma^{M_{4}M_{5}M_{6}}\Theta)\cR_{M_{1}M_{2}M_{3}M_{4}M_{5}M_{6}}\right ]^{4}
\end{equation}
The normalisation constant $c$ is fixed such that we recover $\cJ_{0}$ at order $R^{4}$ as defined in \eqref{eq:SuperInvRF}.
We find that (recall \eqref{eq:T24J0F5})
\begin{equation}
t_{24}\cR^{4}\bigl |_{gR} =\dfrac{1}{2^{5}\cdot 3^{2}} \cJ_{0}\quad \Rightarrow\quad c=2^{5}\cdot 3^{2}
\end{equation}
We are mainly interested in the terms arising to linear order in $\lambda_{i}$ where we find that the CP-even part is given by
\begin{align}\label{eq:ContributionActionT24G2R3Coeff} 
c\, t_{24}\cR^{4}&=\dfrac{\lambda_{1}}{2^{5} } t_{18}|G_{3}|^{2}R^{3}  -2\lambda_{2}\,  T_{18}^{(1)}|G_{3}|^{2}R^{3}-2(2\lambda_{3}+\lambda_{4})\,  T_{16}^{(1)}|G_{3}|^{2}R^{3}\nn\\
&\quad+\dfrac{36\lambda_{5}}{5}\,  T_{18}^{(2)}|G_{3}|^{2}R^{3}
\end{align}
where $T_{N}^{(i)}$ are certain tensor structures carrying $N$ indices.
We summarised the decomposition of the individual kinematical structures in Tab.~\ref{tab:CoeffKinStructGTRCT24}.
We find the following relationships among the different terms
\begin{align}\label{eq:IDsT24Results} 
T_{18}^{(1)}|G_{3}|^{2}R^{3}&=\dfrac{-t_{18}|G_{3}|^{2}R^{3}}{8\cdot 4!}-2T_{18}^{(2)}|G_{3}|^{2}R^{3}\, ,\nn\\
T_{16}^{(1)}|G_{3}|^{2}R^{3}&=\dfrac{-t_{18}|G_{3}|^{2}R^{3}}{8\cdot 4!}-T_{18}^{(2)}|G_{3}|^{2}R^{3}\, ,\nn\\
T_{18}^{(1)}|G_{3}|^{2}R^{3}&=T_{16}^{(1)}|G_{3}|^{2}R^{3}-T_{18}^{(2)}|G_{3}|^{2}R^{3}\, .
\end{align}

\begin{table}[t!]
\centering
{\small
\begin{tabular}{|m{1.45cm}|m{-0.1cm}m{0.25cm}|m{0.05cm}m{0.1cm}m{0.25cm}|m{0.15cm}m{0.15cm}m{0.15cm}m{0.05cm}m{0.15cm}m{0.15cm}m{0.15cm}m{0.05cm}m{0.15cm}m{0.15cm}m{0.3cm}|m{0.05cm}m{0.15cm}m{0.15cm}m{0.15cm}m{0.15cm}m{0.15cm}m{0.05cm}m{0.3cm}|}
\hline 
 & $a_{1}$ & $a_{2}$ & $b_{1}$ & $b_{2}$ & $b_{3}$ & $c_{1}$ & $c_{2}$ & $c_{3}$ & $c_{4}$ & $c_{5}$ & $c_{6}$ & $c_{7}$ & $c_{8}$ & $c_{9}$ & $c_{10}$ & $c_{11}$ & $d_{1}$ & $d_{2}$ & $d_{3}$ & $d_{4}$ & $d_{5}$ & $d_{6}$ & $d_{7}$ & $d_{8}$ \\ 
\hline 
\hline 
 &  &  &  &  &  &  &  &  &  &  &  &  &  &  &  &  &  &  &  &  &  &  &  &  \\ [-0.5em]
{\tiny $T_{18}^{(1)}|G_{3}|^{2}R^{3}$ } & $\frac{1}{72}$ & $\frac{1}{36}$ & $\frac{1}{4}$ & $\frac{-1}{4}$ & $\frac{1}{2}$ & $0$ &  $0$ & $0$ & 0 &  $0$ & $\frac{2}{3}$ & $0$ & $\frac{1}{3}$ & $-2$ & $\frac{1}{6}$ & $\frac{-1}{6}$ & $-1$ & $\frac{1}{2}$ & $\frac{1}{2}$ &  $0$ & $0$ & $0$ & $0$ & $0$  \\ [0.5em]
\hline 
 &  &  &  &  &  &  &  &  &  &  &  &  &  &  &  &  &  &  &  &  &  &  &  &  \\ [-0.5em]
{\tiny $T_{16}^{(1)}|G_{3}|^{2}R^{3}$  }& $\frac{1}{72}$ & $\frac{1}{36}$ & $\frac{1}{4}$ & $\frac{-1}{4}$ & $\frac{1}{2}$ & $0$ &  $0$ & $0$ & 0 &  $0$ & $\frac{1}{3}$ & $0$ & $\frac{2}{3}$ & $-2$ & $\frac{1}{3}$ & $\frac{-1}{3}$ & $0$ & $0$ & $0$ &  $0$ & $0$ & $0$ & $0$ & $0$  \\ [0.5em]
\hline 
 &  &  &  &  &  &  &  &  &  &  &  &  &  &  &  &  &  &  &  &  &  &  &  &  \\ [-0.5em]
{\tiny $T_{18}^{(2)}|G_{3}|^{2}R^{3}$}  & $0$ & $0$ & $0$ & $0$ & $0$ & $0$ &  $0$ & $0$ & 0 &  $0$ & $\frac{-1}{3}$ & $0$ & $\frac{1}{3}$ & $0$ & $\frac{1}{6}$ & $\frac{-1}{6}$ & $1$ & $\frac{-1}{2}$ & $\frac{-1}{2}$ &  $0$ & $0$ & $0$ & $0$ & $0$  \\ [0.5em]
\hline 
\end{tabular} 
}\caption{Decomposition of leading order flux terms obtained from $t_{24}\cR^{4}$ in the 24 component basis for $R^{3}$ defined in App.~\ref{app:BasisRC}.}\label{tab:CoeffKinStructGTRCT24} 
\end{table}

Before we continue,
we highlight the following caveat.
The $\lambda_{i}$ in \eqref{eq:DefForR6ModG3} are not the only non-linear modifications contributing at the level of the $5$-point contact terms in the non-MUV sector.
Terms at order $\Theta^{6}\ov G_{3}R$ in \eqref{eq:NonLinearModTheta4G3} contribute at the same level.
This means we cannot simply expect the $\mathrm{U}(1)$-neutral sector of Eq.~\eqref{eq:QuinticActionFluxComp} to be constructable from $t_{24}\cR^{4}$ alone.
Hence,
we can only compare the value of the $\lambda_{i}$ to the string amplitude result and make a prediction about contributions from higher-order non-linear terms.

From \eqref{eq:PredG2R3FromStringsML},
we expect to find a term of the form $t_{18}|G_{3}|^{2}R^{3}$ upon expanding $t_{24}\cR^{4}$
which, according to \eqref{eq:ContributionActionT24G2R3Coeff}, is trivially achieved by setting $\lambda_{1}=-2^{5}$ and $\lambda_{i}=0$ for all other coefficients.
However,
we generically expect that all $\lambda_{i}$ are non-vanishing and,
given the identities \eqref{eq:IDsT24Results},
a non-trivial combination of values for the $\lambda_{i}$ can also do the job.
In fact, it turns out that the equation $ct_{24}\cR^{4}=-t_{18}|G_{3}|^{2}R^{3} $ has the solution
\begin{equation}
\lambda_{5}=\dfrac{5}{36}\left (24+\dfrac{3}{4}\lambda_{1}-\lambda_{2}\right )\kom \lambda_{4} = -24 - \dfrac{3}{4} \lambda_{1} -  \lambda_{2} - 2 \lambda_{3}\, .
\end{equation}

Initially,
one might hope that by modifying \eqref{eq:DefForR6ModG3} accordingly the additional piece $\sim T(\epsilon_{10},t_{8})|G_{3}|^{2}R^{3}$ (see \eqref{eq:KinematicsUniversalG2R3} below)
can also be reabsorbed into the definition of $t_{24}\cR^{4}$.
However, looking at the coefficients collected in Tab.~\ref{tab:CoeffKinStructGTRC}, this seems not to be the case.
For instance,
$\epsilon_{9}\epsilon_{9}|G_{3}|^{2}R^{3}$ involves terms with $d_{i}\neq 0$ for $i\geq 4$ which cannot arise from $t_{24}$ as observed in Tab.~\ref{tab:CoeffKinStructGTRCT24}.
In principle, a natural way to extend the linearised superfield \eqref{eq:LinSuperfield} would be terms of the form $\Theta^{6}R\overline{G}_{3}$ with the symmetry properties conjectured in \cite{Peeters:2001ub}.
According to these arguments,
one would need terms of the form\footnote{One can argue e.g. for the existence of the former based on the structure of terms found in \cite{Howe:1983sra} following the derivation of $\Theta^{4}$ terms in \cite{deHaro:2002vk}.}
\begin{align}
\Theta^{6}R\ov G&\supset \Theta \Gamma^{M_{1}N_{1}P_{1}}\Theta\, \Theta \Gamma^{M_{2}N_{2}P_{2}}\Theta\, \Theta \Gamma^{M_{3}N_{3}P_{3}}\Theta\biggl (a_{1} g_{P_{1}P_{2}}R_{M_{1}N_{1}[M_{2}N_{2}}\overline{G}_{M_{3}N_{3}P_{3}]}\nn\\
&\quad+a_{2}g_{P_{1}P_{2}}g_{M_{1}P_{3}}R^{p}\,_{(N_{1}|\, [M_{2}N_{2}}\overline{G}_{|M_{3})\, N_{3}]\, p} \biggl )
\end{align}
which would appear as
\begin{equation}
\Delta^{4}\supset (\Theta^{2}G_{3})(\Theta^{6}R\overline{G}_{3})(\Theta^{4}R)^{2}\, .
\end{equation}
However,
this is beyond the scope of the present work.

\subsection{A new perspective on tree and 1-loop kinematics}\label{sec:TreeOneLoopKinFiveP}

To summarise,
the structure of $5$-point contact terms built from the complex $3$-form and the Riemann tensor in \eqref{eq:QuinticActionFluxCompV1} and \eqref{eq:QuinticActionFluxCompV2} is dramatically simplified by introducing $t_{18}$.
Altogether,
we showed that up to five points \eqref{eq:QuinticActionFluxComp}, the effective action can be summarised as (ignoring the terms involving $\nabla G_{3}$)  
\begin{align}\label{eq:FullResultG2R3TreeLoop} 
\cL&=\alpha\biggl\{ f_{0}(\tau,\bar{\tau})\, t_{16}R^{4}+\dfrac{3}{2} \left (f_{1}(\tau,\bar{\tau})\, t_{18}G^{2}_{3}R^{3}+f_{-1}(\tau,\bar{\tau})\, t_{18}\ov G^{2}_{3}R^{3}\right )\nn\\[0.5em]
&\hphantom{=\alpha\biggl\{} +f_{0}(\tau,\bar{\tau})\, \left (T(\epsilon_{10},t_{8})-t_{18}  \right )|G_{3}|^{2}R^{3}\biggl \}
\end{align}
where we defined
\begin{equation}\label{eq:KinematicsUniversalG2R3} 
T(\epsilon_{10},t_{8})=-\dfrac{1 }{3}\epsilon_{9}\epsilon_{9}+2\tilde{t}_{8}\tilde{t}_{8}-\dfrac{1}{2}\epsilon_{8}\epsilon_{8}
\end{equation}
and the appropriate (anti-)symmetrisation and contraction of indices is implied.

For later purposes and to make contact with previous work \cite{Richards:2008sa,Liu:2013dna,Liu:2019ses},
we now extract the tree and $1$-loop kinematics in the respective sectors.
In the NSNS-sector one finds after using the large $\mathrm{Im}(\tau)$ expansion of $f_{w}$ in Eq.~\eqref{eq:ExpansionModFuncLargeImTauK}
\begin{align}\label{eq:T18NSNS} 
f_{0}t_{18}|G_{3}|^{2}R^{3}-\dfrac{3}{2}\left (f_{1}t_{18}G_{3}^{2}R^{3}+f_{-1}t_{18}\overline{G}_{3}^{2}R^{3}\right )\biggl |_{\text{NSNS}}&=\ee^{-\phi}\left (f_{0}+\dfrac{3}{2}(f_{1}+f_{-1})\right )t_{18}H_{3}^{2}R^{3}\nn\\[0.4em]
&=4a_{T}\ee^{-\phi} \, t_{18}H_{3}^{2}R^{3} \, ,
\end{align}
while the corresponding RR-sector expression reads
\begin{align}\label{eq:T18RR} 
f_{0}t_{18}|G_{3}|^{2}R^{3}-\dfrac{3}{2}\left (f_{1}t_{18}G_{3}^{2}R^{3}+f_{-1}t_{18}\overline{G}_{3}^{2}R^{3}\right )\biggl |_{\text{RR}}&=\left (f_{0}-\dfrac{3}{2}(f_{1}+f_{-1})\right )t_{18}F_{3}^{2}R^{3}\nn\\[0.4em]
&=-2\ee^{\phi}(a_{T}-a_{L})t_{18}F_{3}^{2}R^{3}\, .
\end{align}
Hence, the structure of terms in \eqref{eq:PredG2R3FromStringsML} together with \eqref{eq:NonMUVResultActionNT} is such that in the NSNS-sector
\begin{align}
\label{eq:NSNSKinematicsH2R3Tree} \cL_{H_{3}^{2}R^{3}}\biggl |_{\text{tree}}&=a_{T}\alpha\ee^{-\phi}\left (-4 \, t_{18} +T(\epsilon_{10},t_{8})\right )H_{3}^{2}R^{3} \, ,\\
\label{eq:NSNSKinematicsH2R3Loop}\cL_{H_{3}^{2}R^{3}}\biggl |_{\text{1-loop}}&= a_{L}\alpha \ee^{-\phi} \, T(\epsilon_{10},t_{8})H_{3}^{2}R^{3}
\end{align}
and in the RR-sector
\begin{align}
\label{eq:RRKinematicsF2R3Tree}\cL_{F_{3}^{2}R^{3}}\biggl |_{\text{tree}}&=a_{T}\alpha\ee^{\phi}\left (2 \, t_{18} +T(\epsilon_{10},t_{8})\right )F_{3}^{2}R^{3} \, ,\\
\label{eq:RRKinematicsF2R3Loop} \cL_{F_{3}^{2}R^{3}}\biggl |_{\text{1-loop}}&=a_{L}\alpha\ee^{\phi}\left (-2\, t_{18}+T(\epsilon_{10},t_{8})\right )F_{3}^{2}R^{3}\, .
\end{align}
The combination of index structures \eqref{eq:KinematicsUniversalG2R3} appears universally in all contributions
since it is associated with the $\mathrm{U}(1)$-neutral part of \eqref{eq:QuinticActionFluxComp}.
Let us further stress that,
while $t_{18}$ plays a role in both tree and 1-loop kinematics for processes $F_{3}^{2}R^{3}$,
it only appears at tree level in the NSNS sector $H_{3}^{2}R^{3}$.
The absence of $t_{18}H_{3}^{2}R^{3}$ at 1-loop has already been observed in \cite{Peeters:2001ub}.
We showed that this is due to an important interplay of modular forms and the relative coefficients in the MUV and, in particular, the non-MUV sector.
In fact, it turns out that the difference of tree and 1-loop kinematics at the level of the effective action (and equally of amplitudes since the pole structure from additional exchange of massless states are removed) is determined by $t_{18}$ only since
\begin{equation}
\Delta\cL_{H_{3}^{2}R^{3}}= \dfrac{1}{a_{T}}\cL_{H_{3}^{2}R^{3}}\biggl |_{\text{tree}}-\dfrac{1}{a_{L}}\cL_{H_{3}^{2}R^{3}}\biggl |_{\text{1-loop}}\sim t_{18}H_{3}^{2}R^{3}
\end{equation}
which agrees with the first line of table~2 in \cite{Liu:2019ses} by comparing to the corresponding line for $t_{18}$ in Tab.~\ref{tab:CoeffKinStructGTRC}.
We expect this to be special about $5$-point amplitudes since the non-MUV sector is unique in the sense that it consists of terms of vanishing $\mathrm{U}(1)$-charge only.

In the superfield language,
the above observation is actually a highly non-trivial cancellation between linear effects (MUV terms) and non-linear contributions (non-MUV terms).
Indeed,
the latter are encoded by
\begin{equation}\label{eq:NonLinearSFContributionG2R3} 
T_{\text{non-lin.}}=-t_{18}+T(\epsilon_{10},t_{8})\, ,
\end{equation}
but the linear superfield contributes another $t_{18}$ such that it precisely cancels out in \eqref{eq:NSNSKinematicsH2R3Loop}.
Hence,
separating Eqs.~\eqref{eq:NSNSKinematicsH2R3Tree}--\eqref{eq:RRKinematicsF2R3Loop} into linear and non-linear superfield contributions,
the tree (1-loop) kinematics is $\mp 3 \, t_{18} +T_{\text{non-lin.}}$ ($\pm t_{18}+ T_{\text{non-lin.}}$) with the upper (lower) sign for $H_{3}$ ($F_{3}$).
We see that precisely at $1$-loop in the NSNS sector the coefficients conspire to cancel $t_{18}$.

In 10 dimensions, $t_{18}$ contains two CP-odd pieces which we ignored throughout this section.
These CP-odd couplings enter at NSNS tree level \eqref{eq:NSNSKinematicsH2R3Tree}, but not at NSNS 1-loop \eqref{eq:NSNSKinematicsH2R3Loop}.
This is precisely opposite to the expectations of \cite{Liu:2019ses} and the terms summarised in \eqref{eq:QuinticActionFluxCompV3}.
From the string world-sheet point of view, the absence of CP-odd couplings at NSNS tree level is due to missing $\epsilon_{10}$ contributions from only NSNS emission vertex operators \cite{Liu:2019ses}.

This apparent issue is resolved by adding an additional CP-odd piece in the non-MUV sector as in \eqref{eq:KinematicsUniversalG2R3} $\sim\vartheta\,  t_{8}\epsilon_{10}$ which must be such that the CP-odd terms in \eqref{eq:NSNSKinematicsH2R3Tree} cancel, i.e., 
\begin{equation}
\left (-4t_{18}\bigl |_{\text{CP-odd}}+\vartheta\,  t_{8}\epsilon_{10}\right )H_{3}^{2}R^{3}=0\, .
\end{equation}
Further,
agreement with \eqref{eq:QuinticActionFluxCompV3} demands
\begin{equation}
\left (-t_{18}\bigl |_{\text{CP-odd}}+\vartheta\,  t_{8}\epsilon_{10}\right ) |G_{3}|^{2}R^{3}=3^{2}\cdot 2^{4}\, G_{3}\wedge X_{7}(\Omega,\ov G_{3})\bigl |_{\text{lin. in } \ov G_{3}}
\end{equation}
which leads us to conclude
\begin{equation}
t_{18}|G_{3}|^{2}R^{3}\bigl |_{\text{CP-odd}}=3\cdot 2^{4}\, G_{3}\wedge X_{7}(\Omega,\ov G_{3})\bigl |_{\text{lin. in } \ov G_{3}}\, .
\end{equation}
We leave a more thorough investigation of CP-odd couplings for the future.

\section{Effective action beyond five points}\label{sec:EightDerBeyondFivePoint}

\subsection{The maximally $\mathrm{U}(1)$-violating couplings}\label{sec:MUVConjecture}

Restricting our attention to terms involving $G_{3}$ and $R$,
we conjecture that maximally $\mathrm{U}(1)$-violating terms are kinematically captured by the linearised superfield in the sense that
\begin{equation}\label{eq:ConjectureMUV} 
\cL_{G_{3},R}^{\text{max.}}=\alpha \sum_{w=0}^{4}\, C_{w}\, f_{w}(\tau,\bar{\tau})\, t_{16+2w}G_{3}^{2w}R^{4-w}+\text{c.c.}
\end{equation}
or more explicitly
\begin{align}\label{eq:ConjectureMUVExp} 
\cL_{G_{3},R}^{\text{max.}}&=\alpha\biggl (C_{0}f_{0}\, t_{16}R^{4}+C_{1}f_{1}\, t_{18}G_{3}^{2}R^{3}+C_{2}f_{2}\, t_{20}G_{3}^{4}R^{2}\nn\\
&\hphantom{=\alpha\biggl (}+C_{3}f_{3}\, t_{22}G_{3}^{6}R+C_{4}f_{4}\, t_{24}G_{3}^{8}+\text{c.c.}\biggl )\, .
\end{align}
The numerical coefficients $C_{w}$ will be discussed below and $\alpha$ was already defined in \eqref{eq:DefAlpha}.
This is supported by observations made in \cite{Boels:2012zr} and confirmed explicitly for the $5$-point structure $G_{3}^{2}R^{3}+\text{c.c.}$ in \cite{Liu:2019ses}.
The special role of MUV amplitudes is further discussed in \cite{Green:2019rhz}.
Further evidence is provided by the $11$D superparticle calculus for which the relevant vertex operator contributions lead to 9D kinematical structures of the form
\begin{align}
\tilde{K}_{G_{3}^{2n}R^{4-n}}&=\int\mathrm{d}^{16}\theta\, \left ((\theta\Gamma^{M_{1}M_{2}M_{3}}\theta)\, G_{M_{1}M_{2}M_{3}}\right )^{2n}\left [(\theta\Gamma^{M_{1}M_{2}}\theta)(\theta\Gamma^{M_{3}M_{4}}\theta)R_{M_{1}M_{2}M_{3}M_{4}}\right ]^{4-n}\nn\\
&=t_{16+2n}G_{3}^{2n}R^{4-n}
\end{align}
and equivalently for the complex conjugates.
One might therefore formulate the conjecture:
\begin{Boxequ}
\emph{The effective action for maximally $\mathrm{U}(1)$-violating tensor structures involving $G_{3}$, $\ov G_{3}$ and $R$ is fully and equivalently determined by either $11$D superparticle amplitudes or the linearised superfield approximation.}
\end{Boxequ}
The equivalence to proper string amplitudes to all loop orders has only been confirmed at $5$-points,
but we expect this to be true up to $8$-points where we conjecture $f_{4}t_{24}G_{3}^{8}+\text{c.c.}$.
Below, we provide further evidence for higher-point coefficients which would appear at the level of string $7$- and $8$-point amplitudes.
In contrast, amplitudes with less $\mathrm{U}(1)$ charge receive further contributions from A) other components of the superparticle vertex operators\footnote{One should keep in mind that this calculus might not necessarily capture the full kinematics due to the light-cone gauge fixing condition. It is hence imperative to make a direct comparison to string amplitude results.}
or from B) non-linear completions of the superfield.

The coefficients $C_{w}$ are such that the pre-factor for MUV terms satisfy \cite{Green:2019rhz}
\begin{equation}\label{eq:IDForModularFormsCovDerCoeff} 
C_{w}f_{w}=2^{w}\cD_{w-1}\ldots\cD_{0}f_{0}\, .
\end{equation}
As discussed in section 4.4 of \cite{Green:2019rhz},
one expects up to $6$-point tree level closed-string amplitudes
\begin{equation}
C_{0}=1\kom C_{1}=\dfrac{3}{2}\kom C_{2}=\dfrac{15}{4}\, .
\end{equation}
One can easily confirm that with this choice and upon applying \eqref{eq:CovDerIdentitiyModForms} the above identities \eqref{eq:IDForModularFormsCovDerCoeff} are indeed satisfied.
We therefore claim that
\begin{equation}\label{eq:ConjectureMUVCoeff} 
C_{w}=\dfrac{2}{\sqrt{\pi}}\, \Gamma\left (\frac{3}{2}+w\right )=\dfrac{\Gamma\left (\frac{3}{2}+w\right )}{\Gamma\left (\frac{3}{2}\right )}\, .
\end{equation}

While determining the coefficients from the string amplitude perspective seems to be quite challenging,
we can make progress by investigating again superparticle amplitudes.
We limit our attention to amplitudes built from the vertex operator contributions
\begin{align}\label{eq:VertexOpGThreeUpdate2} 
V_{h}&\supset 2h_{ij} \cR^{il}\cR^{jm}k_{l}k_{m}\kom V_{G_{3}}\supset -\sqrt{2}G_{3} P^{z} \cR^{lmn} \, .
\end{align}
The left is the vertex operator for the linearised Riemann tensor, while the right is restricted to contributions involving KK states on the $T^{2}$.
In terms of KK-charges $l_{i}$,
we defined
\begin{equation}
P^{z} =\dfrac{1}{\sqrt{\tau_{2}v_{0}}}\left (l_{1}-\tau l_{2}\right )\kom P_{\bar{z}}=\overline{P_{z}}\, .
\end{equation}
The contributions to the Type IIB couplings $G_{3}^{2w}R^{4-w}$ arise from $P$-point amplitudes with $P=4+w$ where
\begin{equation}
v_{0}\cA_{G_{3}^{2w}R^{4-w}}=\dfrac{2^{4-w}\, 2^{w}}{2^{6}\, \Gamma\left (\frac{3}{2}\right )}\, S(P,w,0)\, t_{16+2w}G_{3}^{2w}R^{4-w}
\end{equation}
in terms of
\begin{equation}
S(P,w,0)=\int\, \dfrac{\dif t}{t} \dfrac{t^{P}}{t^{9/2}}\sum_{l_{1},l_{2}}\, P_{z}^{2w}\mathrm{e}^{-t g^{ab}l_{a}l_{b}}\, .
\end{equation}
Here, the factor $2^{P-w}$ arises from the graviton vertex operators and the normalisation $2^{6}\, \Gamma\left (\frac{3}{2}\right )$ is chosen such that the numerical coefficient of $R^{4}$ is set to $C_{0}=1$.
In App.~\ref{sec:MUVSuperPartGRAmps},
we find that $S(P,w,0)$ is given by \eqref{eq:ResultExpressionMUVAmpSupGen} which implies
\begin{equation}
v_{0}\cA_{G_{3}^{2w}R^{4-w}}=\dfrac{\Gamma\left (\frac{3}{2}+w\right )}{\Gamma\left (\frac{3}{2}\right )\sqrt{v_{0}}}\, f_{w}\, t_{16+2w}G_{3}^{2w}R^{4-w}\xrightarrow{\; v_{0}\raw 0\;}\alpha \, C_{w}\, f_{w}\, t_{16+2w}G_{3}^{2w}R^{4-w}
\end{equation}
as expected from \eqref{eq:ConjectureMUV} and \eqref{eq:ConjectureMUVCoeff}.

\subsection{Non-MUV couplings}

The non-MUV couplings are more difficult to determine given the significantly involved kinematics.
However,
we may be able to at least determine a particular class of contributions based on the structure of superparticle amplitudes.
Let us define
\begin{equation}
\cL_{\text{non-MUV}}^{(P)}=\sum_{w=0}^{2(P-4)-1}\,  C_{w}^{(P)}\,  f_{w}\,  \left (t_{16+2(m-w)}+T_{16+2(m-w)}\right ) G_{3}^{m}\ov G_{3}^{m-2w}R^{4+w-m}+\text{c.c.}
\end{equation}
in terms of $m=P-4+w$.
The kinematics is encoded in some tensor structures $t_{16+2(m-w)}+T_{16+2(m-w)}$ where $T_{16+2(m-w)}$ generalises \eqref{eq:KinematicsUniversalG2R3}.
From the superparticle perspective,
a contribution involving $t_{16+2(m-w)}$ is always guaranteed even at higher points by inspecting the corresponding vertex operators.
In the context of the superfield language,
a non-linear contribution $\sim |G_{3}|^{2}$ at order $\Theta^{4}$ equally ensures the presence of a contribution $\sim t_{16+m+n}$ upon expanding $t_{24}\cR^{4}$, see the conclusions in Sect.~\ref{sec:Conclusions} and specifically Tab.~\ref{tab:ColPic}.

We again use the vertex operator contributions defined in \eqref{eq:VertexOpGThreeUpdate2}.
However,
we now pick up contributions that are non-MUV, namely
\begin{equation}
v_{0}\cA_{G_{3}^{m}\ov G^{n}R^{4-(m+n)/2}}=\dfrac{2^{4-(m+n)/2}\, (-2)^{m+n}}{2^{6}\, \Gamma\left (\frac{3}{2}\right )}\, S(P,m,n)\, t_{16+m+n}G_{3}^{m}\ov G_{3}^{n}R^{4-(m+n)/2}
\end{equation}
in terms of
\begin{equation}
S(P,m,n)=\int\, \dfrac{\dif t}{t} \dfrac{t^{P}}{t^{9/2}}\sum_{l_{1},l_{2}}\, P_{z}^{m}P_{\bar{z}}^{n}\mathrm{e}^{-t g^{ab}l_{a}l_{b}}\, .
\end{equation}
These functions can be computed as in the MUV case of App.~\ref{sec:MUVSuperPartGRAmps}.
We compute a total of $10$ amplitudes which we summarise in App.~\ref{sec:NonMUVSuperParticleResults}.
In the limit $v_{0}\raw 0$,
we recover
\begin{equation}
\cL_{\text{non-MUV}}^{(\text{SP})}=\sum_{P=5}^{8}\, \sum_{w=0}^{2(P-4)-1}\, C_{w}^{(P)}\,  f_{w}\, t_{16+2m-w} G_{3}^{m}\ov G_{3}^{m-w}R^{4-2m+w}+\text{c.c.}\kom P=4+2m-w\, .
\end{equation}
in terms of the coefficients
\begin{equation}\label{eq:nonMUVCoeffs} 
C_{w}^{(P)}=\dfrac{(2|w|+1)(2|w|-1)C_{P-4}}{(2(P-4)+1)(2(P-4)-1)}\kom |w|\leq P-4\, .
\end{equation}
Notice that for MUV amplitudes $|w|=P-4$ we recover $C_{P-4}^{(P)}=C_{P-4}$ as expected.
The fact that the coefficients for $10$ distinct amplitudes can be summarised by a single expression \eqref{eq:nonMUVCoeffs} is quite astonishing and hints at a deeper relationship among the various terms even in the non-MUV sector.

While we are unable to verify the correctness of the above results from string amplitudes,
it certainly provides evidence for the appearance of the index structures $t_{16+N}$ even in the non-MUV sector.
Again, such effects are sourced by non-linear couplings in superfield language.

\section{Five-point contact terms in compactifications}
\label{sec:Compactifications}

Compactifications allow to test new higher-derivative interactions by checking their consistency with the constraints imposed by lower-dimensional supersymmetry. In turn, these interactions have interesting implications for lower-dimensional physics.

\subsection{K3 reductions and $\cN=(2,0)$ supersymmetry in six dimensions}

A non-trivial test of the five point couplings concerns K3 reduction to 6 dimensions.  For NSNS sector couplings, these have been previously studied in \cite{Liu:2013dna,Liu:2019ses}.  Here, we take the opportunity and provide a further check including the RR-sector by working in terms of $G_{3}$ directly.  In particular, we highlight several non-trivial cancellations among the various $5$-point index structures of Sect.~\ref{sec:EightDerFivePointAction} necessary to ensure consistency with $\cN=(2,0)$ supersymmetry in 6D.

The reduction of IIB supergravity on K3 results in six-dimensional $\cN=(2,0)$ supergravity coupled to 21 tensor multiplets.  As shown in \cite{Lin:2015dsa}, supersymmetry restricts the four-derivative couplings to be a $F$-term interaction that is quartic in the tensor multiplets.  In particular, the $\cN=(2,0)$ supergravity multiplet receives corrections only starting at the 8-derivative level, just as in the Type II case in 10 dimensions.

The bosonic components of the $\cN=(2,0)$ supergravity multiplet are comprised of a graviton and five self-dual tensors.  From the IIB perspective, the graviton and two of the self-dual tensors come from the spacetime reduction of the 10-dimensional graviton and $G_3$ with self-dual projection.  The other three self-dual tensors arise from reducing the self-dual $F_5$ on the three self-dual 2-cycles of K3.  The bosonic components of a $\cN=(2,0)$ tensor multiplet are comprised of an anti-self dual tensor and five scalars.  Of the 21 tensor multiplets from the reduction, 19 come directly from the reduction of $G_3$ and $F_5$ on the anti-self-dual 2-cycles of K3 along with the $19\times3$ K3 moduli.  The other two tensor multiplets come from the spacetime reduction of $G_3$ with anti-self dual projection along with $G_3$ reduced on the three self-dual 2-cycles of K3, the IIB axio-dilaton, K3 volume modulus, and $F_5$ reduced fully on K3.

For simplicity, we avoid the fields obtained by reducing on the cohomology of K3.  We also disregard six-dimensional scalars since knowledge of the scalar couplings will necessarily be incomplete in the absence of the full 5-point action involving the axio-dilaton.  Thus we focus only on couplings of the $6$D Riemann tensor to $G_{3}$ and its complex conjugate.  These fields will provide information on the $\cN=(2,0)$ supergravity multiplet along with the two special tensor multiplets.

As in \cite{Liu:2019ses}, we focus on factorised pieces where a piece $\int_{\text{K3}}\, R^{2}$ soaks up four derivatives.  This reduces the eight-derivative couplings in ten dimensions to four-derivative couplings in six.  Schematically, such couplings will take the form $R^2$, $G_{3}^2R$ and $G_{3}^4$ (with possible complex conjugates on some of the fields), corresponding to two-, three- and four-point interactions.  Supersymmetry requires the two- and three-point terms to vanish, and restricts the four-point interactions to the tensor multiplets \cite{Lin:2015dsa}.  Restricted to the NSNS fields only (i.e., taking $G_3\to H_3$), Ref.~\cite{Liu:2019ses} confirmed the vanishing of $R^2$ and $H_{3}^2R$ couplings at both tree and one-loop level and demonstrated that the one-loop $H_{3}^4$ coupling is indeed restricted to the tensor multiplet%
\footnote{Tree-level $H_{3}^4$ was not examined in \cite{Liu:2019ses} as that would require knowledge of the six-point $H_{3}^4R^2$ coupling.  On the other hand, the one-loop test was possible because of heterotic/Type II duality.}.
Decoupling of the gravity sector $H_{3}^4$ required the combination of both CP-even and CP-odd four-derivative terms.

Since we only have knowledge of Riemann and $G_3$ couplings up to five points in ten dimensions, we are unable to probe the quartic $G_{3}^4$ couplings in six dimensions.  At the same time, it is well established that the quadratic $\mbox{(Riemann)}^2$ couplings automatically vanish for the IIB combination $(t_8t_8-\fft14\epsilon_8\epsilon_8)R^4$.  Hence we restrict the K3 analysis to the three-point couplings $|G_{3}|^2R$ and $G_{3}^2R$.  Here $R$ is shorthand for the Riemann tensor as Ricci terms can be removed by using the leading order equations of motion, thereby introducing quartic dilaton and 3-form terms in 6D that no longer contribute to three-point couplings.

The six-dimensional $|G_{3}|^2R$ and $G_{3}^2R$ couplings arise from the $\mathrm{U}(1)$-preserving and MUV sectors, respectively.  We start with the ten-dimensional MUV couplings, which are given by (\ref{eq:QuinticActionFluxCompV2}), or more elegantly by (\ref{eq:PredG2R3FromStringsML}).  These $G_{3}^2R^3$ couplings are reduced by taking two of the Riemann tensors to be on K3, leaving $G_{3}^2R$ in six dimensions.  In our choice of basis of $R^{3}$ as given in App.~\ref{app:BasisRC}, the only coefficient that we are sensitive to is $c_{1}$ multiplying $\tilde X^1_{M_1M_2M_3M_4}$ as this is the only term that can yield a factorized form involving $G_3$ and Riemann.  From the decomposition of index structures in Tab.~\ref{tab:CoeffKinStructGTRC}, we then deduce immediately that $t_{18}$ does not contribute any factorised terms in K3 reductions since $c_{1}=0$.  Thus the three-point MUV couplings $G_{3}^2R$ vanish trivially as required by supersymmetry.

Turning to the $\mathrm{U}(1)$-preserving sector, we need to consider the following CP-even pieces from \eqref{eq:FullResultG2R3TreeLoop}
\begin{align}\label{eq:FullResultG2R3TreeLoopK3} 
\cL\bigl |_{K3}&= f_{0}\alpha \biggl \{ t_{16}R^{4}+6\left (t_{8}t_{8}-\dfrac{1}{4}\epsilon_{8}\epsilon_{8}\right )|\nabla G_{3}|^{2}R^{2}+(T(\epsilon_{10},t_{8}) -t_{18})\,  |G_{3}|^{2}R^{3}\biggl \}\biggl |_{K3}\, .
\end{align}
As mentioned above, in $6$D the $\mathrm{(Riemann)}^2$ term in the factorised part of $t_{16}R^{4}$ cancel, leaving us with only the $|\nabla G_{3}|^2R^2$ and $|G_{3}|^2R^3$ terms to consider.  Furthermore, the $t_{18}$ term vanishes on K3 for the same reason that its MUV counterpart vanishes.

From the definition \eqref{eq:KinematicsUniversalG2R3} of $T(\epsilon_{10},t_{8})$, we notice by using the coefficients $c_{1}$ collected in Tab.~\ref{tab:CoeffKinStructGTRC} that
\begin{equation}
\left (2\tilde{t}_{8}\tilde{t}_{8}-\dfrac{1}{2}\epsilon_{8}\epsilon_{8}\right )|G_{3}|^{2}R^{3}=0+\cdots,
\end{equation}
where $\cdots$ denotes Ricci and non-factorised terms. Recall that this is the piece obtained from generalised geometry.  In addition, resorting to Tab.~\ref{tab:CoeffKinStructGTRC}, we find the factorised terms inside $\epsilon_{9}\epsilon_{9}|G_{3}|^{2}R^{3}$ to be
\begin{equation}
-\dfrac{1}{3}\epsilon_{9}\epsilon_{9}|G_{3}|^{2}R^{3}=-48 R^{N_{1}N_{2}M_{1}M_{2}}\, \ov G_{N_{1}N_{2}}\,^{P}G_{M_{1}M_{2}P}\mbox{(Riemann)}^2+\cdots.
\label{eq:e9e9inT}
\end{equation}
We are thus left with
\begin{equation}
T(\epsilon_{10},t_{8})\,  |G_{3}|^{2}R^{3}=-48 R^{N_{1}N_{2}M_{1}M_{2}}\, \ov G_{N_{1}N_{2}}\,^{P}G_{M_{1}M_{2}P}\mbox{(Riemann)}^2+\cdots.
\end{equation}
Consistency with the lack of three-point interactions in $\cN=(2,0)$ supersymmetry requires that this term vanishes when combined with the factorized $|\nabla G_{3}|^2R^2$ contribution.

We now determine the factorised piece inside $|\nabla G_{3}|^{2}R^{2}$.
After using the Bianchi identity for $G_{3}$, i.e., $\dif G_{3}=0$ up to axio-dilaton terms,
we obtain
\begin{align}
6\left (t_{8}t_{8}-\dfrac{1}{4}\epsilon_{8}\epsilon_{8}\right )|\nabla G_{3}|^{2}R^{2}&=48 \left (\nabla_{M_{1}} \ov G_{M_{2}M_{3}M_{4}}\right )\left (\nabla^{M_{2}}  G^{M_{1}M_{3}M_{4}}\right )\mbox{(Riemann)}^2  +\cdots\nn\\
&=48 R^{M_{1}M_{2}M_{3}M_{4}}\, \ov G_{M_{1}M_{2}}\,^{P}G_{M_{3}M_{4}P} \mbox{(Riemann)}^2  +\cdots,
\end{align}
where we integrated by parts in the second line and ignored terms proportional to the equations of motion for $G_{3}$.  It is now apparent that the factorized piece inside the $\mathrm{U}(1)$-preserving sector cancels, namely
\begin{equation}
6\left (t_{8}t_{8}-\dfrac{1}{4}\epsilon_{8}\epsilon_{8}\right )|\nabla G_{3}|^{2}R^{2}+T(\epsilon_{10},t_{8})\,  |G_{3}|^{2}R^{3}=0+\cdots \, .
\end{equation}
In this way, we tested the coefficient $-1/3$ of $\epsilon_{9}\epsilon_{9}$ implicit in (\ref{eq:e9e9inT}) in the non-MUV sector to which the Calabi-Yau threefold reductions are insensitive.

Finally, note that there are CP-odd couplings in both the $\textrm{U}(1)$-preserving and MUV sector that can potentially lead to three-point couplings.  However, as shown in \cite{Liu:2013dna} and further discussed in \cite{Liu:2019ses}, the CP-odd $H_{3}^2R$ coupling involves the Ricci tensor, and hence does not contribute to the three-point function.  The same argument applies for couplings to the complex three-form $G_{3}$.

\subsection{Calabi-Yau threefold reductions to four dimensions}\label{sec:ScalarPotentialSLTZInvDer} 

We now apply our previous results in the context of Calabi-Yau (CY) threefold reductions to $\cN=2$ 4D SUGRA.
Given that we identified the relevant $5$-point kinematics between the $3$-form and the metric,
we are in the perfect position to directly derive for the first time from 10D the $(\alpha^{\prime})^{3}$-corrected $4$D scalar potential in compactifications on CY threefolds $X_{3}$ with background fluxes.

In the low-energy 4D effective action,
$3$-form fluxes induce a non-trivial $F$-term scalar potential \cite{Giddings:2001yu}.
More precisely,
it is determined by a Kähler potential $\cK$ and superpotential $\cW$ through
\begin{equation}\label{eq:IIB:ScalarPotN1}
V_{F}=\ee^{\cK}\left (\cK^{A\bar{B}}\, D_{A}\cW\, D_{\bar{B}}\overline{\cW}-3|\cW|^{2}\right )\kom D_{A}\cW=\p_{A}\cW+\cK_{A}\cW\, .
\end{equation}
Here, the sum over $A$ includes $h^{1,1}(X_{3})$ Kähler moduli $T_{\alpha}$, $h^{1,2}(X_{3})$ complex structure moduli $Z_{i}$ and the axio-dilaton $\tau$.
We work with the Kähler potential \cite{Antoniadis:1997eg}
\begin{equation}\label{eq:KPCorrectedFullSL} 
\cK=\cK^{(0)}-2\log\left (\cV+\dfrac{\zeta}{4} f_{0}(\tau,\bar{\tau})\right )
\end{equation}
where $\cK^{(0)}$ is the Kähler potential on complex structure moduli space
\begin{equation}\label{eq:KPCSMSL} 
\cK^{(0)}=-\ln(-\I(\tau-\bar{\tau}))-\log\left (-\I\int_{X_{3}}\, \Omega\wedge\overline{\Omega}\right )\, .
\end{equation}
Moreover, the constant $\zeta$ is given by \cite{Antoniadis:1997eg}
\begin{equation}\label{eq:KPCorrectedFullSLExp} 
 \zeta=-\dfrac{\chi(X_{3})}{2(2\pi)^{3}}
\end{equation}
in terms of the Euler characteristic $\chi(X_{3})$ of $X_{3}$.
In the case of Type IIB flux compactifications,
the superpotential entering \eqref{eq:IIB:ScalarPotN1} is the \emph{Gukov-Vafa-Witten superpotential} \cite{Gukov:1999ya,Giddings:2001yu} (recall \eqref{eq:TauPGT})
\begin{equation}\label{eq:FC:Superpot} 
\cW\equiv\cW_{\text{GVW}}(\tau,Z)=  \int_{X_{3}}\,  \tilde{G}_{3}\wedge\Omega_{3}\, .
\end{equation}
This superpotential depends on the axio-dilaton $\tau$ through the complexified $3$-form flux as well as on the complex structure moduli $Z_{i}$, $i=1,\ldots , h^{1,2}(X_{3})$, due to the presence of the holomorphic $3$-form $\Omega_{3}=\Omega_{3}(Z_{i})$.


Despite many efforts, the derivation of  $(\alpha^{\prime})^{3}$ corrections to the $4$D flux scalar potential from first principles is still lacking.
It was already noted in \cite{Becker:2002nn} that this requires the presence of both non-MUV and MUV higher derivative terms in 10D of the form $|G_{3}|^{2}R^{3}$, $G_{3}^{2}R^{3}$ and $\ov G_{3}^{2}R^{3}$.
One might also expect $|\nabla G_{3}|^{2}R^{2}$ to contribute at the same level based on dimensional grounds \cite{Conlon:2005ki,Burgess:2020qsc,Cicoli:2021rub}.
Of course, a major bottleneck has been the construction of the $10$D higher derivative action for $G_{3}$ up to $5$-points which has finally been achieved through \cite{Policastro:2006vt,Policastro:2008hg,Richards:2008sa,Liu:2013dna,Liu:2019ses} and further concretised in this paper.

Below,
we show that the structure of the $F$-term scalar potential \eqref{eq:IIB:ScalarPotN1} beyond leading order depends only on a single 10D kinematical structure, while the relative coefficients between RR-flux and NSNS-flux contributions can already be identified from 10D $\mathrm{SL}(2,\mathbb{Z})$ invariance.
The remaining overall coefficient can only be identified upon constructing the corrected flux background which is beyond the scope of the current work, though this should be feasible by employing similar strategies to those of \cite{Becker:2001pm,Grimm:2014xva}.
Instead, we deduce novel relationships between 4D supersymmetry and the kinematical structure of 10D higher derivative terms where once again $t_{18}$ plays a very prominent role.
More specifically,
we argue that the absence of RR-flux contributions to \eqref{eq:IIB:ScalarPotN1} at order $(\alpha^{\prime})^{3}$ at string tree level already observed in \cite{Becker:2002nn} highly constrains the non-MUV kinematics when put entirely on a CY threefold.
The strategy is as follows: By investigating the form of the corrected scalar potential \eqref{eq:IIB:ScalarPotN1} obtained from \eqref{eq:KPCorrectedFullSL} and \eqref{eq:FC:Superpot},
we trace constraints from $4$D supersymmetry back to the $10$D action \eqref{eq:FullResultG2R3TreeLoop}.

\subsubsection{The 4D perspective}

In this first part,
we compute \eqref{eq:IIB:ScalarPotN1} by plugging in the Kähler potential \eqref{eq:KPCorrectedFullSL} and the superpotential \eqref{eq:FC:Superpot}.
Here, it is convenient to expand $V_{F}$ as follows
\begin{equation}\label{eq:FtermScalarPotAP3} 
V_{F}=\dfrac{V_{\text{Flux}}}{\cV^{2}}+(\alpha^{\prime})^{3}V^{(3)}+\cO((\alpha^{\prime})^{4})\, .
\end{equation}
The first term corresponds to the standard no-scale flux scalar potential given by \cite{Polchinski:1995sm,Michelson:1996pn,Gukov:1999ya,Dasgupta:1999ss}\footnote{
To arrive at the RHS,
one decomposes $\tilde{G}_{3}^{+}$ in a basis $\lbrace \Omega,\ov\chi_{i}\rbrace$ of $H^{(3,0)}\oplus H^{(1,2)}$ (see e.g. appendix B of \cite{Grimm:2004uq})
\begin{equation}
\tilde{G}_{3}^{+}=-\dfrac{1}{\int_{X_{3}}\, \Omega\wedge\overline{\Omega}}\left (\Omega \int_{X_{3}}\, \overline{\Omega}\wedge  \tilde{G}_{3}+\cK^{i\bar{j}}\overline{\chi}_{j}\int_{X_{3}}\, \chi_{i}\wedge  \tilde{G}_{3}\right )\, .
\end{equation}
The other term $\sim \int_{X_{3}}\,  {G}_{3}\wedge \overline{{G}}_{3}$ is cancelled through the integrated Bianchi identity for $F_{5}$ \cite{Giddings:2001yu}.
}
\begin{equation}
V_{\text{Flux}}=\dfrac{1}{2\mathrm{Im}(\tau)}\int_{X_{3}}\,  \tilde{G}_{3}^{+}\wedge\star_{6} \overline{ \tilde{G}}_{3}^{+}=\ee^{\cK^{(0)}}\left (\cK^{i\bar{\jmath}}\, D_{i}\cW\, D_{\bar{\jmath}}\overline{\cW}+\cK^{\tau\bar{\tau}}\, D_{\tau}\cW\, D_{\bar{\tau}}\overline{\cW} \right )
\end{equation}
in terms of (A)ISD flux $\star_{6} \tilde{G}_{3}^{\pm}=\mp\I \tilde{G}_{3}^{\pm}$.
As explained in \cite{Becker:2002nn}, the $(\alpha^{\prime})^{3}$ corrections encoded in $V^{(3)}$ fall into two classes, i.e.,
\begin{equation}\label{eq:ScalarPotentialAPTCorrectionSL} 
V^{(3)}=-\dfrac{\zeta f_{0}(\tau,\bar{\tau})}{2\cV^{3}}\, V_{\text{Flux}}+V_{\zeta}\, .
\end{equation}
The first term simply originates from a $4$D Weyl rescaling $g_{4}^{E}=(\cV+\zeta f_{0}/4)g_{4}$ by expanding to linear order in $\zeta$.

The second term $V_{\zeta}$ is more interesting because it is obtained from direct dimensional reduction of \eqref{eq:FullResultG2R3TreeLoop}. Using \eqref{eq:CovDerMoFunc} for the $4$D axio-dilaton, one finds to linear order in $\zeta$ 
\begin{align}\label{eq:CorrectedScalarPotentialSLTZInvDWC} 
V_{\zeta}&=\dfrac{3\zeta \mathrm{e}^{\cK^{(0)}}}{8\cV^{3}}\biggl \{f_{0}\left [|\cW|^{2}-(\tau-\bar{\tau})^{2}\, |D_{\tau}\cW|^{2}\right ]+(\tau-\bar{\tau})\left [f_{-1}\overline{\cW}D_{\tau}\cW-f_{1}{\cW}\overline{D_{\tau}\cW}\right ] \biggl\}
\end{align}
where the classical Kähler covariant derivative with respect to $\tau$ is given by
\begin{equation}\label{eq:KCVDer} 
D_{\tau}\cW=(\p_{\tau}+\cK_{\tau}^{(0)})\cW=\dfrac{-1}{\tau-\bar{\tau}} \int_{X_{3}}\, \overline{\tilde{G}}_{3}\wedge\Omega\, .
\end{equation}
As shown in \cite{Becker:2002nn}, the coefficient on the right hand side cannot be reproduced by simply considering the flux kinetic term in the corrected $10$D background. It was therefore argued that additional higher-derivative terms must contribute in the reduction.

To make contact with expressions obtained from direct dimensional reduction,
it is even more instructive to rewrite \eqref{eq:CorrectedScalarPotentialSLTZInvDWC} in terms of fundamental integrals using \eqref{eq:FC:Superpot} and \eqref{eq:KCVDer} such that
\begin{align}\label{eq:CorrectedScalarPotentialSLTZInvIntegralsGOV1} 
V_{\zeta}&= \dfrac{3}{8}\;\dfrac{\zeta \mathrm{e}^{\cK^{(0)}}}{\cV^{3}}\biggl \{f_{0}\left [ \int_{X_{3}}\,   \tilde{G}_{3}\wedge\Omega\int_{X_{3}}\,  \overline{\tilde{G}}_{3}\wedge\overline{\Omega}+\int_{X_{3}}\, \overline{\tilde{G}}_{3}\wedge\Omega\int_{X_{3}}\,  {\tilde{G}}_{3}\wedge\overline{\Omega}\right ]\nn\\
&\quad-\left [f_{1}\int_{X_{3}}\,   \tilde{G}_{3}\wedge\Omega \int_{X_{3}}\,  {\tilde{G}}_{3}\wedge\overline{\Omega}+f_{-1}\int_{X_{3}}\,  \overline{\tilde{G}}_{3}\wedge\overline{\Omega}\int_{X_{3}}\, \overline{\tilde{G}}_{3}\wedge\Omega \right ] \biggl\}\, .
\end{align}
The above expressions makes clear the way the three different kinematical structures appear in the reduction to $4$D.
In particular, the $10$D $\mathrm{U}(1)$-violating terms appear prominently in the scalar potential as already anticipated in \cite{Becker:2002nn}.
The $\mathrm{U}(1)$-neutral contribution is obtained from $|G_{3}|^{2}R^{3}$ and $|\nabla G_{3}|^{2}R^{2}$, but is also affected by backreaction effects from warping.

For the subsequent arguments,
it is actually instructive to separate the above integrals into their NSNS- and RR-flux contributions at tree and 1-loop level.
Plugging in the expansion \eqref{eq:ExpansionModFuncTL} for the modular functions at large $\mathrm{Im}(\tau)$, we find (ignoring non-perturbative terms $\cO(\ee^{-\text{Im}(\tau)})$)
\begin{align}\label{eq:CorrectedScalarPotentialSLTZInvIntegralsGO} 
V_{\zeta}&= \dfrac{\zeta \mathrm{e}^{\cK^{(0)}}}{\cV^{3}}  \left (-\dfrac{1}{4}\right ) \biggl \{ \left (-6a_{T}-2a_{L}\right )\ee^{-2\phi_{0}}  \int_{X_{3}}\,   H_{3}\wedge\Omega\int_{X_{3}}\,  H_{3}\wedge\overline{\Omega} \nn\\
&\quad +\left ( -4a_{L}\right ) \int_{X_{3}}\,   F_{3}\wedge\Omega \int_{X_{3}}\,  F_{3}\wedge\overline{\Omega}+\ldots  \biggl\}\, .
\end{align}
Here, the tree-level term only depends on NSNS-flux and is in agreement with \cite{Becker:2002nn} after using $\xi=\zeta(3)\, \zeta$.
In the subsequent section,
we argue that the structure of \eqref{eq:CorrectedScalarPotentialSLTZInvIntegralsGO} is directly accessible from our 10D expressions for the NSNS- and RR-flux terms \eqref{eq:NSNSKinematicsH2R3Tree} -- \eqref{eq:RRKinematicsF2R3Loop} derived in Sect.~\ref{sec:TreeOneLoopKinFiveP}.

\subsubsection{Dimensional reduction}

Beforehand,
let us collect all of the relevant pieces potentially contributing to \eqref{eq:CorrectedScalarPotentialSLTZInvIntegralsGO}.
Initially,
we write the $10$D action as
\begin{equation}\label{eq:FullActionReduction4DSch} 
S=S^{(0)}+\alpha S^{(3)}+\cO((\alpha^{\prime})^{5})
\end{equation}
where $S^{(0)}$ is the classical action \eqref{eq:ch1:ActionIIB10D} (setting $2\kappa_{10}^{2}=1$ in what follows).
After solving the equations of motion to order $(\alpha^{\prime})^{3}$,
the solutions will be of the form
\begin{equation}\label{eq:CorrectedBackgroundParam} 
\varphi=\varphi^{(0)}+\alpha \left (f_{0}\varphi^{(1)}_{0}+f_{1}\varphi^{(1)}_{1}+\text{c.c.}\right )+\cO((\alpha^{\prime})^{5})
\end{equation}
for $\varphi\in \lbrace g,\tau,G_{3},F_{5},\cA\rbrace$.
In particular, in the presence of non-trivial $G_{3}$, the background becomes warped as parametrised by $\cA$.
We provide a few more details on the corrected background in App.~\ref{sec:BackgroundCYReduction}.
In \eqref{eq:CorrectedBackgroundParam},
we allow backreaction effects that carry non-trivial $\mathrm{U}(1)$-charge which we expect to appear for $G_{3}$ and $\cP$, though the latter are irrelevant in our Type IIB background where $\cP$ vanishes internally.

\paragraph*{The scalar potential}

For the subsequent discussion,
we collect the relevant terms in \eqref{eq:FullActionReduction4DSch} (after evaluation on the corrected background) contributing to \eqref{eq:CorrectedScalarPotentialSLTZInvIntegralsGOV1} in the following action
\begin{equation}
S_{\text{Flux}}=S_{\text{backreact.}}+S_{|\nabla G_{3}|^{2}R^{2}}+S_{|G_{3}|^{2}R^{3}}+S_{G_{3}^{2}R^{3}+\text{c.c.}}\, .
\end{equation}
Here, $S_{\text{backreact.}}$ arises from evaluating the classical action on the corrected background.
Then,
the scalar potential can very schematically be written as
\begin{equation}
M_{P}^{4}\int\, V_{F}\,\sqrt{-g^{(4)}}\dif^{4}x=S_{\text{Flux}}\biggl |_{X_{3}}
\end{equation}
where $\ldots\bigl |_{X_{3}}$ indicates that all legs of each tensor are taken along the internal CY directions.

At the level of the above discussion,
we may limit our attention to terms that have the same form as $|G_{3}|^{2}R^{3}$ (at least in the reduction) for which we write
\begin{align}
\label{eq:BackReactQuarticRedFD} 
S_{\text{backreact.}}+S_{|\nabla G_{3}|^{2}R^{2}}&=\alpha \int\,   \left (f_{0}\,  \delta_{0} |G_{3}|^{2}R^{3}+f_{1}\,  \delta_{1} G_{3}^{2}R^{3}+f_{-1}\,  \delta_{1} \ov G_{3}^{2}R^{3}\right ) \,\star_{10}1
\end{align}
in terms of some index structures $\delta_{0}, \delta_{1}$.
We comment further on the corrected background and contributions to $\delta_{0}, \delta_{1}$ in Sect.~\ref{sec:BackgroundCYReduction}.
We stress however that their actual form is completely irrelevant for our argument.
We essentially rely on $\cN=2$ supersymmetry in $4$D: We know that in the absence of D-branes and O-planes there are no additional contributions coming from the above reduction.
That is,
we argue that all such terms must have the form given in \eqref{eq:BackReactQuarticRedFD}.

We know due to $4$D supersymmetry that the total contribution from higher-derivative terms and backreaction effects involving the RR-flux has to vanish at tree level.
In particular,
this implies that the reduction of the tree level $F_{3}^{2}R^{3}$ terms \eqref{eq:RRKinematicsF2R3Tree} has to cancel against backreaction effects, i.e.,
\begin{equation}\label{eq:ReductionF2R3TreeLevelCYT} 
S_{\text{Flux}} \biggl |_{\text{RR, tree, }X_{3}}=\alpha \int\,  a_{T}\ee^{\phi}\left (2 \, t_{18} +T(\epsilon_{10},t_{8})+\delta_{0}+2\delta_{1} \right )F_{3}^{2}R^{3}\biggl |_{X_{3}}=0\, .
\end{equation}
Even more importantly,
this means that kinematically
\begin{equation}
 \int\,  T(\epsilon_{10},t_{8}) F_{3}^{2}R^{3}\biggl |_{X_{3}}= - \int\,   (2 \, t_{18}+\delta_{0}+2\delta_{1}) F_{3}^{2}R^{3}\biggl |_{X_{3}}
\end{equation}
and by trivial extension this also holds for $H_{3}^{2}R^{3}$ (it does not matter what the $3$-form is in our background).
It is crucial to notice that this kinematical constraint based on the requirement of $4$D $\cN=2$ supersymmetry is the main ingredient for our argumentation.
It relates the complicated kinematics in the non-MUV sector encoded by $T(\epsilon_{10},t_{8})$ to $t_{18}$ as well as further uknown effects in the reduction through requiring the absence of RR-flux in the $4$D scalar potential.\footnote{Clearly, it would have been great to prove the absence from first principles via direct dimensional reduction, but this is beyond the scope of the present work.}
Overall,
the contribution to the $\cN=1$ $4$D scalar potential from \eqref{eq:NSNSKinematicsH2R3Tree} -- \eqref{eq:RRKinematicsF2R3Loop} are given by
\begin{align}
\label{eq:NSNSKinematicsH2R3TreeMod}  S_{\text{Flux}} \biggl |_{\text{NSNS, tree, }X_{3}}&=-6\,  \alpha \int\,  a_{T}\ee^{-\phi}\,   \left [t_{18}+\dfrac{2}{3}\delta_{1}\right ]\, H_{3}^{2}R^{3}\biggl |_{X_{3}} \, ,\\
\label{eq:NSNSKinematicsH2R3LoopMod} S_{\text{Flux}} \biggl |_{\text{NSNS, 1-loop, }X_{3}}&=  -2\, \alpha\int\,   a_{L} \ee^{-\phi}\, \left [t_{18}+\dfrac{2}{3}\delta_{1}\right ]\, H_{3}^{2}R^{3}\biggl |_{X_{3}}\, ,\\
\label{eq:RRKinematicsF2R3TreeMod} S_{\text{Flux}} \biggl |_{\text{RR, tree, }X_{3}}&=0\, ,\\
\label{eq:RRKinematicsF2R3LoopMod} S_{\text{Flux}} \biggl |_{\text{RR, 1-loop, }X_{3}}&=-4\,  \alpha\int\,  a_{L}\ee^{\phi}\, \left [ t_{18}+\dfrac{2}{3}\delta_{1}\right ]\, F_{3}^{2}R^{3}\biggl |_{X_{3}}\, .
\end{align}
The relative coefficients are \emph{exactly the ones found in 4D} in \eqref{eq:CorrectedScalarPotentialSLTZInvIntegralsGO}.
In order to derive also the overall coefficient as well as the structure of terms in \eqref{eq:CorrectedScalarPotentialSLTZInvIntegralsGO},
it remains to show that
\begin{equation}\label{eq:IntegralT18CYThreefold} 
\alpha \int_{X_{3}}\, \ee^{-\phi} \left [t_{18}+\dfrac{2}{3}\delta_{1}\right ]\, H^{2}_{3}R^{3}=-\dfrac{\zeta \ee^{\cK^{(0)}}}{4 \cV}\, \int_{X_{3}}\, H_{3}\wedge\Omega\, \int_{X_{3}}\, H_{3}\wedge\ov \Omega
\end{equation}
and equivalently for $F_{3}^{2}R^{3}$.

Let us highlight the importance of the above result:
Imposing only the absence of RR-flux in 4D through \eqref{eq:ReductionF2R3TreeLevelCYT} gave rise to a single relevant kinematical structure depending on $t_{18}$ and possible backreaction effects entering in the MUV sector.
Stated differently, only the MUV kinematics is relevant which is a rather unexpected observation from the 10D point of view.
The remaining relative coefficients are fixed through $\mathrm{SL}(2,\mathbb{Z})$ invariance.
Clearly,
these arguments only apply to the scalar potential, though the non-MUV kinematics will be checked at the level of $4$D kinetic terms for hypermultiplets below.

Having derived the above results,
we may actually come back to the MUV expression in 10D \eqref{eq:PredG2R3FromStringsML} which must reduce to the second line of \eqref{eq:CorrectedScalarPotentialSLTZInvIntegralsGOV1}.
Using \eqref{eq:IntegralT18CYThreefold} for $G_{3}$, one verifies that reducing \eqref{eq:PredG2R3FromStringsML} together with MUV backreaction effects leads to
\begin{equation}\label{eq:IntegralT18CYThreefoldMUV} 
\alpha \int_{X_{3}}\, \ee^{-\phi} \left [\dfrac{3}{2}t_{18}+\delta_{1}\right ]\, G^{2}_{3}R^{3}=-\dfrac{3\zeta \ee^{\cK^{(0)}}}{8 \cV}\, \int_{X_{3}}\, G_{3}\wedge\Omega\, \int_{X_{3}}\, G_{3}\wedge\ov \Omega
\end{equation}
with $-3/8$ being precisely the coefficient in \eqref{eq:CorrectedScalarPotentialSLTZInvIntegralsGOV1}.
Let us stress that, if we had not gone through the argument for NSNS- and RR-flux separately,
the kinematic constraints arising in the reduction would have been totally obscure in the non-MUV sector.

While it will generically be hard to identify $\delta_{1}$ explicitly,
reducing $t_{18}$ should be feasible.
Depending on the result, one might be able to identify $\delta_{1}$ indirectly through \eqref{eq:IntegralT18CYThreefoldMUV}. 
Naively, given that in the MUV sector the complete kinematics is determined by $t_{18}$ only,
one could speculate that $\delta_{1}$ is kinematically related to $t_{18}$, i.e., $\delta_{1}=a_{1}t_{18}$ for some numerical constant $a_{1}$.

Clearly,
there remain several interesting future directions.
For once,
the absence of $(2,1)$-form flux in \eqref{eq:CorrectedScalarPotentialSLTZInvDWC} (i.e., no couplings involving $D_{Z}\cW$) as already observed in \cite{Becker:2002nn} requires a delicate cancellation among higher-derivative terms.
We essentially reduced this problem to proving that \eqref{eq:IntegralT18CYThreefold} contains no such terms.
We leave a derivation of \eqref{eq:CorrectedScalarPotentialSLTZInvDWC} as well as \eqref{eq:IntegralT18CYThreefold} from direct dimensional reduction for future works.

\paragraph{The kinetic terms}

As a final application of our five-point results,
we derive the moduli space metrics for the hypermultiplets in Type IIB reductions to 4D.
Initially,
we recall that the non-chirality of Type IIA implies that both sign combinations of the odd-odd $\epsilon_{8}\epsilon_{8}R^{4}$ structure appear in the 10D action.
In CY threefold reductions to four dimensions,
this implies that the Einstein Hilbert term is corrected as $(a_{T}-a_{L})\chi(X_{3})R^{(4)}$.
Ultimately, this ensures that the vectormultiplets are only corrected at tree level, while hypermultiplets are corrected at 1-loop \cite{Antoniadis:2003sw}.
In contrast,
the hypermultiplets in Type IIB are corrected at both tree and 1-loop level, while the vectormultiplets remain uncorrected, see e.g. \cite{Grana:2014vva}.

In Type IIB, the hypermultiplet scalars consist of Kähler moduli $t^{\alpha}$ as well as $p$-form axions $(c^{\alpha}, b^{\alpha},\rho_{\alpha})$.
Let $\omega_{\alpha}\in H^{1,1}(X_{3})$ be a basis of $(1,1)$-forms.
We express the CY volume in terms of the Kähler moduli as follows
\begin{equation}
\cV=\int_{X_{3}}\, J\wedge J\wedge J=\dfrac{1}{3!}k_{\alpha\beta\gamma}\, t^{\alpha}t^{\beta}t^{\gamma}\kom k_{\alpha\beta\gamma}=\int_{X_{3}}\, \omega_{\alpha}\wedge\omega_{\beta}\wedge\omega_{\gamma}
\end{equation}
where we expanded the Kähler form as $J=t^{\alpha}\omega_{\alpha}$.
Below,
we make use of the following
\begin{equation}
G_{3}=\omega_{\alpha}G^{\alpha}\kom G^{\alpha}=\dif c^{\alpha}-\tau\dif b^{\alpha}\kom k_{\alpha}=\dfrac{1}{2}k_{\alpha\beta\gamma}\, t^{\beta}t^{\gamma}\kom k_{\alpha\beta}=k_{\alpha\beta\gamma}\, t^{\gamma}\, .
\end{equation}

In the following, we reduce the relevant higher-derivative terms in \eqref{eq:FullResultG2R3TreeLoop} and \eqref{eq:FourPointCouplingsAP3} together with contributions from the corrected background \eqref{eq:BackReact10DWeyl}, see Sect.~\ref{sec:BackgroundCYReduction} for details.
We partially use the tree level results of \cite{Bonetti:2016dqh,Grimm:2017okk} and provide the full reduction in App.~\ref{sec:Reduction4DCaclDet}.
Overall,
we obtain the 4D action
\begin{align}\label{eq:Action4DKinTFin} 
S^{(4)}&=\int \biggl \{\left [R^{(4)}-V_{\text{F}}\right ]\star_{4}1-\ee^{2\phi}\left (\dfrac{1}{2}- \dfrac{3 f_{0}\zeta}{16\cV^{2}}\right ) \dif\tau\wedge\star_{4}\dif\ov \tau\nn\\
&\hphantom{=\int \biggl \{}- \I \dfrac{3\ee^{\phi}\zeta k_{\alpha}}{4\cV^{2}}\dif  t^{\alpha}\wedge\star_{4}(f_{1}\dif\tau-f_{-1}\dif \ov \tau)\nn\\
&\hphantom{=\int \biggl \{}  +\left (\dfrac{1}{\cV}\left [\dfrac{1}{2}k_{\alpha\beta}-\dfrac{k_{\alpha}k_{\beta}}{\cV}\right ]-\dfrac{ f_{0} \zeta}{8\cV^{2}}\left [k_{\alpha\beta}-4\dfrac{k_{\alpha}k_{\beta}}{\cV}\right ]\right )\dif t^{\alpha}\wedge\star_{4}\dif  t^{\beta}\nn\\
&\hphantom{=\int \biggl \{}+\ee^{\phi}\left (\dfrac{1}{2\cV}\left [k_{\alpha\beta}-\dfrac{k_{\alpha}k_{\beta}}{\cV}\right ]-\dfrac{ f_{0}\zeta}{8\cV^{2}}\left [k_{\alpha\beta}-\dfrac{5}{4}\dfrac{k_{\alpha}k_{\beta}}{\cV}\right ]  \right ) G^{\alpha}\wedge\star_{4}\ov G^{\beta}\nn\\
&\hphantom{=\int \biggl \{}-\dfrac{3\zeta\ee^{\phi}k_{\alpha}k_{\beta}}{64\cV^{3}}  \left [f_{1}\,  G^{\alpha}\wedge\star_{4} G^{\beta}+f_{-1}\,  \ov G^{\alpha}\wedge\star_{4} \ov G^{\beta}\right ]\biggl \}\, .
\end{align}
Here,
$V_{\text{F}}$ is the $(\alpha^{\prime})^{3}$-corrected scalar potential defined in \eqref{eq:FtermScalarPotAP3}.
The terms without $G^{\alpha}$ restricted to tree level are equivalent to \cite{Bonetti:2016dqh}.
What is left to be done is to find suitable coordinates to bring the above into a canonical form.

To match with 4D SUSY,
let us first compute the Kähler metric from the Kähler potential $\cK$ defined in \eqref{eq:KPCorrectedFullSL}, namely
\begin{align}
\cK_{\alpha\beta}&=-\dfrac{2}{\cV}\left [k_{\alpha\beta}-\dfrac{k_{\alpha}k_{\beta}}{\cV}\right ]+\dfrac{\zeta f_{0}}{2\cV^{2}}\left [k_{\alpha\beta}-2\dfrac{k_{\alpha}k_{\beta}}{\cV}\right ]\, ,\nn\\[0.5em]
\cK_{\alpha\tau}&=-\I\ee^{\phi}\dfrac{3f_{1}}{8}\dfrac{k_{\alpha}\zeta}{\cV^{2}} \kom \cK_{\tau\ov\tau}=\dfrac{\ee^{2\phi}}{2}\left (\dfrac{1}{2}-\dfrac{3\zeta f_{0}}{16\cV}\right )\, .
\end{align}
Let us ignore all terms $\sim k_{\alpha}$ in the above action for a moment (these are affected by field redefinitions).
Then we have
\begin{align}
S^{(4)}&=\int \biggl \{\left [R^{(4)}-V_{\text{F}}\right ]\star_{4}1-2\cK_{\tau\bar{\tau}} \dif\tau\wedge\star_{4}\dif\ov \tau\nn\\
&\quad  -\dfrac{1}{4}\tilde{\cK}_{\alpha\beta} \left [\dif t^{\alpha}\wedge\star_{4}\dif  t^{\beta}+\ee^{\phi} G^{\alpha}\wedge\star_{4}\ov G^{\beta}\right ]\biggl \}
\end{align}
where
\begin{equation}
\tilde{\cK}_{\alpha\beta}=-\dfrac{2k_{\alpha\beta}}{\cV}\left [1-\dfrac{\zeta f_{0}}{4\cV^{2}}\right ]\, .
\end{equation}
Notice that this result is essentially trivial: the terms $k_{\alpha\beta}$ at order $(\alpha^{\prime})^{3}$ all come from the 4D Weyl rescaling and as such must all have the same coefficient.

In \cite{Grimm:2017okk}, it was argued that the Type IIA terms $\sim k_{\alpha}k_{\beta}\dif b^{\alpha}\wedge\star_{4}\dif b^{\beta}$ cannot be absorbed into a redefinition of the (universal) hypermultiplet scalar.
This led to the prediction of a tree level $-2t_{8}t_{8}H_{3}^{2}R^{3}$ operator in Type IIA.
To make contact with these results,
let us write out \eqref{eq:Action4DKinTFin} in terms of NSNS fields at tree and 1-loop level by using \eqref{eq:ExpansionModFuncLargeImTauK}.
At tree level NSNS,
we obtain
\begin{align}
S^{(4)}&\supset \int\ee^{-\phi} \biggl \{\dfrac{1}{2\cV}\left [k_{\alpha\beta}-\dfrac{k_{\alpha}k_{\beta}}{\cV}\right ]-\dfrac{ a_{T}\zeta}{8\cV^{2}} \left [k_{\alpha\beta}-2\dfrac{k_{\alpha}k_{\beta}}{\cV} \right ]   \biggl \}\dif b^{\alpha}\wedge\star_{4}\dif b^{\beta}\nn\\
&=\dfrac{-1}{4}\int\ee^{-\phi} \, \cK_{\alpha\beta}\biggl |_{\text{tree}}\dif b^{\alpha}\wedge\star_{4}\dif b^{\beta}  \, .
\end{align}
This is in agreement with the Type IIA analysis of \cite{Grimm:2017okk} as expected.
Looking at NSNS 1-loop,
we find
\begin{align}
S^{(4)}&\supset \ee^{-\phi}\left (-\dfrac{ a_{L}\zeta}{8\cV^{2}} \left [k_{\alpha\beta}-\dfrac{k_{\alpha}k_{\beta}}{\cV} \right ]  \right ) \dif b^{\alpha}\wedge\star_{4}\dif b^{\beta}\, .
\end{align}
This does not match the 4D Kähler metric which is not at all surprising: the 10D NSNS 1-loop action \eqref{eq:NSNSKinematicsH2R3Loop} does not contain the $-2t_{8}t_{8}$ piece which would lead to the 4D replacement $k_{\alpha}k_{\beta}\raw 2k_{\alpha}k_{\beta}$.
The above analysis can be repeated for the kinetic terms for RR $C_{2}$-axions to find
\begin{align}
S^{(4)}&\supset -\int\,  \dfrac{ \zeta\ee^{\phi}}{8\cV^{2}} \left [ \left (a_{T}+a_{L}\right )k_{\alpha\beta}-\dfrac{1}{2}\left (a_{T}+3a_{L}\right )\dfrac{k_{\alpha}k_{\beta}}{\cV} \right ]  \dif c^{\alpha}\wedge\star_{4}\dif c^{\beta}  \, .
\end{align}
Here, both tree and 1-loop result are in disagreement with the 4D Kähler metric.

There are two ways in which this issue could be alleviated. The first option is a suitable redefinition of the 4D coordinates modifying only the piece $\sim k_{\alpha}k_{\beta}$.
The second possibility would be additional contributions from a more careful treatment of backreaction effects.



\section{Open questions and outlook}\label{sec:Conclusions}

The main result of this paper is in  revealing the structure of the 10D Type IIB effective action involving $G_{3}$ and $R$ in the maximally R-symmetry-violating sector. We examined it with two different approaches, the superfield and the 11D superparticle, and compared our findings to the expectations from string-theory amplitudes. There are a number of open questions and venues for further research concerning the ten-dimensional effective actions.

\medskip

The way how the couplings of the form $|G_{3}|^{2}R^{3}$, $G_{3}^{2}R^{3}+\text{c.c.}$ computed  in \cite{Liu:2019ses} are repackaged using elementary tensor structures should motivate the study of non-linear extensions in the superspace approach of \cite{Howe:1983sra}.

\medskip

We have argued that  the entire eight-derivative action in the MUV sector for couplings of the form $G_{3}^{2w}R^{4-w}$ is determined by a single index structure obtained from a sixteen fermion integral. Regarding the non-maximally R-symmetry-violating sectors, we only provided concrete evidence of the existence of certain kinematical structures and of their outstanding role in compactifications to lower dimensions, but much work remains to be done to determine the full effective action. In particular, if we were to replace the curvature tensor $R\raw \cR$ (defined in \eqref{eq:CalR}) as dictated by non-linear SUSY,  a whole tower of kinematical structures listed in Tab.~\ref{tab:ColPic} will be generated.\footnote{
Note that here we have collected only various $t_{N}$ as defined by standard fermionic integrals in \eqref{eq:FermIntegrals}. There are however further contributions similar to the ones listed in Tab.~\ref{tab:CoeffKinStructGTRCT24} at 5-points.}

\begin{table}[t!]
\centering
\begin{tabular}{|c||l|l|l|l|l|}
\hline 
&  &  &  &  &  \\ [-0.8em]
$\mathrm{U}(1)$ & 0 & 2 & 4 & 6 & 8 \\ [0.2em]
\hline 
\hline 
 &  &  &  &  &  \\ [-0.8em]
MUV & $t_{24}\cR^{4}$ & $t_{24}G_{3}^{2}\cR^{3}$ & $t_{24}G_{3}^{4}\cR^{2}$ & $t_{24}G_{3}^{6}\cR$ & $t_{24}G_{3}^{8}$ \\ [0.2em]
\hline 
\hline 
&  &  &  &  &  \\ [-0.8em]
5-point & $t_{18}|G_{3}|^{2}R^{3}$ &  &  &  &  \\ [0.2em]
\hline 
 &  &  &  &  &  \\ [-0.8em]
6-point & $t_{20}(|G_{3}|^{2})^{2}R^{2}$ & $t_{20}G_{3}^{2}|G_{3}|^{2}R^{2}$ &  &  &  \\ [0.2em]
\hline 
 &  &  &  &  &  \\ [-0.8em]
7-point & $t_{22}(|G_{3}|^{2})^{3}R$ & $t_{22}G_{3}^{2}(|G_{3}|^{2})^{2}R$ & $t_{22}G_{3}^{4}|G_{3}|^{2}R$ &  &  \\ [0.2em]
\hline 
 &  &  &  &  &  \\ [-0.8em]
8-point & $t_{24}(|G_{3}|^{2})^{4}$ & $t_{24}G_{3}^{2}(|G_{3}|^{2})^{3}$ & $t_{24}G_{3}^{4}(|G_{3}|^{2})^{2}$ & $t_{24}G_{3}^{6}|G_{3}|^{2}$ &  \\ [0.2em]
\hline 
\end{tabular} 
\caption{3-form contact terms that are captured by $t_{24}$ and a suitable definition of $\cR$.}\label{tab:ColPic} 
\end{table}

In Sect.~\ref{sec:ExpandingRiemannTorsion},
we illustrated the way $t_{24}\cR^{4}$ contributes at the level of 5-point contact terms corresponding to the third line of Tab.~\ref{tab:ColPic}. Fixing the coefficients will however require a more detailed calculation of the non-linear superfield following \cite{deHaro:2002vk,Green:2003an}.

\medskip

We recall that the tree-level effective action should be T-duality invariant. Imposing this invariance apparently allows to determine the eight-derivative NSNS action to all powers in $H_{3}$ \cite{Garousi:2020lof}. Moreover, the recent work \cite{Wulff:2021fhr} discusses $5$-point structures constrained by $\mathrm{O}(d,d)$ invariance.
Comparing these results to ours, which requires the extensive use of field redefinitions, and eventual use of T-duality as a way of constraining higher-order interactions, is left to future work.

\medskip

We have moved closer  to a full completion of the five-point effective action at order $(\alpha')^3$.
Terms of the form $H_{3}^{2}(\nabla H_{3})^{2}R$ and their RR counterparts which are in principle obtainable from the results of \cite{Richards:2008sa,Richards:2008jg,Liu:2019ses} have not been analysed here. Moreover, the relation of the CP-odd couplings \eqref{eq:QuinticActionFluxCompV3} to the elementary tensor structures used here needs further clarifications. Finally, the dilaton couplings continue to be a top challenge. In the MUV sector,
dilaton amplitudes were systematically analysed in \cite{Green:2019rhz}.  The non-MUV part however remains largely unexplored.
In this context, the F-theory lifts along the lines of \cite{Minasian:2015bxa} could offer a geometric principle underlying these couplings and eventually provide a key to determining scalar couplings also beyond four points.

\medskip

Even though the notion of the higher-dimensional tensors we use here is quite established at the eight-derivative level through e.g.~the linearised superfield \cite{Howe:1983sra},
their role in the 10D effective action in the presence of a non-trivial background with $\cP$ and $G_{3}$ remains to large extent unexplored. Also their generalisation to orders $(\alpha^{\prime})^{5}$ and higher remains unclear. Given that they correspond to $1/4$-BPS and $1/8$-BPS interactions as opposed to $1/2$-BPS for dimension-8 operators, they are certainly less constrained by supersymmetry. For instance, while the coefficients of $1/2$-BPS and $1/4$-BPS interactions satisfy Laplace eigenvalue equations, the pre-factors of $1/8$-BPS terms satisfy an inhomogeneous Laplace equation instead \cite{Green:2005ba}, see also \cite{Green:2019rhz} for a more recent discussion.
It would be interesting to understand the appearance of potentially novel index structures at higher orders in $\alpha^{\prime}$.

\acknowledgments
We would like to thank Michele Cicoli, Michael Green, Tom Pugh, Fernando Quevedo, Oliver Schlotterer, Gary Shiu, Roberto Valandro, and Pierre Vanhove for useful discussions.
AS acknowledges support by the German Academic Scholarship Foundation and by DAMTP through an STFC studentship. This work was supported in part by the US Department of Energy under grant DE-SC0007859 (JTL) and by ERC grants 772408-Stringlandscape and 787320-QBH Structure (RM).

\appendix

\section{Definitions and conventions}

\subsection{$\mathrm{SL}(2,\bZ)$-covariant modular forms}\label{app:ModFunctions} 

Throughout this paper, we make heavy use of special modular forms and their properties. The relevant functions are all generalisations of the non-holomorphic Eisenstein series of weight $3/2$ denoted as $f_{0}(\tau,\bar{\tau})$. More generally, we define
\begin{equation}\label{eq:ModFormsDef} 
f_{w}(\tau,\bar{\tau})=\sum_{(\hat{l}_{1},\hat{l}_{2})\neq (0,0)}\, \dfrac{\im(\tau)^{\frac{3}{2}}}{(\hat{l}_{1}+\tau\hat{l}_{2})^{\frac{3}{2}+w}(\hat{l}_{1}+\bar{\tau}\hat{l}_{2})^{\frac{3}{2}-w}}\, .
\end{equation}
They transform covariantly under $\mathrm{SL}(2,\mathbb{Z})$
\begin{equation}
f_{w}\left (\dfrac{a\tau+b}{c\tau+d},\dfrac{a\bar{\tau}+b}{c\bar{\tau}+d}\right )=\left (\dfrac{c\tau+d}{c\bar{\tau}+d}\right )^{w}f_{w}(\tau,\bar{\tau})\, .
\end{equation}
Further, these functions satisfy
\begin{equation}\label{eq:DefModFuncExp} 
(\tau-\bar{\tau})\dfrac{\p}{\p \tau}f_{w}=\left (w+\dfrac{3}{2}\right )f_{w+1}-wf_{w}\quad ,\quad (\tau-\bar{\tau})\dfrac{\p}{\p\bar{\tau}}f_{w}=\left (w-\dfrac{3}{2}\right )f_{w-1}-wf_{w}\, .
\end{equation}
More elegantly, this can be written in terms of a covariant derivative $\cD_{w}$ where
\begin{equation}\label{eq:CovDerMoFunc} 
\cD_{w}f_{w}=i\left (\tau_{2}\dfrac{\p}{\p\tau}-i\dfrac{w}{2}\right )f_{w}=\dfrac{3+2w}{4}f_{w+1}
\end{equation}
so that (see (2.11) in \cite{Green:2019rhz})
\begin{equation}\label{eq:CovDerIdentitiyModForms} 
f_{w}=\dfrac{2^{w-1}\sqrt{\pi}}{\Gamma\bigl (\frac{3}{2}+w\bigl )}\cD_{w-1}\ldots\cD_{0}f_{0}=\dfrac{2^{3w+1}\, (w+1)!}{(2(w+1))!}\cD_{w-1}\ldots\cD_{0}f_{0}\, .
\end{equation}

Last but not least, we expand $f_{w}$ in the large $\mathrm{Im}(\tau)\gg 1$ (small string coupling) regime where
\begin{equation}\label{eq:ExpansionModFuncLargeImTauK} 
f_{w}(\tau,\bar{\tau})=a_{T}+\dfrac{a_{L}}{(1-4w^{2})}+\cO\left (\ee^{-\mathrm{Im}(\tau)}\right )
\end{equation}
in terms of
\begin{equation}\label{eq:TLCMFEXP} 
a_{T}=2\zeta(3)\mathrm{Im}(\tau)^{\frac{3}{2}}\kom a_{L}=\dfrac{2\pi^{2}}{3}\mathrm{Im}(\tau)^{-\frac{1}{2}}\, .
\end{equation}
The first term is associated with closed string tree level \cite{Gross:1986iv}, whereas the second term with $1$-loop effects \cite{Green:1981ya}. The final piece encodes contributions from non-perturbative D-instanton states \cite{Green:1997tv}.
For the lowest order modular functions, we can write
\begin{align}\label{eq:ExpansionModFuncTL} 
f_{0}(\tau,\bar{\tau})&=a_{T}+a_{L}+\cO(\mathrm{e}^{-\mathrm{Im}(\tau)})\kom f_{\pm 1}(\tau,\bar{\tau})=a_{T}-\dfrac{1}{3}a_{L}+\cO(\mathrm{e}^{-\mathrm{Im}(\tau)})\, .
\end{align}

\subsection{A basis for $R^{3}$}\label{app:BasisRC}

As shown in~\cite{Liu:2019ses}, the decomposition of $H_{3}^{2}R^{3}$ requires $24$ independent Lorentz singlets. We therefore introduce a $24$-dimensional basis for contractions of $H_{3}^{2}R^{3}$ built from $R^{3}$ invariants $\lbrace \tilde{S}^{i},\tilde{W}^{i}_{M_{1}M_{2}},\tilde{X}^{i}_{M_{1}M_{2}M_{3}M_{4}},\tilde{Q}^{i}_{M_{1}M_{2}M_{3}M_{4}M_{5}M_{6}}\rbrace$ transforming \emph{reducibly} under $\mathrm{SO}(1,9)$ (with the exception of $\tilde{S}^{i}$). First, there are the following two singlets
\begin{align}
\tilde{S}^{1}&=R_{MN}\,^{RS}R^{MNOP}R_{OPRS}\kom \tilde{S}^{2}=R^{M}\,_{NP}\,^{Q}R^{R}\,_{MQ}\,^{S}R^{N}\,_{RS}\,^{P}\, .
\end{align}
One infers that in this basis  the $6$d Euler density may be written as
\begin{align}\label{eq:EDF1BasisS} 
Q&=\dfrac{1}{12}\left (\tilde{S}^{1}+2\tilde{S}^{2}\right )\, .
\end{align}
Furthermore, we define the $2$-tensors
\begin{align}
\tilde{W}_{M_{1}M_{2}}^{1}&=R_{M_{1}NM_{2}P}R^{N}\,_{QRS}R^{PQRS}\kom \tilde{W}_{M_{1}M_{2}}^{2}=R_{M_{1}NPQ}R_{M_{2}}\,^{NRS}R^{PQ}\,_{RS}\, ,\nn\\
\tilde{W}_{M_{1}M_{2}}^{3}&=R_{M_{1}NPQ}R_{M_{1}}\,^{RPS}R^{N}\,_{R}\,^{Q}\,_{S}\, .
\end{align}
There are $11$ independent $4$-index tensors
\begin{align}
\tilde{X}^{1}_{M_{1}M_{2}M_{3}M_{4}}&=R_{M_{1}M_{2}M_{3}M_{4}}R_{NPQR}R^{NPQR}\kom \tilde{X}^{2}_{M_{1}M_{2}M_{3}M_{4}}=R_{M_{1}M_{2}M_{3}N}R_{M_{4}PQR}R^{NPQR}\, ,\nn\\
\tilde{X}^{3}_{M_{1}M_{2}M_{3}M_{4}}&=R_{M_{1}M_{2}NP}R_{M_{3}M_{4}QR}R^{NPQR}\kom \tilde{X}^{4}_{M_{1}M_{2}M_{3}M_{4}}=R_{M_{1}M_{3}NP}R_{M_{2}M_{4}QR}R^{NPQR}\, ,\nn\\
\tilde{X}^{5}_{M_{1}M_{2}M_{3}M_{4}}&=R_{M_{1}NM_{3}P}R_{M_{2}QM_{4}R}R^{NPQR}\kom \tilde{X}^{6}_{M_{1}M_{2}M_{3}M_{4}}=R_{M_{1}M_{3}NP}R_{M_{2}}\,^{N}\,_{QR}R_{M_{4}}\,^{PQR}\, ,\nn\\
\tilde{X}^{7}_{M_{1}M_{2}M_{3}M_{4}}&=R_{M_{1}NM_{3}P}R_{M_{2}}\,^{N}\,_{QR}R_{M_{4}}\,^{PQR}\kom \tilde{X}^{8}_{M_{1}M_{2}M_{3}M_{4}}=R_{M_{1}M_{3}NP}R_{M_{2}Q}\,^{N}\,_{R}R_{M_{4}}\,^{QPR}\, ,\nn\\
\tilde{X}^{9}_{M_{1}M_{2}M_{3}M_{4}}&=R_{M_{1}NM_{3}P}R_{M_{2}Q}\,^{N}\,_{R}R_{M_{4}}\,^{QPR}\kom \tilde{X}^{10}_{M_{1}M_{2}M_{3}M_{4}}=R_{M_{1}M_{2}}\,^{NP}R_{M_{3}NQR}R_{M_{4}P}\,^{QR}\, ,\nn\\
\tilde{X}^{11}_{M_{1}M_{2}M_{3}M_{4}}&=R_{M_{1}M_{2}NP}R_{M_{3}Q}\,^{N}\,_{R}R_{M_{4}}\,^{QPR}
\end{align}
and another $8$ combinations of $6$-tensors
\begin{align}
\tilde{Q}^{1}_{M_{1}M_{2}M_{3}M_{4}M_{5}M_{6}}&=R_{M_{1}M_{4}N}\,^{P}R_{M_{2}M_{5}P}\,^{Q}R_{M_{3}QM_{6}}\,^{N}\, ,\nn\\
\tilde{Q}^{2}_{M_{1}M_{2}M_{3}M_{4}M_{5}M_{6}}&=R_{M_{1}M_{2}N}\,^{P}R_{M_{4}M_{5}P}\,^{Q}R_{M_{3}QM_{6}}\,^{N}\, ,\nn\\
\tilde{Q}^{3}_{M_{1}M_{2}M_{3}M_{4}M_{5}M_{6}}&=R_{M_{1}M_{2}N}\,^{P}R_{M_{3}M_{4}P}\,^{Q}R_{M_{5}QM_{6}}\,^{N}\, ,\nn\\
 \tilde{Q}^{4}_{M_{1}M_{2}M_{3}M_{4}M_{5}M_{6}}&=R_{M_{1}NM_{4}}\,^{P}R_{M_{2}PM_{5}}\,^{Q}R_{M_{3}QM_{6}}\,^{N}\, ,\nn\\
\tilde{Q}^{5}_{M_{1}M_{2}M_{3}M_{4}M_{5}M_{6}}&=R_{M_{1}NPQ}R_{M_{2}M_{4}}\,^{PQ}R_{M_{3}M_{5}M_{6}}\,^{N}\, ,\nn\\
\tilde{Q}^{6}_{M_{1}M_{2}M_{3}M_{4}M_{5}M_{6}}&=R_{M_{1}NPQ}R_{M_{4}M_{5}}\,^{PQ}R_{M_{2}M_{3}M_{6}}\,^{N}\, ,\nn\\
\tilde{Q}^{7}_{M_{1}M_{2}M_{3}M_{4}M_{5}M_{6}}&=R_{M_{1}NPQ}R_{M_{2}}\,^{N}\,_{M_{4}}\,^{Q}R_{M_{3}M_{5}M_{6}}\,^{P}\, ,\nn\\
 \tilde{Q}^{8}_{M_{1}M_{2}M_{3}M_{4}M_{5}M_{6}}&=R_{M_{1}M_{2}M_{4}M_{5}}R_{M_{3}NPQ}R_{M_{6}}\,^{NPQ}\, .
\end{align}

\section{Superfield calculus}\label{app:Superfield}

\subsection{The linearised description}

This section closely follows the conventions in Appendix~C of \cite{Green:2019rhz}.
Let $\theta_{i}$ be two sixteen-component chiral spinors of $\mathrm{Spin}(1,9)$ which we combine into the complex supercharge $\Theta=\theta_{1}+\I\theta_{2}$.
Linearised effective interactions preserving half of the original $32$ supersymmetries are derived from a constrained superfield $\Phi(x^{\mu}-\I\bar{\Theta}\gamma^{\mu}\Theta,\Theta)$ which satisfies the holomorphic condition
\begin{equation}\label{eq:LSC:Cond1} 
\overline{D}_{\Theta}\Phi=0\kom (\overline{D}_{\Theta})_{A}=-\dfrac{\p}{\p\bar{\Theta}^{A}}\quad ,\; A=1,\ldots ,16\, .
\end{equation}
It is further constrained by
\begin{equation}\label{eq:LSC:Cond2} 
(D_{\Theta})^{4}\Phi=(\overline{D}_{\Theta})^{4}\overline{\Phi}
\end{equation}
where
\begin{equation}
(D_{\Theta})_{A}=\dfrac{\p}{\p\Theta^{A}}+2\I (\gamma^{\mu}\bar{\Theta})_{A}\p_{\mu}\kom (\ov D_{\Theta})_{A}=-\dfrac{\p}{\p\ov \Theta^{A}}\, .
\end{equation}
The two operators $D_{\Theta}$ and $\overline{D}_{\Theta}$ are the (anti-)holomorphic covariant derivatives and commute with
\begin{equation}
Q_{A}=\dfrac{\p}{\p\Theta^{A}}\kom \bar{Q}_{A}=-\dfrac{\p}{\p\bar{\Theta}^{A}}+2\I(\bar{\Theta}\gamma^{\mu})_{A}\p_{\mu}
\end{equation}
corresponding to the rigid supersymmetries.

The two conditions \eqref{eq:LSC:Cond1} and \eqref{eq:LSC:Cond2} imply that the expansion of $\Phi$ in powers of $\Theta$ terminates at $\Theta^{8}$.
The scalar superfield components are completely specified by choosing the lowest order scalar component.
It turns out to be convenient to work in a parametrisation of scalar fluctuations as $\hat{\tau}=\I\delta\tau/2\tau_{2}^{0}$ for $\delta\tau=\tau-\tau^{0}$ \cite{Schwarz:1983qr,Green:2019rhz}.
The corresponding superfield $\Phi$ was previously discussed in \cite{Howe:1983sra,deHaro:2002vk} and is defined as
\begin{align}
\Phi&=\tau_{2}^{0}+\tau_{2}^{0}\Delta
\end{align}
where $\tau_{2}^{0}\Delta$ parametrises the linearised fluctuations around a constant, purely imaginary flat background $\tau_{2}^{0}=g_{s}^{-1}$ with (see also eq. (5.26) in \cite{Green:1998by})
\begin{align}\label{eq:LinSuperfield} 
\Delta&=\sum_{r=0}^{8}\, \Theta^{r}\Phi^{(r)}\nn\\
&=\hat{\tau}+\Theta\lambda+\Theta^{2}G_{3}+\Theta^{3}\p\psi+\Theta^{4}(R+\p F_{5})\nn\\[0.3em]
&\quad+\Theta^{5}\p^{2}\psi^{*}+\Theta^{6}\p^{2}\bar{G}_{3}+\Theta^{7}\p^{3}\lambda^{*}+\Theta^{8}\p^{4}\hat{\bar{\tau}}\, .
\end{align}
Here, $\lambda$ and $\psi$ are the complex dilatino and gravitino respectively.
For our purposes below,
it suffices to note that
\begin{align}
\Theta^{2}G_{3}&=(\Theta\Gamma^{i_{1}i_{2}i_{3}}\Theta)\, G_{i_{1}i_{2}i_{3}}\kom \Theta^{4}R=(\Theta\Gamma^{i_{1}i_{2}k}\Theta)(\Theta\Gamma_{k}\,^{i_{3}i_{4}}\Theta)R_{i_{1}i_{2}i_{3}i_{4}}\, .
\end{align}

Terms encoded in $\Phi^{(r)}$ have $\mathrm{U}(1)$ R-symmetry charge
\begin{equation}
q_{r}=-2+\dfrac{r}{2}
\end{equation}
where we assigned charge $-1/2$ to $\Theta$ and $-2$ to $\Phi$.
This leads to
\begin{equation}
q_{\hat{\tau}}=-2\kom q_{\lambda}=-\dfrac{3}{2}\kom q_{G_{3}}=-1\kom q_{\psi}=-\dfrac{1}{2}\kom q_{R}=q_{F_{5}}=0\, .
\end{equation}

Even though the linearised approximation gives only partial results for the structure of terms in the effective action,
it is still useful to find and relate various terms in the weak coupling limit $\tau_{2}^{0}=g_{1}^{-1}\raw \infty$.
Generally,
interactions are constructed from a function $\textbf{F}[\Phi]$ of $\Phi$ by integrating over the $16$ components of $\Theta$, that is,
\begin{equation}\label{eq:LSC:ActionExpectation} 
S_{\text{linear}}=\int\dif^{10}x\dif^{16}\Theta\, \det(e)\, \mathbf{F}[\Phi]+\text{c.c.}\, .
\end{equation}
Here, $\det(e)$ is the determinant of the zehnbein and the total expression is invariant under the rigid supersymmetries.
In an expansion in powers of $\Theta$, we find
\begin{equation}\label{eq:ExpanSuperfieldLinApproxModF} 
\textbf{F}[\Phi]=F(\tau_{2}^{0})+\sum_{n=1}^{\infty}\, \dfrac{1}{n!}\Delta^{n}\left (\dfrac{\p}{\p\tau_{2}^{0}}\right )^{n}F(\tau_{2}^{0})
\end{equation}
Substituting this expansion into \eqref{eq:LSC:ActionExpectation} and keeping the terms with $\Theta^{16}$,
we recover all the interactions at order $(\alpha^{\prime})^{3}$
\begin{align}\label{eq:ActionAtOrderAP3} 
S^{(3)}&=\int\dif^{10}x\,\biggl \{f^{(12,-12)}\lambda^{16}+f^{(11,-11)}{G}_{3}\lambda^{14}+\ldots+f^{(4,-4)}{G}_{3}^{8}+\ldots+f^{(1,-1)}{G}_{3}^{2}R^{3}\nn\\[-0.4em]
&\quad+f^{(0,0)} (|{G}_{3}|^{2}+|{\cP}|^{2}) R^{3}+\ldots+f^{(0,0)}R^{4}+\ldots+f^{(-12,12)}(\lambda^{*})^{16}
\biggl \}\, .
\end{align}
Here, the modular forms $f_{w}=f^{(w,-w)}(\tau,\bar{\tau})$ have holomorphic and anti-holomorphic weights $(w,-w)$ and are eigenfunctions of the $\mathrm{SL}(2,\mathbb{Z})$-Laplacian \cite{Green:1998by}.
The presence of these coefficient functions is required by $\mathrm{SL}(2,\mathbb{Z})$ invariance,
see \cite{Green:2019rhz} and references therein.
The functions $f^{(w,-w)}$ carry $\mathrm{U}(1)$ charge $q_{f_{w}}=2w$ (in our convention) and appear generically as
\begin{equation}
\int\dif^{10}x\, \mathrm{det}(e)\, f^{(w,-w)}\prod_{n=1}^{P}\Phi^{(r_{n})}\, .
\end{equation}
The value $w$ is fixed by the sum of $\mathrm{U}(1)$ charges of the $\Phi^{(r_{n})}$:
\begin{equation}\label{eq:UOneViolRelPoints} 
\sum_{n=1}^{P}\, q_{r_{n}}=8-2P\overset{!}{=}-2w
\end{equation}
where we used that $\sum_{n}\, r_{n}=16$ to ensure the presence of $16$ powers of $\Theta$.

As far as the index structures in \eqref{eq:ActionAtOrderAP3} are concerned,
one finds contributions like
\begin{align}
t_{16}R^{4}&=\int\dif^{16}\Theta \left [(\Theta\Gamma^{i_{1}i_{2}k}\Theta)(\Theta\Gamma_{k}\,^{i_{3}i_{4}}\Theta)R_{i_{1}i_{2}i_{3}i_{4}}\right ]^{4}\, ,\nn\\
t_{18}G_{3}^{2}R^{3}&=\int\dif^{16}\Theta\, \left ((\Theta\Gamma^{i_{1}i_{2}i_{3}}\Theta)\, G_{i_{1}i_{2}i_{3}}\right )^{2}\left [(\Theta\Gamma^{i_{1}i_{2}k}\Theta)(\Theta\Gamma_{k}\,^{i_{3}i_{4}}\Theta)R_{i_{1}i_{2}i_{3}i_{4}}\right ]^{3}\, .
\end{align}

Several comments are in order.
In the linearised approximation,
we work in a regime where we neglect the inhomogeneous part of the modular covariant derivative $D_{w}=w+2\I \tau_{2}^{0}\p_{\tau_{2}^{0}}$, that is,
\begin{equation}
2\I \tau_{2}^{0}\p_{\tau_{2}^{0}} f_{w}\gg wf_{w}\, .
\end{equation}
This is clearly violated for terms in $f_{w}$ that are powers of $\tau_{2}^{0}$.
In contrast,
D-instanton contributions $\sim (\tau_{2}^{0})^{n} \mathrm{e}^{-2\pi |N|\tau_{2}^{0}}$ satisfy the above inequality in the limit $\tau_{2}^{0}\raw\infty$.
Thus, the linearised description contains the exact leading multi-instanton effects.
In a non-linearly completed theory, the $\mathrm{SL}(2,\mathbb{Z})$ symmetry requires the $f_{w}$ to become the familiar modular forms.
Then, the relative coefficients for interactions of differing $\mathrm{U}(1)$ charge can be computed from supersymmetry.

\subsection{Non-linear superfield}\label{sec:NonLinSuperfieldApp} 

The non-linear superfield completion is generally cumbersome in the presence of non-trivial $\cP$ and $G_{3}$ backgrounds.
Looking at the $\Theta^{4}$ term in \eqref{eq:LinSuperfield},
it was already proposed in \cite{Paulos:2008tn} (and even earlier in \cite{deHaro:2002vk,Rajaraman:2005up}) that full graviton and $F_{5}$ kinematics is encoded in
\begin{equation}
\Theta^{4}\tilde{\cR}=(\Theta\Gamma^{i_{1}i_{2}i_{3}}\Theta)(\Theta\Gamma^{i_{4}i_{5}i_{6}}\Theta)\tilde{\cR}_{i_{1}i_{2}i_{3}i_{4}i_{5}i_{6}}
\end{equation}
where $\tilde{\cR}$ was defined in \eqref{eq:DefForR6ModF5}.
Then, the single index structure
\begin{align}
t_{24}\tilde{\cR}^{4}&=\int\dif^{16}\Theta \left [(\Theta\Gamma^{i_{1}i_{2}i_{3}}\Theta)(\Theta\Gamma^{i_{4}i_{5}i_{6}}\Theta)\tilde{\cR}_{i_{1}i_{2}i_{3}i_{4}i_{5}i_{6}}\right ]^{4}
\end{align}
includes tensor contractions of the form $(\nabla F_{5})^{n}F_{5}^{2m}R^{4-n-m}$ \cite{Paulos:2008tn} which are in agreement with string amplitude calculations \cite{Peeters:2003pv}.

This clearly implies that $\tilde{\cR}$ as defined in \eqref{eq:DefForR6ModF5} enjoys the following symmetries to be imposed implicitly further below:
\begin{enumerate}
\item Invariance under the exchange of fermion bilinears in \eqref{eq:NonLinSuperTheta4F5Rev} implies that only the part of $\tilde{\cR}_{i_{1}i_{2}i_{3}i_{4}i_{5}i_{6}}$ symmetric under the exchange $(i_{1},i_{2},i_{3})\leftrightarrow(i_{4},i_{5},i_{6})$ contributes.
\item Anti-symmetry of the $\Gamma$-matrices in \eqref{eq:NonLinSuperTheta4F5Rev} means that we have to anti-symmetrise in both $(i_{1},i_{2},i_{3})$ and $(i_{4},i_{5},i_{6})$.
\item The fermion bilinears in \eqref{eq:NonLinSuperTheta4F5Rev} enjoy further Fierz identities which essentially project onto certain tensor representations.
Applying considerations from representation theory,
it turns out that \cite{Peeters:2001ub}
\begin{equation}
\left ( \mathbf{16}\otimes\mathbf{16}\otimes\mathbf{16}\otimes\mathbf{16} \right )_{\text{anti-sym}}=\mathbf{770}\oplus\mathbf{1050}^{+}
\end{equation}
where $\textbf{770}=[0,2,0,0,0]$ and $\mathbf{1050}^{+}=[1,0,0,0,2]$ in terms of their Dynkin labels under $D_{5}$.
We define the following two projection operators
\begin{align}
\cT_{i_{1}i_{2}i_{3}\, ,\, i_{4}i_{5}i_{6}}\bigl |_{\mathbf{1050}^{+}}&=\dfrac{1}{2} \biggl \{\dfrac{1}{2}\bigl (\cT_{i_{1}i_{2}i_{3}\, ,\, i_{4}i_{5}i_{6}}-3\, \cT_{i_{1}i_{2}i_{6}\, ,\, i_{4}i_{5}i_{3}}-\cT_{i_{1}i_{2}k\, ,\, i_{4}i_{5}}\,^{k}\, g_{i_{3}i_{6}}\nn\\
&\quad +2\, \cT_{i_{1}i_{5} k\, ,\, i_{4}i_{2}}\,^{k}\, g_{i_{3}i_{6}} \bigl )\pm \dfrac{1}{4!}\epsilon_{i_{1}i_{2}i_{3} i_{4}i_{5}}\,^{k_{1}k_{2}k_{3}k_{4}k_{5}}\cT_{k_{1}k_{2}k_{3}\, ,\, k_{4}k_{5}i_{6}} \biggl \}\, ,\\
\cT_{i_{1}i_{2}\, ,\, i_{4}i_{5}}\bigl |_{\mathbf{770}}&=\dfrac{2}{3}\left (\cT_{i_{1}i_{2}\, ,\, i_{4}i_{5}}+\cT_{i_{1}i_{5}\, ,\, i_{4}i_{2}} \right )-\dfrac{1}{2}\cT_{i_{1}k\, ,\, i_{4}}\,^{k}\, g_{i_{2}i_{5}}+\dfrac{1}{36}\cT_{jk\, ,}\,^{jk}\, g_{i_{1}i_{4}}g_{i_{2}i_{5}} \nn\, .
\end{align}
One easily verifies applying the projector onto $\mathbf{770}$ to the Riemann tensor that
\begin{equation}
R_{i_{1}i_{2}i_{4}i_{5}} \bigl |_{\textbf{770}}=C_{i_{1}i_{2}i_{4}i_{5}}
\end{equation}
which implies that only the Weyl tensor enters \eqref{eq:DefForR6ModF5}.

Both $F_{5}$ terms in \eqref{eq:DefForR6ModF5} do not contain any $\mathbf{770}$ piece, though a $\mathbf{1050}^{+}$ part.
For $F_{5}^{2}$,
one uses self duality to remove the $\epsilon$-tensor, thereby finding \cite{Paulos:2008tn}
\begin{equation}
\left ( F_{i_{1}i_{2}i_{3}kl}F_{i_{4}i_{5}i_{6}}\,^{kl} \right )\bigl |_{\mathbf{1050}^{+}}=\dfrac{1}{2}\left (F_{i_{1}i_{2}i_{3}kl}F_{i_{4}i_{5}i_{6}}\,^{kl}-3F_{i_{1}i_{2}i_{6}kl}F_{i_{4}i_{5}i_{3}}\,^{kl}\right )
\end{equation}
which is already imposed in \eqref{eq:DefForR6ModF5}.
For $\nabla F_{5}$,
the $\mathbf{1050}^{+}$ component is obtained by imposing
\begin{equation}
\nabla_{k}F^{k}\,_{i_{1}i_{2}i_{3}i_{4}}=0\kom F_{5}=\star_{10}F_{5}\, .
\end{equation}
\end{enumerate}

\section{11D superparticle amplitudes}

We compute $11$D amplitudes in the superparticle formalism compactified on a $2$-torus \cite{Green:1997di,Green:1997as,Green:1997me,Green:1999by}.
We start from the vertex operator
\begin{equation}
V_{G_{4}}=4k_{[I}C_{LMN]}\left (\dot{X}^{I}-\dfrac{2}{3}\cR^{IJ}k_{J}\right )\cR^{LMN}\mathrm{e}^{-\I k\cdot X}
\end{equation}
for the $3$-form $C_{3}$ in terms of $11$d indices $I,J,K,L,\ldots$ using the conventions of \cite{Green:1999by} for the fermion bilinears $\cR^{IJ},\cR^{LMN}$.
Once we compactify the vertex operator on a $T^{2}$, we split the $11$d indices as $\lbrace I,J,K,L,\ldots\rbrace$ into $T^{2}$-indices $\alpha,\beta,\ldots=1,2$ and $9$d indices $ i,j,k,l,\ldots =0,3,\ldots,10$.
Schematically, we distinguish the following types of terms in the reduction
\begin{equation}
k_{[i}C_{lmn]}\raw  F_{ilmn1}\kom k_{[i}C_{lm]1}\raw F_{ilm}\kom k_{[i}C_{lm]2}\raw H_{ilm}\kom k_{[i}C_{l]1\, 2}\raw F_{il}
\end{equation}
where $k_{[i}C^{(4)}_{lmn]1}= F_{ilmn1}$ and $F_{il}$ is the field-strength tensor of the $9$d Type IIB gravi-photon $A_{i}$.
We will only be interested in the Type IIB 3-forms $F_{3}$ and $H_{3}$ in 9D.

For our purposes (on the Type IIB side), it is more convenient to work with a complexified basis for the two $T^{2}$ direction for which $G_{3}$ is obtained from \cite{Green:1999by}
\begin{align}
k_{[l}C_{mn]z}&=\dfrac{1}{\sqrt{v_{0}\tau_{2}}}\left (k_{[l}C_{mn]1}-\tau k_{[l}C_{mn]2}\right )\, .
\end{align}
This amounts to the following set of vertex operators for $G_{3}$
\begin{align}\label{eq:VertexOpGThreeUpdate3} 
V_{G_{3}}&=3\sqrt{2}\, k_{[i}C_{mn]z}\left (\dot{X}^{i}-\dfrac{2}{3}\cR^{ij}k_{j}\right ) \cR^{z mn}\mathrm{e}^{-\I k\cdot X}\, ,\nn\\
&\quad-\sqrt{2}\, k_{[l}C_{ mn]z}\left (\dot{X}^{z}-\dfrac{2}{3}\cR^{z j}k_{j}\right )\cR^{lmn}\mathrm{e}^{-\I k\cdot X}\, .
\end{align}
Notice that the terms in the first line were not present in \cite{Green:1999by} which are however important to provide additional contributions in the $\mathrm{U}(1)$-preserving sector at 5-points.

Next, the 11D graviton vertex operator reads
\begin{equation}
V_{h}=h_{IJ}\left (\dot{X}^{I}\dot{X}^{J}-2\dot{X}^{I}\cR^{JM}k_{M}+2\cR^{IL}\cR^{JM}k_{L}k_{M}\right )\mathrm{e}^{-\I k\cdot X}\, .
\end{equation}
In 9D,
we obtain the graviton vertex operator
\begin{equation}
V_{h}=h_{ij}\left (\dot{X}^{i}\dot{X}^{j}-2\dot{X}^{i}\cR^{jm}k_{m}+2\cR^{il}\cR^{jm}k_{l}k_{m}\right ) \mathrm{e}^{-\I k\cdot X}
\end{equation}
as well as the axio-dilaton vertex operator
\begin{align}
V_{P}&=h_{z{z}}\left (\dot{X}^{z}\dot{X}^{z}-2\dot{X}^{{z}}\cR^{zm}k_{m}+2\cR^{zl}\cR^{{z}m}k_{l}k_{m}\right )  \mathrm{e}^{-\I k\cdot X}\, .
\end{align}

\subsection{General amplitudes}

In this section, we discuss the general form of amplitudes to be encounter below.
To begin with, we stress that there are essentially two classes of amplitudes.
If the fields are neutral under the Type IIB $\mathrm{U}(1)$, then the general result can be written as
\begin{equation}\label{eq:NeutralAmp} 
\cA_{\text{neutral}}=\cN_{1}\, \tilde{K}_{\text{neutral}}\left (C+\cN_{2}\dfrac{f_{0}(\tau,\bar{\tau})}{v_{0}^{3/2}}\right )
\end{equation}
for some kinematical structure $\tilde{K}$.
The constant $C$ generally remains undetermined in this formalism without a proper microscopic description, but can be determined from e.g. duality considerations.
In contrast, a $\mathrm{U}(1)$-violating combination of fields results in an amplitude of the form
\begin{equation}\label{eq:ChargedAmp} 
\cA_{\text{viol.}}^{(w)}=\cN\, \tilde{K}_{\text{viol.}} \, \dfrac{f_{w}(\tau,\bar{\tau})}{v_{0}^{3/2}}\, .
\end{equation}
This is expected simply because there is not associated analogue in $11$d and the result must disappear in the limit $v_{0}\raw \infty$!

The factor of $v_{0}^{-3/2}$ in \eqref{eq:NeutralAmp} and \eqref{eq:ChargedAmp} is reminiscent of the terms coming with $\alpha^{\prime}/R^{2}=\alpha^{\prime}/(r_{A}^{(s)})^{2}$ in \cite{Liu:2010gz} since
\begin{equation}
\dfrac{g_{B}^{1/2}}{v_{0}^{3/2}}=(r_{B}^{(s)})^{2}=\dfrac{1}{(r_{A}^{(s)})^{2}}\, .
\end{equation}
Hence, $\mathrm{U}(1)$ uncharged amplitudes of the form \eqref{eq:NeutralAmp} naturally appear with a $1\pm v_{0}^{-3/2}\sim 1\pm \alpha^{\prime}/R^{2}$ prefactor, whereas charged amplitudes \eqref{eq:ChargedAmp} only come with a $v_{0}^{-3/2}\sim \alpha^{\prime}/R^{2}$ term.

More explicitly, we will be interested in the following types of $P$-point amplitudes
\begin{align}
\cA_{P}(m,n)&=\tilde{K}(P)\int\dfrac{\dif t}{t} t^{P}\int\dif^{9}\mathbf{p}\sum_{l_{1},l_{2}}\, P_{z}^{m}P_{\bar{z}}^{n}\mathrm{e}^{-t(\mathbf{p}^{2}+g^{ab}l_{a}l_{b})}\, .
\end{align}
For the moment, we keep the index structures in $\tilde{K}$ implicit.
Clearly, for $m=n$, $\cA^{m,n}$ will be of the form \eqref{eq:NeutralAmp}.
After integrating out the $9$d momenta, we obtain
\begin{align}
\cA_{P}(m,n)&=\pi^{\frac{9}{2}}\tilde{K}(P)\, S(P,m,n)
\end{align}
in terms of
\begin{equation}
S(P,m,n)=\int\, \dfrac{\dif t}{t} \dfrac{t^{P}}{t^{9/2}}\sum_{l_{1},l_{2}}\, P_{z}^{m}P_{\bar{z}}^{n}\mathrm{e}^{-t g^{ab}l_{a}l_{b}}\, .
\end{equation}
These functions can be computed systematically for any number of points and KK-momenta.
Throughout this work,
we require only the following explicit results
\begin{align}\label{eq:SPMNEx} 
S(5,0,0)&=C+4\,\sqrt{\pi} \dfrac{f_{0}(\tau,\bar{\tau})}{v_{0}^{{3}/{2}}}\kom S(5,1,1)= C-\sqrt{\pi}\dfrac{f_{0}(\tau,\bar{\tau})}{v_{0}^{{3}/{2}}}  \, ,\nn\\
S(5,2,0)&=3\sqrt{\pi}\dfrac{ f_{1}(\tau,\bar{\tau})}{v_{0}^{3/2}}  \kom S(5,0,2)=3\sqrt{\pi}\dfrac{ f_{-1}(\tau,\bar{\tau})}{v_{0}^{3/2}}  \, .
\end{align}
Here, $C$ is typically a divergent constant which can be identified through duality considerations \cite{Green:1997as}.

With the above formulas, the open task remains to determine kinematical structures.
In contrast to string amplitudes, we do \emph{not} impose a priori that we compute even/even, odd/odd or even/odd sector couplings.
This comes about naturally from the higher-dimensional index structures which we define as
\begin{equation}
t_{N+M}^{i_{1}\ldots i_{N+M}}=\tr\left (\cR^{i_{1}i_{2}}\ldots \cR^{i_{N-1}i_{N}}\cR^{i_{N+1}i_{N+2}i_{N+3}}\ldots \cR^{i_{M-2}i_{M-1}i_{M}}\right )\, .
\end{equation}
In $9$ (or any number of odd) spacetime dimensions, index structures $t_{N}$ with $N$ odd are associated with parity-odd couplings.
The reason is simple: if $N$ is even, then there is a chance to find terms of the form $t_{8}t_{8}$ or $\epsilon_{D}\epsilon_{D}$. However, if $N$ is odd, then there must be always a single $\epsilon_{D}$ of odd dimensions be involved.
This is to be contrasted with the situation discussed in \cite{Green:2005qr}.
Here, they provide the decomposition of $t_{24}$ in $10$ dimensions which involves both parity-even and parity-odd contributions.

Another comment concerns the situation where the index structure carries torus indices, that is
\begin{equation}
t_{N}^{z\bar{z}}=\tr\left (\cR^{z i_{1}i_{2}}\cR^{\bar{z}i_{3}i_{4}} \cR^{i_{5}i_{6}}\ldots \right )\, .
\end{equation}
In particular, structures of this type either appear alone as in the case of $1$ $b_{\mu 9}$ and $4$ $h_{\mu\nu}$ or in combination with $t_{N}$.
The latter scenario appears frequently in the amplitudes discussed below.
These instances can be understood from the $11$d perspective where one might find an index structure of the form
\begin{equation}
t_{N}^{ijklmn\ldots} G_{ijka}G_{lmn}\,^{a}\ldots\raw \left (t_{N}^{ijklmn\ldots} G_{ijk\color{red}{z}}G_{lmn}\,^{\color{red}{z}} +t_{N}^{{\color{red}{z}}jk\bar{\color{red}{z}}mn\ldots} G_{{\color{red}{z}}jka}G_{\bar{\color{red}{z}}mn}\,^{a}+\ldots\right )\ldots
\end{equation}
As we will see below, this happens for instance when studying $9$d amplitudes for $|G_{3}|^{2}R^{3}$.

\subsection{Maximally $\mathrm{U}(1)$-violating amplitudes}\label{sec:MUVSuperPartGRAmps} 

We derive the coefficients for MUV terms in the Type IIB action involving only the $3$-form and the metric.
The general $9$D superparticle amplitude for such contributions is given by
\begin{equation}
v_{0}\cA_{G_{3}^{2w}R^{4-w}}=\dfrac{2^{4-w}\, 2^{w}}{2^{6}\, \Gamma\left (\frac{3}{2}\right )}\, S(P,w,0)\, t_{16+2w}G_{3}^{2w}R^{4-w}
\end{equation}
in terms of
\begin{equation}
S(P,w,0)=\int\, \dfrac{\dif t}{t} \dfrac{t^{P}}{t^{9/2}}\sum_{l_{1},l_{2}}\, P_{z}^{2w}\mathrm{e}^{-t g^{ab}l_{a}l_{b}}\, .
\end{equation}
One easily verifies that after Poisson resummation
\begin{equation}
\sum_{l_{1},l_{2}}\, P_{z}^{2w}\mathrm{e}^{-t g^{ab}l_{a}l_{b}}=\dfrac{v_{0}^{w+1}}{\tau_{2}^{w}(2t)^{2w+1}}\sum_{\hat{l}_{1},\hat{l}_{2}}\,  (\hat{l}_{1}+\tau\hat{l}_{2})^{2w} \mathrm{e}^{- g_{ab}\hat{l}_{a}\hat{l}_{b}/(4t)}
\end{equation}
For $w>0$,
the zero winding term with $(\hat{l}_{1},\hat{l}_{2})=(0,0)$ simply drops out.
Next, we substitute $t\raw  (4\tilde{t})^{-1}\, g_{ab}\hat{l}_{a}\hat{l}_{b}$ to find
\begin{equation}
\int\, \dfrac{\dif t}{t} \dfrac{t^{P}}{t^{9/2+2w+1}} =\int t^{P-\frac{13}{2}-2w}\, \dif t\raw \int \left (\dfrac{g_{ab}\hat{l}_{a}\hat{l}_{b}}{4\tilde{t}}\right )^{P-\frac{11}{2}-2w}\, \dfrac{1}{\tilde{t}}\dif \tilde{t}
\end{equation}
where we used
\begin{equation}
\dif t= -\dfrac{g_{ab}\hat{l}_{a}\hat{l}_{b}}{4\tilde{t}^{2}}\dif\tilde{t}\, .
\end{equation}

Putting everything together,
we recover
\begin{align}
S(P,w,0)&= \dfrac{v_{0}^{w+1}}{\tau_{2}^{w}2^{2w+1}} \left (\dfrac{4}{v_{0}}\right )^{\frac{3}{2}+w}\sum_{(\hat{l}_{1},\hat{l}_{2})\neq (0,0)}\,  (\hat{l}_{1}+\tau\hat{l}_{2})^{2w} \left (\tilde{g}_{ab}\hat{l}_{a}\hat{l}_{b}\right )^{-\frac{3}{2}-w}\, \int\,  \tilde{t}^{\frac{1}{2}+w} \mathrm{e}^{- \tilde{t}} \dif \tilde{t}\nn\\
&=\dfrac{4\Gamma\left (\frac{3}{2}+w\right )}{\tau_{2}^{w}\sqrt{v_{0}}}\sum_{(\hat{l}_{1},\hat{l}_{2})\neq (0,0)}\,  (\hat{l}_{1}+\tau\hat{l}_{2})^{2w} \left (\tilde{g}_{ab}\hat{l}_{a}\hat{l}_{b}\right )^{-\frac{3}{2}-w}\, .
\end{align}
Since
\begin{equation}
\tilde{g}_{ab}\hat{l}_{a}\hat{l}_{b}=\dfrac{ (\hat{l}_{2}+\tau \hat{l}_{1})(\hat{l}_{2}+\bar{\tau} \hat{l}_{1})}{\tau_{2}}\, ,
\end{equation}
we can use the definition \eqref{eq:ModFormsDef} for modular forms $f_{w}(\tau,\bar{\tau})$ to obtain
\begin{equation}\label{eq:ResultExpressionMUVAmpSupGen} 
S(P,w,0)=\dfrac{4\Gamma\left (\frac{3}{2}+w\right )}{\sqrt{v_{0}}}\, f_{w}(\tau,\bar{\tau})\, .
\end{equation}

\subsection{Special non-MUV amplitudes}\label{sec:NonMUVSuperParticleResults}

The superparticle amplitudes in the non-MUV sector giving rise to contributions involving the higher-dimensional index structures $t_{18}, t_{20},\ldots$ are given by
\begin{equation}
v_{0}\cA_{G_{3}^{m}\ov G^{n}R^{4-(m+n)/2}}=\dfrac{2^{4-(m+n)/2}\, (-2)^{m+n}}{2^{6}\, \Gamma\left (\frac{3}{2}\right )}\, S(P,m,n)\, t_{16+m+n}G_{3}^{m}\ov G_{3}^{n}R^{4-(m+n)/2}
\end{equation}
in terms of
\begin{equation}
S(P,m,n)=\int\, \dfrac{\dif t}{t} \dfrac{t^{P}}{t^{9/2}}\sum_{l_{1},l_{2}}\, P_{z}^{m}P_{\bar{z}}^{n}\mathrm{e}^{-t g^{ab}l_{a}l_{b}}\, .
\end{equation}
The objects $S(P,m,n)$ can be computed as before.
The final expressions will be of the form
\begin{equation}\label{eq:ResNonMUVAmplitudeSp} 
v_{0}\cA_{G_{3}^{m}\ov G^{n}R^{4-(m+n)/2}}=\left (v_{0}C_{\infty} \delta_{w,0}+C_{w}^{(P)}\dfrac{f_{w}(\tau,\bar{\tau})}{\sqrt{v_{0}}}\right ) t_{16+m+n}G_{3}^{m}\ov G_{3}^{n}R^{4-(m+n)/2}
\end{equation}
where
\begin{equation}
w=\dfrac{m-n}{2}\, .
\end{equation}
The first zero winding piece only appears at the $\mathrm{U}(1)$-neutral level $m=n$ which contributes in the limit $v_{0}\raw \infty$ to the 11D M-theory action.
However, the constant $C_{\infty}$ is generically divergent because the superparticle picture does not provide a microscopic description of M-theory.
Such constants can be inferred though e.g. via dualities to Type IIA as discussed in \cite{Green:1997as} for $R^{4}$.

The second term in \eqref{eq:ResNonMUVAmplitudeSp} encodes as usual the contributions to the Type IIB effective action upon taking the limit $v_{0}\raw 0$.
Overall, the corresponding coefficients can be expressed in the following compact way
\begin{equation}
C_{w}^{(P)}=\dfrac{(2|w|+1)(2|w|-1)C_{P-4}}{(2(P-4)+1)(2(P-4)-1)}\, .
\end{equation}
Notice that for MUV amplitudes $|w|=P-4$ we recover $C_{P-4}^{(P)}=C_{P-4}$ as expected.

\section{Details on the reduction to 4D}\label{sec:CYReduction}

\subsection{Comment on the corrected background}\label{sec:BackgroundCYReduction}

To give some intuition on effects contributing to $\delta_{0}, \delta_{1}$,
the corrected metric background in Einstein frame involves an overall Weyl rescaling (see e.g. \cite{Bonetti:2016dqh}) which in string frame is associated with the corrected dilaton \cite{Becker:2002nn}, that is,
\begin{equation}\label{eq:MetricAnsatz10DEFG} 
\dif s_{10}^{2}=\mathrm{e}^{\Phi}\left [\mathrm{e}^{2\cA}\eta_{\mu\nu}\dif x^{\mu}\dif x^{\nu}+\mathrm{e}^{-2\cA}g_{mn}\dif y^{m}\dif y^{n}\right ]
\end{equation}
with
\begin{align}
\Phi=\alpha\Phi^{(1)}+\cO(\alpha^{2})\kom \cA=\cA^{(0)}+\alpha\cA^{(1)}+\cO(\alpha^{2})\kom g_{mn}=g_{mn}^{(0)}+\alpha g_{mn}^{(1)}+\cO(\alpha^{2})
\end{align}
In 10D Einstein frame,
neither $\phi$ nor $ G_{3}$ are corrected at order $(\alpha^{\prime})^{3}$,
\begin{align}
\phi=\phi_{0}+\cO((\alpha^{\prime})^{4})\kom  G_{3}=(\alpha^{\prime})^{1} { G}_{3}^{(0)}+(\alpha^{\prime})\, \alpha G_{3}^{(1)}+\cO((\alpha^{\prime})^{7})\, .
\end{align}
From the Bianchi identity for ${ F}_{5}$, one deduces that also ${ F}_{5}\sim \cO((\alpha^{\prime})^{2})$.
The remaining leading order solutions to the equations of motion can be determined from the results of \cite{Bonetti:2016dqh}. From Einstein's equations, one infers that the internal Ricci tensor receives a correction of the form
\begin{equation}\label{eq:IntRicciTensorCorrectedSLTZ} 
R^{(1)}_{mn}=-3\cdot 2^{9}\, f_{0}(\tau,\bar{\tau})\, \nabla_{m}^{(0)}\nabla_{n}^{(0)}Q^{(0)}\, .
\end{equation}
Finally, the $10$D Weyl rescaling $\Phi$ is determined to be
\begin{equation}\label{eq:WeylRescTenDBackground} 
\Phi^{(1)}=-3\cdot 2^{6}f_{0}(\tau,\bar{\tau})\, Q^{(0)}\, .
\end{equation}

When reducing the classical action \eqref{eq:ch1:ActionIIB10D},
we perform the Weyl rescaling
\begin{equation}
g_{MN}=\mathrm{e}^{\alpha \Phi^{(1)}}\tilde{g}_{MN}\, .
\end{equation}
Then,
we obtain
\begin{align}\label{eq:BackReact10DWeyl} 
S^{(0)}(g)&=S^{(0)}(\tilde{g})+\dfrac{\alpha}{2\kappa_{10}^{2}}\,\int\,  \Phi^{(1)}\left ( 4 R-8 |\cP|^{2}-\dfrac{|G_{3}|^{2}}{6}\right )\tilde{\star}_{10}\mathds{1}
\end{align}
where $S^{(0)}(\tilde{g})$ is the classical action evaluated on the new metric.
The term $\sim \Phi^{(1)} |G_{3}|^{2}$ contributes to $\delta_{0}$ in \eqref{eq:BackReactQuarticRedFD}.

\subsection{Details on the derivation of 4D kinetic terms}\label{sec:Reduction4DCaclDet}

In this section,
we provide further details on the reduction to 4D.
We reduce \eqref{eq:BackReact10DWeyl} as well as the relevant terms in \eqref{eq:FullResultG2R3TreeLoop} and \eqref{eq:FourPointCouplingsAP3}.
For the moment,
we ignore terms involving $F_{5}$ which complement the hypermultiplets in 4D as well as contribute warping terms.
They are known explicitly by means of \eqref{eq:T24DefinitionF5} and will be studied in more detail in the future.

For the Einstein Hilbert term,
the reduction of \eqref{eq:BackReact10DWeyl} to 4D leads to
\begin{align}
\int_{X_{3}}\,  4\Phi^{(1)} R\, \tilde{\star}_{10}1&=-384\, \biggl [f_{0}2(2\pi)^{3}\chi(X_{3})R^{(4)}+f_{0}\left (\mathscr{R}_{\alpha\beta}+2\mathscr{I}_{\alpha\beta}\right )\dif t^{\alpha}\wedge\star_{4}\dif t^{\beta}\nn\\
&\hphantom{=-384\, \biggl [}-6 \mathscr{I}_{\alpha}\dif t^{\alpha}\wedge\star_{4}(f_{1}\cP+f_{-1}\ov \cP)\biggl]
\end{align}
where we defined
\begin{align}
\label{eq:DefIA} \mathscr{I}_{\alpha}&=-\I (2\pi)^{3}\int_{X_{3}}\, \omega_{\alpha\, i}\,^{i}\, c_{3}(X_{3})=(2\pi)^{3}\chi(X_{3})\dfrac{k_{\alpha}}{\cV}\, ,\\
\mathscr{R}_{\alpha\beta}&=(2\pi)^{3}\int_{X_{3}}\, \omega_{\alpha\, i\bar{\jmath}}\, \omega_{\beta}\,^{\bar{\jmath}i}\, c_{3}(X_{3})\, ,\\
\label{eq:DefIAB}\mathscr{I}_{\alpha\beta}&=(2\pi)^{3}\int_{X_{3}}\, \omega_{\alpha\, i}\,^{i}\, \omega_{\beta\, j}\,^{j}\, c_{3}(X_{3})=-\dfrac{k_{\alpha}k_{\beta}}{\cV}(2\pi)^{3}\chi(X_{3})\, .
\end{align}
As observed in \cite{Grimm:2017okk},
$\mathscr{R}_{\alpha\beta}$ cancels in the reduction and only integrals of the form $\mathscr{I}_{\alpha}$, $\mathscr{I}_{\alpha\beta}$ appear which can be evaluated explicitly given that the trace of $(1,1)$-forms is constant.

From $R^{4}$,
we obtain in the reduction
\begin{equation}
\int_{X_{3}}\, \left (t_{8}t_{8}\pm \dfrac{1}{4}\epsilon_{8}\epsilon_{8}\right )R^{4}\tilde{\star}_{10}1=\pm 768(2\pi)^{3}\chi(X_{3})R^{(4)}+ 384\mathscr{R}_{\alpha\beta}\dif t^{\alpha}\wedge\star_{4}\dif t^{\beta}\, .
\end{equation}
As a remark, recall that in Type IIA both sign combinations appear giving rise to $(a_{T}-a_{L})\chi(X_{3})R^{(4)}$ in the reduction.
Ultimately, this ensures that the vectormultiplets are only corrected at tree level, while hypermultiplets are corrected at 1-loop.
In contrast,
we find in Type IIB only a single sign corresponding to $t_{16}$ defined in \eqref{eq:J0T16R4} so that
\begin{align}
\int_{X_{3}}\,   \left [4\Phi^{(1)}  R+f_{0} t_{16}R^{4}\right ]\tilde{\star}_{10}1&=-1536(2\pi)^{3}\chi(X_{3})\, f_{0}\, R^{(4)}-768\, f_{0} \mathscr{I}_{\alpha\beta}\dif t^{\alpha}\wedge\star_{4}\dif t^{\beta}\nn\\
&\quad+6\cdot 384 \mathscr{I}_{\alpha}\dif t^{\alpha}\wedge\star_{4}(f_{1}\cP+f_{-1}\ov \cP)\, .
\end{align}
As we will see below,
this implies that the hypermultiplets are corrected at both tree and 1-loop level, while the vectormultiplets remain uncorrected.

Next, let us look at the contribution from the 3-form.
The backreaction from the metric gives rise to
\begin{equation}
\int_{X_{3}}\, \Phi^{(1)}\dfrac{|G_{3}|^{2}}{6}\tilde{\star}_{10}1=384\ee^{\phi}  f_{0}\,  \mathscr{R}_{\alpha\beta} G^{\alpha}\wedge\star_{4}\ov G^{\beta}
\end{equation}
in terms of $G^{\alpha}=\dif c^{\alpha}-\tau\dif b^{\alpha}$.
From the higher derivative terms,
we find from the torsionful Riemann tensor (essentially equivalent to \cite{Grimm:2017okk})
\begin{equation}
\int_{X_{3}}\, 2f_{0}\tilde{t}_{8}\tilde{t}_{8} \left (|G_{3}|^{2}R^{3}+3|\nabla G_{3}|^{2}R^{2}\right )= 384\ee^{\phi}  f_{0}\,  \mathscr{R}_{\alpha\beta} G^{\alpha}\wedge\star_{4}\ov G^{\beta}
\end{equation}
as well as from $t_{18}$ (this is the piece proposed by \cite{Grimm:2017okk} at NSNS tree level)
\begin{equation}
\int_{X_{3}}\,  \dfrac{1}{2}t_{8}t_{8}|G_{3}|^{2}R^{3}=- 192\ee^{\phi}  f_{0}\,  \mathscr{I}_{\alpha\beta} G^{\alpha}\wedge\star_{4}\ov G^{\beta}\, .
\end{equation}
Altogether,
this amounts to
\begin{align}
\int_{X_{3}}\,  \left [-\Phi^{(1)}\dfrac{|G_{3}|^{2}}{6}+f_{0}\left (2\tilde{t}_{8}\tilde{t}_{8}-\dfrac{1}{2}t_{8}t_{8}\right )|G_{3}|^{2}R^{3}\right ]\tilde{\star}_{10}1=192\ee^{\phi} f_{0}\,  \mathscr{I}_{\alpha\beta}G^{\alpha}\wedge\star_{4}\ov G^{\beta}\, .
\end{align}
Notice that $\mathscr{R}_{\alpha\beta}$ cancels out exactly which is actually necessary to perform the remaining integrals explicitly as we will see below.
In addition,
we also have contributions from the 10D MUV sector which are of the form
\begin{equation}
\int_{X_{3}}\,  \left [\dfrac{3f_{1}}{4}t_{8}t_{8}G_{3}^{2}R^{3}+\text{c.c.}\right ]\tilde{\star}_{10}1=-288\ee^{\phi}  f_{1}\,  \mathscr{I}_{\alpha\beta} G^{\alpha}\wedge\star_{4} G^{\beta}+\text{c.c.}\, .
\end{equation}

To complete the argument,
we also have to add terms involving the dilaton.
At the 5-point level,
contact terms with two dilatons and three gravitons can only be $\mathrm{U}(1)$-preserving.\footnote{In fact,
as we will see below, we can turn this argument around by arguing that terms like $f_{2}\cP^{2}R^{3}$ in 10D are actually forbidden by 4D SUSY.}
They remain to large extent unspecified, see however \cite{Minasian:2015bxa} for a proposal based on 12D convariance.
Here,
we make an ansatz similar to \cite{Bonetti:2016dqh} by adding a term proportional to the $6$D Euler density, namely
\begin{equation}
\int_{X_{3}}\, \left [-8\Phi^{(1)} |\cP|^{2}-3\cdot 2^{10}|\cP|^{2}Q \right ]\tilde{\star}_{10}1=-1536 (2\pi)^{3}\chi(X_{3})\, f_{0}\,  \cP\wedge\star_{4}\ov \cP\, .
\end{equation}
To summarise,
we obtain the 4D action
\begin{align}
S^{(4)}&=\dfrac{1}{2\kappa_{10}^{2}}\int \biggl \{\left (\cV-1536\alpha(2\pi)^{3}\chi(X_{3})\, f_{0}\right )R^{(4)}\star_{4}1-(V_{\text{Flux}}+V_{\zeta})\star_{4}1\nn\\
&\quad-\left (2\cV+1536\alpha (2\pi)^{3}\chi(X_{3})\, f_{0}\right ) |\cP|^{2}\star_{4}1+6\cdot 384\alpha \mathscr{I}_{\alpha}\dif  t^{\alpha}\wedge\star_{4}(f_{1}\cP+f_{-1}\ov \cP)\nn\\
&\quad  +\left (\dfrac{1}{2}\left [k_{\alpha\beta}+\dfrac{k_{\alpha}k_{\beta}}{\cV}\right ]-768\alpha\, f_{0} \mathscr{I}_{\alpha\beta}\right )\dif t^{\alpha}\wedge\star_{4}\dif  t^{\beta}\nn\\
&\quad+\left (\dfrac{1}{2}\left [k_{\alpha\beta}-\dfrac{k_{\alpha}k_{\beta}}{\cV}\right ]+ 192\alpha\ee^{\phi} f_{0}\,  \mathscr{I}_{\alpha\beta}\right ) G^{\alpha}\wedge\star_{4}\ov G^{\beta}\nn\\
&\quad-288\alpha\ee^{\phi}  f_{1}\,  \mathscr{I}_{\alpha\beta} G^{\alpha}\wedge\star_{4} G^{\beta}+\text{c.c.}\biggl \}\, .
\end{align}

Up to this point,
we collected all the relevant contributions at the 2-derivative level in 4D.
The final step is to perform a Weyl rescaling of the 4D metric to arrive at 4D Einstein frame.
To this end,
we define
\begin{equation}
g_{\mu\nu}=\ee^{\kappa/2}\tilde{g}_{\mu\nu}\kom \kappa=-2\log(\cY)\kom \cY=\cV- (2\pi)^{3} \chi(X_{3})\dfrac{f_{0}}{8}
\end{equation}
and expand to liner order in $\chi$.
In string units,
we set
\begin{equation}
\ell_{s}=2\pi\sqrt{\alpha^{\prime}}=1\quad\Rightarrow\quad (\alpha^{\prime})^{3}=\dfrac{1}{(2\pi)^{6}}\kom \zeta=-\dfrac{\chi(X_{3})}{2(2\pi)^{3}}
\end{equation}
to arrive at (dropping the tilde on $\tilde{g}$ again) \eqref{eq:Action4DKinTFin}.

\vskip 1cm

%
\bibliographystyle{JHEP}
\bibliography{Literatur}

\providecommand{\href}[2]{#2}\begingroup\raggedright\begin{thebibliography}{10}

\bibitem{Gross:1986mw}
D.~J. Gross and J.~H. Sloan, \emph{{The Quartic Effective Action for the
  Heterotic String}},
  \href{https://doi.org/10.1016/0550-3213(87)90465-2}{\emph{Nucl. Phys. B}
  {\bfseries 291} (1987) 41--89}.

\bibitem{Gross:1986iv}
D.~J. Gross and E.~Witten, \emph{{Superstring Modifications of Einstein's
  Equations}}, \href{https://doi.org/10.1016/0550-3213(86)90429-3}{\emph{Nucl.
  Phys. B} {\bfseries 277} (1986) 1}.

\bibitem{Peeters:2001ub}
K.~Peeters, P.~Vanhove and A.~Westerberg, \emph{{Chiral splitting and world
  sheet gravitinos in higher derivative string amplitudes}},
  \href{https://doi.org/10.1088/0264-9381/19/10/312}{\emph{Class. Quant. Grav.}
  {\bfseries 19} (2002) 2699--2716},
  [\href{https://arxiv.org/abs/hep-th/0112157}{{\ttfamily hep-th/0112157}}].

\bibitem{Liu:2013dna}
J.~T. Liu and R.~Minasian, \emph{{Higher-derivative couplings in string theory:
  dualities and the B-field}},
  \href{https://doi.org/10.1016/j.nuclphysb.2013.06.002}{\emph{Nucl. Phys.}
  {\bfseries B874} (2013) 413--470},
  [\href{https://arxiv.org/abs/1304.3137}{{\ttfamily 1304.3137}}].

\bibitem{Liu:2019ses}
J.~T. Liu and R.~Minasian, \emph{{Higher-derivative couplings in string theory:
  five-point contact terms}},
  \href{https://doi.org/10.1016/j.nuclphysb.2021.115386}{\emph{Nucl. Phys. B}
  {\bfseries 967} (2021) 115386},
  [\href{https://arxiv.org/abs/1912.10974}{{\ttfamily 1912.10974}}].

\bibitem{Policastro:2006vt}
G.~Policastro and D.~Tsimpis, \emph{{R**4, purified}},
  \href{https://doi.org/10.1088/0264-9381/23/14/012}{\emph{Class. Quant. Grav.}
  {\bfseries 23} (2006) 4753--4780},
  [\href{https://arxiv.org/abs/hep-th/0603165}{{\ttfamily hep-th/0603165}}].

\bibitem{Policastro:2008hg}
G.~Policastro and D.~Tsimpis, \emph{{A Note on the quartic effective action of
  type IIB superstring}},
  \href{https://doi.org/10.1088/0264-9381/26/12/125001}{\emph{Class. Quant.
  Grav.} {\bfseries 26} (2009) 125001},
  [\href{https://arxiv.org/abs/0812.3138}{{\ttfamily 0812.3138}}].

\bibitem{Green:2003an}
M.~B. Green and C.~Stahn, \emph{{D3-branes on the Coulomb branch and
  instantons}},
  \href{https://doi.org/10.1088/1126-6708/2003/09/052}{\emph{JHEP} {\bfseries
  09} (2003) 052}, [\href{https://arxiv.org/abs/hep-th/0308061}{{\ttfamily
  hep-th/0308061}}].

\bibitem{Paulos:2008tn}
M.~F. Paulos, \emph{{Higher derivative terms including the Ramond-Ramond
  five-form}}, \href{https://doi.org/10.1088/1126-6708/2008/10/047}{\emph{JHEP}
  {\bfseries 10} (2008) 047},
  [\href{https://arxiv.org/abs/0804.0763}{{\ttfamily 0804.0763}}].

\bibitem{deHaro:2002vk}
S.~de~Haro, A.~Sinkovics and K.~Skenderis, \emph{{On a supersymmetric
  completion of the R4 term in 2B supergravity}},
  \href{https://doi.org/10.1103/PhysRevD.67.084010}{\emph{Phys. Rev. D}
  {\bfseries 67} (2003) 084010},
  [\href{https://arxiv.org/abs/hep-th/0210080}{{\ttfamily hep-th/0210080}}].

\bibitem{Rajaraman:2005ag}
A.~Rajaraman, \emph{{On the supersymmetric completion of the R**4 term in
  M-theory}}, \href{https://doi.org/10.1103/PhysRevD.72.125008}{\emph{Phys.
  Rev. D} {\bfseries 74} (2006) 085018},
  [\href{https://arxiv.org/abs/hep-th/0512333}{{\ttfamily hep-th/0512333}}].

\bibitem{Howe:1983sra}
P.~S. Howe and P.~C. West, \emph{{The Complete N=2, D=10 Supergravity}},
  \href{https://doi.org/10.1016/0550-3213(84)90472-3}{\emph{Nucl. Phys. B}
  {\bfseries 238} (1984) 181--220}.

\bibitem{Melo:2020amq}
J.~a.~F. Melo and J.~E. Santos, \emph{{Stringy corrections to the entropy of
  electrically charged supersymmetric black holes with $\mathrm{AdS}_5\times
  S^5$ asymptotics}},
  \href{https://doi.org/10.1103/PhysRevD.103.066008}{\emph{Phys. Rev. D}
  {\bfseries 103} (2021) 066008},
  [\href{https://arxiv.org/abs/2007.06582}{{\ttfamily 2007.06582}}].

\bibitem{Garousi:2020lof}
M.~R. Garousi, \emph{{On NS-NS couplings at order \ensuremath{\alpha}'3}},
  \href{https://doi.org/10.1016/j.nuclphysb.2021.115510}{\emph{Nucl. Phys. B}
  {\bfseries 971} (2021) 115510},
  [\href{https://arxiv.org/abs/2012.15091}{{\ttfamily 2012.15091}}].

\bibitem{Green:1997as}
M.~B. Green, M.~Gutperle and P.~Vanhove, \emph{{One loop in
  eleven-dimensions}},
  \href{https://doi.org/10.1016/S0370-2693(97)00931-3}{\emph{Phys. Lett.}
  {\bfseries B409} (1997) 177--184},
  [\href{https://arxiv.org/abs/hep-th/9706175}{{\ttfamily hep-th/9706175}}].

\bibitem{Green:1997me}
M.~B. Green, M.~Gutperle and H.-h. Kwon, \emph{{Sixteen fermion and related
  terms in M theory on T**2}},
  \href{https://doi.org/10.1016/S0370-2693(97)01551-7}{\emph{Phys. Lett. B}
  {\bfseries 421} (1998) 149--161},
  [\href{https://arxiv.org/abs/hep-th/9710151}{{\ttfamily hep-th/9710151}}].

\bibitem{Green:1999by}
M.~B. Green, M.~Gutperle and H.~H. Kwon, \emph{{Light cone quantum mechanics of
  the eleven-dimensional superparticle}},
  \href{https://doi.org/10.1088/1126-6708/1999/08/012}{\emph{JHEP} {\bfseries
  08} (1999) 012}, [\href{https://arxiv.org/abs/hep-th/9907155}{{\ttfamily
  hep-th/9907155}}].

\bibitem{Green:1997tv}
M.~B. Green and M.~Gutperle, \emph{{Effects of D instantons}},
  \href{https://doi.org/10.1016/S0550-3213(97)00269-1}{\emph{Nucl. Phys. B}
  {\bfseries 498} (1997) 195--227},
  [\href{https://arxiv.org/abs/hep-th/9701093}{{\ttfamily hep-th/9701093}}].

\bibitem{Kehagias:1997jg}
A.~Kehagias and H.~Partouche, \emph{{D instanton corrections as (p,q) string
  effects and nonrenormalization theorems}},
  \href{https://doi.org/10.1142/S0217751X98002365}{\emph{Int. J. Mod. Phys. A}
  {\bfseries 13} (1998) 5075--5092},
  [\href{https://arxiv.org/abs/hep-th/9712164}{{\ttfamily hep-th/9712164}}].

\bibitem{Alessio:2021krn}
F.~Alessio, G.~Barnich and M.~Bonte, \emph{{Notes on massless scalar field
  partition functions, modular invariance and Eisenstein series}},
  \href{https://doi.org/10.1007/JHEP12(2021)211}{\emph{JHEP} {\bfseries 12}
  (2021) 211}, [\href{https://arxiv.org/abs/2111.03164}{{\ttfamily
  2111.03164}}].

\bibitem{Sen:2021tpp}
A.~Sen, \emph{{Normalization of type IIB D-instanton amplitudes}},
  \href{https://doi.org/10.1007/JHEP12(2021)146}{\emph{JHEP} {\bfseries 12}
  (2021) 146}, [\href{https://arxiv.org/abs/2104.11109}{{\ttfamily
  2104.11109}}].

\bibitem{Sen:2021jbr}
A.~Sen, \emph{{Muti-instanton amplitudes in type IIB string theory}},
  \href{https://doi.org/10.1007/JHEP12(2021)065}{\emph{JHEP} {\bfseries 12}
  (2021) 065}, [\href{https://arxiv.org/abs/2104.15110}{{\ttfamily
  2104.15110}}].

\bibitem{Green:1998by}
M.~B. Green and S.~Sethi, \emph{{Supersymmetry constraints on type IIB
  supergravity}}, \href{https://doi.org/10.1103/PhysRevD.59.046006}{\emph{Phys.
  Rev. D} {\bfseries 59} (1999) 046006},
  [\href{https://arxiv.org/abs/hep-th/9808061}{{\ttfamily hep-th/9808061}}].

\bibitem{Green:2019rhz}
M.~B. Green and C.~Wen, \emph{{Modular Forms and $SL(2, {\mathbb
  Z})$-covariance of type IIB superstring theory}},
  \href{https://doi.org/10.1007/JHEP06(2019)087}{\emph{JHEP} {\bfseries 06}
  (2019) 087}, [\href{https://arxiv.org/abs/1904.13394}{{\ttfamily
  1904.13394}}].

\bibitem{Caron-Huot:2010nes}
S.~Caron-Huot and D.~O'Connell, \emph{{Spinor Helicity and Dual Conformal
  Symmetry in Ten Dimensions}},
  \href{https://doi.org/10.1007/JHEP08(2011)014}{\emph{JHEP} {\bfseries 08}
  (2011) 014}, [\href{https://arxiv.org/abs/1010.5487}{{\ttfamily 1010.5487}}].

\bibitem{Boels:2012ie}
R.~H. Boels and D.~O'Connell, \emph{{Simple superamplitudes in higher
  dimensions}}, \href{https://doi.org/10.1007/JHEP06(2012)163}{\emph{JHEP}
  {\bfseries 06} (2012) 163},
  [\href{https://arxiv.org/abs/1201.2653}{{\ttfamily 1201.2653}}].

\bibitem{Boels:2012zr}
R.~H. Boels, \emph{{Maximal R-symmetry violating amplitudes in type IIB
  superstring theory}},
  \href{https://doi.org/10.1103/PhysRevLett.109.081602}{\emph{Phys. Rev. Lett.}
  {\bfseries 109} (2012) 081602},
  [\href{https://arxiv.org/abs/1204.4208}{{\ttfamily 1204.4208}}].

\bibitem{Boels:2013jua}
R.~H. Boels, \emph{{On the field theory expansion of superstring five point
  amplitudes}},
  \href{https://doi.org/10.1016/j.nuclphysb.2013.08.009}{\emph{Nucl. Phys. B}
  {\bfseries 876} (2013) 215--233},
  [\href{https://arxiv.org/abs/1304.7918}{{\ttfamily 1304.7918}}].

\bibitem{Wang:2015jna}
Y.~Wang and X.~Yin, \emph{{Constraining Higher Derivative Supergravity with
  Scattering Amplitudes}},
  \href{https://doi.org/10.1103/PhysRevD.92.041701}{\emph{Phys. Rev. D}
  {\bfseries 92} (2015) 041701},
  [\href{https://arxiv.org/abs/1502.03810}{{\ttfamily 1502.03810}}].

\bibitem{Nahm:1977tg}
W.~Nahm, \emph{{Supersymmetries and their Representations}},
  \href{https://doi.org/10.1016/0550-3213(78)90218-3}{\emph{Nucl. Phys. B}
  {\bfseries 135} (1978) 149}.

\bibitem{Green:1982tk}
M.~B. Green and J.~H. Schwarz, \emph{{Extended Supergravity in
  Ten-Dimensions}},
  \href{https://doi.org/10.1016/0370-2693(83)90781-5}{\emph{Phys. Lett. B}
  {\bfseries 122} (1983) 143--147}.

\bibitem{Schwarz:1983wa}
J.~H. Schwarz and P.~C. West, \emph{{Symmetries and Transformations of Chiral
  N=2 D=10 Supergravity}},
  \href{https://doi.org/10.1016/0370-2693(83)90168-5}{\emph{Phys. Lett. B}
  {\bfseries 126} (1983) 301--304}.

\bibitem{Schwarz:1983qr}
J.~H. Schwarz, \emph{{Covariant Field Equations of Chiral N=2 D=10
  Supergravity}},
  \href{https://doi.org/10.1016/0550-3213(83)90192-X}{\emph{Nucl. Phys. B}
  {\bfseries 226} (1983) 269}.

\bibitem{Douglas:1996yp}
M.~R. Douglas, D.~N. Kabat, P.~Pouliot and S.~H. Shenker, \emph{{D-branes and
  short distances in string theory}},
  \href{https://doi.org/10.1016/S0550-3213(96)00619-0}{\emph{Nucl. Phys. B}
  {\bfseries 485} (1997) 85--127},
  [\href{https://arxiv.org/abs/hep-th/9608024}{{\ttfamily hep-th/9608024}}].

\bibitem{Green:1997di}
M.~B. Green and P.~Vanhove, \emph{{D instantons, strings and M theory}},
  \href{https://doi.org/10.1016/S0370-2693(97)00785-5}{\emph{Phys. Lett. B}
  {\bfseries 408} (1997) 122--134},
  [\href{https://arxiv.org/abs/hep-th/9704145}{{\ttfamily hep-th/9704145}}].

\bibitem{Green:1997tn}
M.~B. Green and M.~Gutperle, \emph{{D Particle bound states and the D instanton
  measure}}, \href{https://doi.org/10.1088/1126-6708/1998/01/005}{\emph{JHEP}
  {\bfseries 01} (1998) 005},
  [\href{https://arxiv.org/abs/hep-th/9711107}{{\ttfamily hep-th/9711107}}].

\bibitem{Green:1998yf}
M.~B. Green and M.~Gutperle, \emph{{D instanton partition functions}},
  \href{https://doi.org/10.1103/PhysRevD.58.046007}{\emph{Phys. Rev. D}
  {\bfseries 58} (1998) 046007},
  [\href{https://arxiv.org/abs/hep-th/9804123}{{\ttfamily hep-th/9804123}}].

\bibitem{deRoo:1992sm}
M.~de~Roo, H.~Suelmann and A.~Wiedemann, \emph{{Supersymmetric R**4 actions in
  ten-dimensions}},
  \href{https://doi.org/10.1016/0370-2693(92)90769-Z}{\emph{Phys. Lett. B}
  {\bfseries 280} (1992) 39--46}.

\bibitem{Minasian:2015bxa}
R.~Minasian, T.~G. Pugh and R.~Savelli, \emph{{F-theory at order $\alpha'^3$}},
  \href{https://doi.org/10.1007/JHEP10(2015)050}{\emph{JHEP} {\bfseries 10}
  (2015) 050}, [\href{https://arxiv.org/abs/1506.06756}{{\ttfamily
  1506.06756}}].

\bibitem{Schwarz:1982jn}
J.~H. Schwarz, \emph{{Superstring Theory}},
  \href{https://doi.org/10.1016/0370-1573(82)90087-4}{\emph{Phys. Rept.}
  {\bfseries 89} (1982) 223--322}.

\bibitem{Sakai:1986bi}
N.~Sakai and Y.~Tanii, \emph{{One Loop Amplitudes and Effective Action in
  Superstring Theories}},
  \href{https://doi.org/10.1016/0550-3213(87)90114-3}{\emph{Nucl. Phys. B}
  {\bfseries 287} (1987) 457}.

\bibitem{Green:2005qr}
M.~B. Green, K.~Peeters and C.~Stahn, \emph{{Superfield integrals in high
  dimensions}},
  \href{https://doi.org/10.1088/1126-6708/2005/08/093}{\emph{JHEP} {\bfseries
  08} (2005) 093}, [\href{https://arxiv.org/abs/hep-th/0506161}{{\ttfamily
  hep-th/0506161}}].

\bibitem{Richards:2008sa}
D.~M. Richards, \emph{{The One-Loop H**2 R**3 and H**2(Delta H)R-2 Terms in the
  Effective Action}},
  \href{https://doi.org/10.1088/1126-6708/2008/10/043}{\emph{JHEP} {\bfseries
  10} (2008) 043}, [\href{https://arxiv.org/abs/0807.3453}{{\ttfamily
  0807.3453}}].

\bibitem{Nilsson:1981bn}
B.~E.~W. Nilsson, \emph{{Simple Ten-dimensional Supergravity in Superspace}},
  \href{https://doi.org/10.1016/0550-3213(81)90111-5}{\emph{Nucl. Phys. B}
  {\bfseries 188} (1981) 176--192}.

\bibitem{Nilsson:1986rh}
B.~E.~W. Nilsson and A.~K. Tollsten, \emph{{Supersymmetrization of Zeta (3) (R
  $\mu \nu \rho \sigma$)**4 in Superstring Theories}},
  \href{https://doi.org/10.1016/0370-2693(86)91255-4}{\emph{Phys. Lett. B}
  {\bfseries 181} (1986) 63--66}.

\bibitem{Kallosh:1987mb}
R.~Kallosh, \emph{{Strings and Superspace}},
  \href{https://doi.org/10.1088/0031-8949/1987/T15/015}{\emph{Phys. Scripta T}
  {\bfseries 15} (1987) 118--125}.

\bibitem{Peeters:2000qj}
K.~Peeters, P.~Vanhove and A.~Westerberg, \emph{{Supersymmetric higher
  derivative actions in ten-dimensions and eleven-dimensions, the associated
  superalgebras and their formulation in superspace}},
  \href{https://doi.org/10.1088/0264-9381/18/5/307}{\emph{Class. Quant. Grav.}
  {\bfseries 18} (2001) 843--890},
  [\href{https://arxiv.org/abs/hep-th/0010167}{{\ttfamily hep-th/0010167}}].

\bibitem{Peeters:2005tb}
K.~Peeters, J.~Plefka and S.~Stern, \emph{{Higher-derivative gauge field terms
  in the M-theory action}},
  \href{https://doi.org/10.1088/1126-6708/2005/08/095}{\emph{JHEP} {\bfseries
  08} (2005) 095}, [\href{https://arxiv.org/abs/hep-th/0507178}{{\ttfamily
  hep-th/0507178}}].

\bibitem{Dasgupta:2000df}
A.~Dasgupta, H.~Nicolai and J.~Plefka, \emph{{Vertex operators for the
  supermembrane}},
  \href{https://doi.org/10.1088/1126-6708/2000/05/007}{\emph{JHEP} {\bfseries
  05} (2000) 007}, [\href{https://arxiv.org/abs/hep-th/0003280}{{\ttfamily
  hep-th/0003280}}].

\bibitem{Gran:2001yh}
U.~Gran, \emph{{GAMMA: A Mathematica package for performing gamma matrix
  algebra and Fierz transformations in arbitrary dimensions}},
  \href{https://arxiv.org/abs/hep-th/0105086}{{\ttfamily hep-th/0105086}}.

\bibitem{Bonetti:2016dqh}
F.~Bonetti and M.~Weissenbacher, \emph{{The Euler characteristic correction to
  the Kähler potential — revisited}},
  \href{https://doi.org/10.1007/JHEP01(2017)003}{\emph{JHEP} {\bfseries 01}
  (2017) 003}, [\href{https://arxiv.org/abs/1608.01300}{{\ttfamily
  1608.01300}}].

\bibitem{Becker:2002nn}
K.~Becker, M.~Becker, M.~Haack and J.~Louis, \emph{{Supersymmetry breaking and
  alpha-prime corrections to flux induced potentials}},
  \href{https://doi.org/10.1088/1126-6708/2002/06/060}{\emph{JHEP} {\bfseries
  06} (2002) 060}, [\href{https://arxiv.org/abs/hep-th/0204254}{{\ttfamily
  hep-th/0204254}}].

\bibitem{Garousi:2020mqn}
M.~R. Garousi, \emph{{Minimal gauge invariant couplings at order $\alpha '^3$:
  NS\textendash{}NS fields}},
  \href{https://doi.org/10.1140/epjc/s10052-020-08662-9}{\emph{Eur. Phys. J. C}
  {\bfseries 80} (2020) 1086},
  [\href{https://arxiv.org/abs/2006.09193}{{\ttfamily 2006.09193}}].

\bibitem{Garousi:2020gio}
M.~R. Garousi, \emph{{Effective action of type II superstring theories at order
  $\alpha'^{3}$: NS-NS couplings}},
  \href{https://doi.org/10.1007/JHEP02(2021)157}{\emph{JHEP} {\bfseries 02}
  (2021) 157}, [\href{https://arxiv.org/abs/2011.02753}{{\ttfamily
  2011.02753}}].

\bibitem{Conlon:2005ki}
J.~P. Conlon, F.~Quevedo and K.~Suruliz, \emph{{Large-volume flux
  compactifications: Moduli spectrum and D3/D7 soft supersymmetry breaking}},
  \href{https://doi.org/10.1088/1126-6708/2005/08/007}{\emph{JHEP} {\bfseries
  08} (2005) 007}, [\href{https://arxiv.org/abs/hep-th/0505076}{{\ttfamily
  hep-th/0505076}}].

\bibitem{Cicoli:2021rub}
M.~Cicoli, F.~Quevedo, R.~Savelli, A.~Schachner and R.~Valandro,
  \emph{{Systematics of the $\alpha'$ Expansion in F-Theory}},
  \href{https://doi.org/10.1007/JHEP08(2021)099}{\emph{JHEP} {\bfseries 2021}
  (2021) 99}, [\href{https://arxiv.org/abs/2106.04592}{{\ttfamily
  2106.04592}}].

\bibitem{Grimm:2017okk}
T.~W. Grimm, K.~Mayer and M.~Weissenbacher, \emph{{Higher derivatives in Type
  II and M-theory on Calabi-Yau threefolds}},
  \href{https://doi.org/10.1007/JHEP02(2018)127}{\emph{JHEP} {\bfseries 02}
  (2018) 127}, [\href{https://arxiv.org/abs/1702.08404}{{\ttfamily
  1702.08404}}].

\bibitem{van1992lie}
M.~A. Van~Leeuwen, A.~M. Cohen and B.~Lisser, \emph{Lie: A package for lie
  group computations}, .

\bibitem{van1994lie}
M.~Van~Leeuwen, \emph{L{\i}e, a software package for lie group computations},
  {\emph{Euromath Bull} {\bfseries 1} (1994) 83--94}.

\bibitem{Green:2020eyj}
M.~B. Green and C.~Wen, \emph{{Maximal U(1)$_{Y}$-violating n-point correlators
  in $ \mathcal{N} $ = 4 super-Yang-Mills theory}},
  \href{https://doi.org/10.1007/JHEP02(2021)042}{\emph{JHEP} {\bfseries 02}
  (2021) 042}, [\href{https://arxiv.org/abs/2009.01211}{{\ttfamily
  2009.01211}}].

\bibitem{Lin:2015dsa}
Y.-H. Lin, S.-H. Shao, Y.~Wang and X.~Yin, \emph{{Supersymmetry Constraints and
  String Theory on K3}},
  \href{https://doi.org/10.1007/JHEP12(2015)142}{\emph{JHEP} {\bfseries 12}
  (2015) 142}, [\href{https://arxiv.org/abs/1508.07305}{{\ttfamily
  1508.07305}}].

\bibitem{Giddings:2001yu}
S.~B. Giddings, S.~Kachru and J.~Polchinski, \emph{{Hierarchies from fluxes in
  string compactifications}},
  \href{https://doi.org/10.1103/PhysRevD.66.106006}{\emph{Phys. Rev. D}
  {\bfseries 66} (2002) 106006},
  [\href{https://arxiv.org/abs/hep-th/0105097}{{\ttfamily hep-th/0105097}}].

\bibitem{Antoniadis:1997eg}
I.~Antoniadis, S.~Ferrara, R.~Minasian and K.~S. Narain, \emph{{R**4 couplings
  in M and type II theories on Calabi-Yau spaces}},
  \href{https://doi.org/10.1016/S0550-3213(97)00572-5}{\emph{Nucl. Phys. B}
  {\bfseries 507} (1997) 571--588},
  [\href{https://arxiv.org/abs/hep-th/9707013}{{\ttfamily hep-th/9707013}}].

\bibitem{Gukov:1999ya}
S.~Gukov, C.~Vafa and E.~Witten, \emph{{CFT's from Calabi-Yau four folds}},
  \href{https://doi.org/10.1016/S0550-3213(00)00373-4}{\emph{Nucl. Phys. B}
  {\bfseries 584} (2000) 69--108},
  [\href{https://arxiv.org/abs/hep-th/9906070}{{\ttfamily hep-th/9906070}}].

\bibitem{Burgess:2020qsc}
C.~P. Burgess, M.~Cicoli, D.~Ciupke, S.~Krippendorf and F.~Quevedo, \emph{{UV
  Shadows in EFTs: Accidental Symmetries, Robustness and No-Scale
  Supergravity}}, \href{https://doi.org/10.1002/prop.202000076}{\emph{Fortsch.
  Phys.} {\bfseries 68} (2020) 2000076},
  [\href{https://arxiv.org/abs/2006.06694}{{\ttfamily 2006.06694}}].

\bibitem{Becker:2001pm}
K.~Becker and M.~Becker, \emph{{Supersymmetry breaking, M theory and fluxes}},
  \href{https://doi.org/10.1088/1126-6708/2001/07/038}{\emph{JHEP} {\bfseries
  07} (2001) 038}, [\href{https://arxiv.org/abs/hep-th/0107044}{{\ttfamily
  hep-th/0107044}}].

\bibitem{Grimm:2014xva}
T.~W. Grimm, T.~G. Pugh and M.~Weissenbacher, \emph{{On M-theory fourfold vacua
  with higher curvature terms}},
  \href{https://doi.org/10.1016/j.physletb.2015.02.047}{\emph{Phys. Lett.}
  {\bfseries B743} (2015) 284--289},
  [\href{https://arxiv.org/abs/1408.5136}{{\ttfamily 1408.5136}}].

\bibitem{Polchinski:1995sm}
J.~Polchinski and A.~Strominger, \emph{{New vacua for type II string theory}},
  \href{https://doi.org/10.1016/S0370-2693(96)01219-1}{\emph{Phys. Lett. B}
  {\bfseries 388} (1996) 736--742},
  [\href{https://arxiv.org/abs/hep-th/9510227}{{\ttfamily hep-th/9510227}}].

\bibitem{Michelson:1996pn}
J.~Michelson, \emph{{Compactifications of type IIB strings to four-dimensions
  with nontrivial classical potential}},
  \href{https://doi.org/10.1016/S0550-3213(97)00184-3}{\emph{Nucl. Phys. B}
  {\bfseries 495} (1997) 127--148},
  [\href{https://arxiv.org/abs/hep-th/9610151}{{\ttfamily hep-th/9610151}}].

\bibitem{Dasgupta:1999ss}
K.~Dasgupta, G.~Rajesh and S.~Sethi, \emph{{M theory, orientifolds and G -
  flux}}, \href{https://doi.org/10.1088/1126-6708/1999/08/023}{\emph{JHEP}
  {\bfseries 08} (1999) 023},
  [\href{https://arxiv.org/abs/hep-th/9908088}{{\ttfamily hep-th/9908088}}].

\bibitem{Grimm:2004uq}
T.~W. Grimm and J.~Louis, \emph{{The Effective action of N = 1 Calabi-Yau
  orientifolds}},
  \href{https://doi.org/10.1016/j.nuclphysb.2004.08.005}{\emph{Nucl. Phys. B}
  {\bfseries 699} (2004) 387--426},
  [\href{https://arxiv.org/abs/hep-th/0403067}{{\ttfamily hep-th/0403067}}].

\bibitem{Antoniadis:2003sw}
I.~Antoniadis, R.~Minasian, S.~Theisen and P.~Vanhove, \emph{{String loop
  corrections to the universal hypermultiplet}},
  \href{https://doi.org/10.1088/0264-9381/20/23/009}{\emph{Class. Quant. Grav.}
  {\bfseries 20} (2003) 5079--5102},
  [\href{https://arxiv.org/abs/hep-th/0307268}{{\ttfamily hep-th/0307268}}].

\bibitem{Grana:2014vva}
M.~Grana, J.~Louis, U.~Theis and D.~Waldram, \emph{{Quantum Corrections in
  String Compactifications on SU(3) Structure Geometries}},
  \href{https://doi.org/10.1007/JHEP01(2015)057}{\emph{JHEP} {\bfseries 01}
  (2015) 057}, [\href{https://arxiv.org/abs/1406.0958}{{\ttfamily 1406.0958}}].

\bibitem{Wulff:2021fhr}
L.~Wulff, \emph{{Completing $R^4$ using $O(d,d)$}},
  \href{https://arxiv.org/abs/2111.00018}{{\ttfamily 2111.00018}}.

\bibitem{Richards:2008jg}
D.~M. Richards, \emph{{The One-Loop Five-Graviton Amplitude and the Effective
  Action}}, \href{https://doi.org/10.1088/1126-6708/2008/10/042}{\emph{JHEP}
  {\bfseries 10} (2008) 042},
  [\href{https://arxiv.org/abs/0807.2421}{{\ttfamily 0807.2421}}].

\bibitem{Green:2005ba}
M.~B. Green and P.~Vanhove, \emph{{Duality and higher derivative terms in M
  theory}}, \href{https://doi.org/10.1088/1126-6708/2006/01/093}{\emph{JHEP}
  {\bfseries 01} (2006) 093},
  [\href{https://arxiv.org/abs/hep-th/0510027}{{\ttfamily hep-th/0510027}}].

\bibitem{Green:1981ya}
M.~B. Green and J.~H. Schwarz, \emph{{Supersymmetrical Dual String Theory. 3.
  Loops and Renormalization}},
  \href{https://doi.org/10.1016/0550-3213(82)90334-0}{\emph{Nucl.\ Phys.\ B}
  {\bfseries 198} (1982) 441--460}.

\bibitem{Rajaraman:2005up}
A.~Rajaraman, \emph{{On a supersymmetric completion of the R**4 term in type
  IIB supergravity}},
  \href{https://doi.org/10.1103/PhysRevD.74.085018}{\emph{Phys. Rev. D}
  {\bfseries 72} (2005) 125008},
  [\href{https://arxiv.org/abs/hep-th/0505155}{{\ttfamily hep-th/0505155}}].

\bibitem{Peeters:2003pv}
K.~Peeters and A.~Westerberg, \emph{{The Ramond-Ramond sector of string theory
  beyond leading order}},
  \href{https://doi.org/10.1088/0264-9381/21/6/022}{\emph{Class. Quant. Grav.}
  {\bfseries 21} (2004) 1643--1666},
  [\href{https://arxiv.org/abs/hep-th/0307298}{{\ttfamily hep-th/0307298}}].

\bibitem{Liu:2010gz}
J.~T. Liu and R.~Minasian, \emph{{Computing 1/$N^{2}$ corrections in AdS/CFT}},
   \href{https://arxiv.org/abs/1010.6074}{{\ttfamily 1010.6074}}.

\end{thebibliography}\endgroup
\end{document}